\PassOptionsToPackage{main=brazil,english}{babel}

\documentclass[
	final,				
	12pt,				
	openright,			
	oneside,			
	a4paper,			
	hyphens,            
	sumario=tradicional,
	english,			
	french,				
	spanish,			
	brazil				
	]{abntex2}

\usepackage{lmodern}			
\usepackage[T1]{fontenc}		
\usepackage[utf8]{inputenc}		
\usepackage{lastpage}			
\usepackage{indentfirst}		
\usepackage{color}			    
\usepackage{graphicx}			
\usepackage{microtype} 			
\usepackage{amsthm,thmtools}	
\usepackage{abntex2lncc}		

\usepackage[brazilian,hyperpageref]{backref}	 
\usepackage[alf]{abntex2cite}	
		
\usepackage{lipsum}				
\usepackage{amsfonts}
\usepackage{amsmath}
\usepackage{braket}
\usepackage{tikz}
\usetikzlibrary{decorations.pathreplacing}
\usetikzlibrary{calc}
\usepackage[portuguese,ruled,lined]{algorithm2e}
\usepackage{algorithmic}

\usepackage{cancel}

\usepackage{pdfpages} 

\renewcommand{\bra}[1]{\Bra{#1}}
\renewcommand{\ket}[1]{\Ket{#1}}
\renewcommand{\braket}[1]{\Braket{#1}}
\newcommand{\ii}{\textnormal{i}}
\newcommand{\F}[2]{\mathcal{F}_{#1}\pr{#2}}
\newcommand{\BF}{B_{\mathcal{F}}}
\newcommand{\QFT}{\textnormal{QFT}}
\newcommand{\QFTrec}{\textnormal{QFT}^\textnormal{rec}}
\newcommand{\eps}{\varepsilon}

\newcommand{\matrx}[1]{\left[\begin{matrix} #1 \end{matrix}\right]}
\newcommand{\pr}[1]{\left( #1 \right)} 
\newcommand{\paren}[1]{\left( #1 \right)} 
\newcommand{\cosp}[1]{\textrm{cos}\pr{#1}}
\newcommand{\sinp}[1]{\textrm{sin}\pr{#1}}

\newcommand{\hilb}{\mathcal{H}}
\newcommand{\floor}[1]{\left\lfloor #1 \right\rfloor}
\newcommand{\ceil}[1]{\left\lceil #1 \right\rceil}
\newcommand{\prob}[1]{\textnormal{prob}\pr{#1}}
\newcommand{\RF}{R^\mathcal{F}}
\newcommand{\control}{\mathcal{C}}

\newcommand{\pctrl}[1]{\control_\textnormal{pot}\pr{#1}}
\newcommand{\bctrl}[3]{\control_{#1, #2}\pr{#3}}
\newcommand{\SWAP}{\textnormal{SWAP}}
\newcommand{\card}[1]{\left|#1\right|}
\newcommand{\U}{\mathcal{U}}
\newcommand{\C}{\mathbb{C}}
\newcommand{\N}{\mathbb{N}}
\newcommand{\R}{\mathbb{R}}
\newcommand{\CNOT}{\textnormal{CNOT}}
\newcommand{\cor}[1]{\textnormal{c}\pr{#1}}

\renewcommand*{\backrefalt}[4]{
	\ifcase #1 %
		Nenhuma citação no texto.%
	\or
		Citado na página #2.%
	\else
		Citado #1 vezes nas páginas #2.%
	\fi}%

\declaretheorem[style=definition,name=Definição, parent=chapter, qed=\textemdash]{definition}

\declaretheorem[style=plain,name=Teorema, qed=\textnormal{\textemdash}]{theorem}

\declaretheorem[style=plain,name=Corolário, qed=\textnormal{\textemdash}]{corollary}

\declaretheorem[style=plain,name=Lema, qed=\textnormal{\textemdash}]{lemma}

\graphicspath{{fig/}}
\logoLNCC{lncc}

\dissertacaoMestrado

\titulo{Algoritmo de Contagem Quântico Aplicado ao Grafo Bipartido Completo}
\nomeAutor{Gustavo Alves}{Bezerra}
\nomeOrientador{Renato}{Portugal}
\coorientador{Raqueline Azevedo Medeiros Santos}

\local{Petrópolis, RJ - Brasil}
\data{Setembro de 2021}
\instituicao{%
  Laboratório Nacional de Computação Científica
  \par
  Programa de Pós-Graduação em Modelagem Computacional}

\preambulo{\tipoTrabalho submetida ao corpo docente do Laboratório Nacional de Computação Científica como parte dos requisitos necessários para a obtenção do grau de \grau em Ciências em Modelagem Computacional.}

\codebib{XXX.XXX}
\codetese{XXXX}

\definecolor{blue}{RGB}{41,5,195}

\makeatletter
\hypersetup{
		pdftitle={\@title},
		pdfauthor={\@author},
    	pdfsubject={\imprimirpreambulo},
	    pdfcreator={LaTeX with abnTeX2},
		pdfkeywords={abnt}{latex}{abntex}{abntex2}{trabalho acadêmico},
		colorlinks=true,       		
    	linkcolor=blue,          	
    	citecolor=blue,        		
    	filecolor=magenta,      		
		urlcolor=blue,
		bookmarksdepth=4
}
\makeatother

\makeindex

\begin{document}

\selectlanguage{brazil}

\frenchspacing


\imprimircapa

\imprimirfolhaderosto*
\includepdf{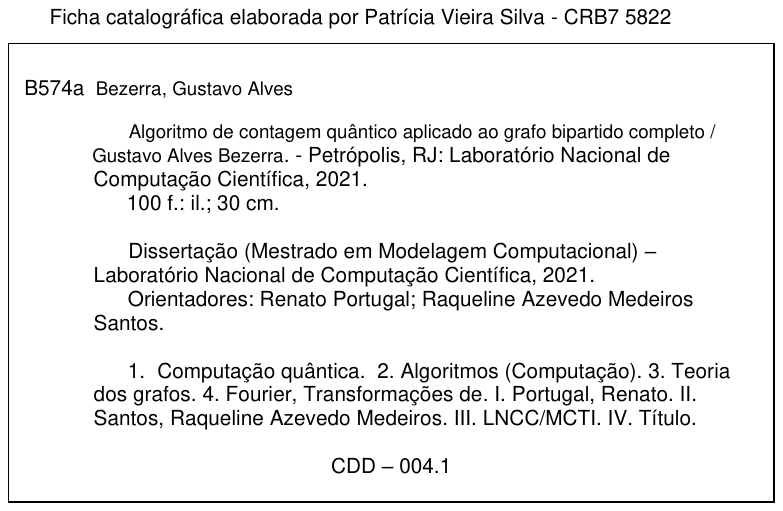}


%
%
\begin{folhadeaprovacao}

  \begin{center}
    {\ABNTEXchapterfont\large\imprimirautor}

    \vspace*{\fill}\vspace*{\fill}
    \begin{center}
      \ABNTEXchapterfont\bfseries\Large\imprimirtitulo
    \end{center}
    \vspace*{\fill}

    \hspace{.45\textwidth}
    \begin{minipage}{.5\textwidth}
        \imprimirpreambulo
    \end{minipage}%
    \vspace*{\fill}
   \end{center}

   \aprovadaPor

   \assinatura{\textbf{Prof. \imprimirorientador, D.Sc.} \\ \presidenteDaBanca}
   \assinatura{\textbf{Prof. Paulo César Marques Vieira, D.Sc.}}
   \assinatura{\textbf{Prof. Franklin de Lima Marquezino, D.Sc.}}

   \begin{center}
    \vspace*{0.5cm}
    {\large\imprimirlocal}
    \par
    {\large\imprimirdata}
    \vspace*{1cm}
  \end{center}

\end{folhadeaprovacao}

\begin{dedicatoria}
   \vspace*{\fill}
   \vspace*{10cm}
   \flushright
   \noindent
   \textbf{\dedicatorianame\\}
   \textit{A todos que me apoiaram \\ durante minha trajetória.\\} \vspace*{\fill}
\end{dedicatoria}

\begin{agradecimentos}

Agradeço à minha família pelo apoio contínuo ao longo de toda minha trajetória educacional.
Agradeço a Victor Santos e a Yuri Messias pelos CiViKs e Blue Days,
contribuindo para a manutenção da minha sanidade mental.
Agradeço a Raul Silva, Breno Viana, Felipe Barbalho, Débora Emili e Patrícia Cruz por motivos similares.

Agradeço também à todas as amizades que fiz ao longo da minha vivência no
Laboratório Nacional de Computação Científica --
pessoas também essenciais para minha sobrevivência nesses tempos pandêmicos.
Em especial,
ressalto os nomes de Dudu Hutter e Nana Grassi (por me fazerem sentir em casa em Petrópolis);
Douglas Terra (pela companhia na república);
Ana Néri e Roberta Machado (por aturarem minhas dúvidas na secretaria);
Cauê Teixeira e Jalil Moqadam (pela recepção no grupo de pesquisa);
Edlaine Fernandes, Haron Calegari, Ítalo Messias, João Vitor de Oliveira,
    Lucas dos Anjos e Weslley Pereira (membros do RPGzim);
Alonso Alvarez, Ana Luiza Karl, Andressa Machado, Luís Cury, Matheus Müller e
    Rafael Terra (membros da comissão discente e organização do EAMC);
Dayana Cristine, Felipe Otávio dos Santos, Frederico Cabral, Natanael Júnior, Pedro Lugão e Renato Borseti
    (pelos perrengues e contribuições compartilhados);
e a Gabriele Iwashima (por iluminar meus dias desde que a conheci).
Peço desculpas a todos aqueles que não foram mencionados pela falta de memória e de espaço.

Por fim,
presto minha homenagem a algumas pessoas próximas que pereceram diante da COVID-19:
Artur Ziviani, Flávio Bezerra e Gabriel Rocha.
Agradeço também à CAPES e à FAPERJ pelo apoio financeiro.

\end{agradecimentos}

\begin{epigrafe}
    \vspace*{\fill}
	\begin{flushright}
		\textit{``I saw an old man sitting with his head in hands \\
            His eyes reflect the wisdom of his life \\
            His words painted a new world \\
            And my thoughts just followed him''\\
    		(Eloy)}
	\end{flushright}
\end{epigrafe}


\setlength{\absparsep}{18pt} 
\begin{resumo}
 
 Estudos na Computação Quântica têm avançado desde a década de 1980,
 numa busca incessante por algoritmos melhores que qualquer algoritmo clássico
 concebível.
 Um exemplo desses algoritmos é o algoritmo de Grover,
 capaz de encontrar $k$ elementos (marcados)
 num banco de dados desordenado com $N$ elementos em $O\pr{\sqrt{N/k}}$ passos.
 O algoritmo de Grover também pode ser interpretado como um passeio quântico
 num grafo completo (com laços) com $N$ vértices dos quais $k$ são marcados.
 Essa interpretação estimulou a análise de algoritmos de busca em
 outros tipos de grafo -- e.g. grafo bipartido completo, malha e hipercubo.
 Utilizando o operador linear que descreve o algoritmo de Grover,
 o algoritmo de contagem quântico resulta numa estimativa do valor $k$
 com erro da ordem de $O\pr{\sqrt k}$ e em $O\pr{\sqrt{N}}$ passos.
 Neste trabalho, analisa-se o problema de usar o algoritmo de contagem quântico
 para estimar a quantidade $k$ de elementos marcados em outros tipos de grafos;
 em particular no grafo bipartido completo.
 De fato, conclui-se que para um subcaso desse tipo de grafo,
 ao executar o algoritmo proposto no máximo $t$ vezes,
 é possível obter uma estimativa de $k$ com erro da ordem de $O\pr{\sqrt k}$
 em $O\pr{t\sqrt{N}}$ passos e
 probabilidade de sucesso maior ou igual a $\pr{1 - 2^{-t}}8/\pi^2$.


 \textbf{\palavrasChave}: Passeios quânticos. Grafo bipartido completo. Algoritmo de contagem.
\end{resumo}

\begin{resumo}[Abstract]
 \begin{otherlanguage*}{english}
 Studies on Quantum Computing have been developed since the 1980s,
 motivating researches on quantum algorithms better than any classical algorithm possible.
 An example of such algorithms is Grover's algorithm,
 capable of finding $k$ (marked) elements in 
 an unordered database with $N$ elements using $O\pr{\sqrt{N/k}}$ steps.
 Grover's algorithm can be interpreted as a quantum walk in a complete graph (with loops)
 containing $N$ vertices from which $k$ are marked.
 This interpretation motivated search algorithms in other graphs --
 complete bipartite graph, grid, and hypercube.
 Using Grover's algorithm's linear operator, the quantum counting algorithm estimates
 the value of $k$ with an error of $O\pr{\sqrt k}$ using $O\pr{\sqrt{N}}$ steps.
 This work tackles the problem of using the quantum counting algorithm for
 estimating the value $k$ of marked elements in other graphs;
 more specifically, the complete bipartite graph.
 It is concluded that for a particular case,
 running the proposed algorithm at most $t$ times
 wields an estimation of $k$ with an error of $O\pr{\sqrt k}$ using $O\pr{t\sqrt{N}}$ steps and
 success probability of at least $\pr{1 - 2^{-t}}8/\pi^2$.

   \textbf{Keywords}: Quantum walks. Complete bipartite graphs. Counting algorithm.
 \end{otherlanguage*}
\end{resumo}

%

%

%

\pdfbookmark[0]{\listfigurename}{lof}
\listoffigures*
\cleardoublepage

\pdfbookmark[0]{\listtablename}{lot}
\listoftables*
\cleardoublepage
\pdfbookmark[0]{\contentsname}{toc}
\tableofcontents*
\cleardoublepage

\textual



\chapter{Introdução}\label{cap_intro}

Os computadores clássicos foram um grande avanço na humanidade possibilitando automação de tarefas como cálculos complexos.
A capacidade de processamento dos computadores depende dos transistores:
quão maior a quantidade de transistores, mais cálculos um computador é capaz de fazer.
Logo, diminuir o tamanho dos transistores é essencial para se obter maior poder computacional numa área cada vez menor.
Os avanços tecnológicos até então permitiram que os transistores (logo processadores) evoluíssem de tal forma a obedecer a Lei de Moore,
que afirma que o poder computacional dobra a cada 18 meses
\cite{moore1965cramming,mack2011fifty}.
Tais avanços fizeram com que os transistores ficassem em escala nanométrica
\cite{yu2002finfet},
a ponto de serem influenciados pelas Leis da Mecânica Quântica
\cite{nielsen2002quantum}.
Tal influencia impede que os transistores diminuam ainda mais de tamanho e realizem computação com baixa margem de erro \cite{nielsen2002quantum}.
Há duas formas de contornar esse problema:
utilizar computação paralela,
não sendo necessário diminuir o tamanho dos transistores para melhorar o desempenho
\cite{patt1997one};
ou criar computadores que se aproveitem das propriedades da Mecânica Quântica,
os computadores quânticos \cite{nielsen2002quantum}.

Estudos na Computação Quântica têm avançado desde a década de 1980
\cite{feynman1982simulating,deutsch1985quantum},
numa busca incessante por algoritmos melhores que qualquer
computador clássico é capaz de executar.
Exemplos desses algoritmos são os algoritmos que realizam consultas a um oráculo,
\textit{e.g.} o algoritmo de Deutsch-Jozsa \cite{deutsch1992rapid},
o algoritmo de Bernstein-Vazirani \cite{bernstein1997quantum},
e o algoritmo de (busca de) Grover
\cite{grover1996fast, grover1997quantum}.
Esse tipo de algoritmo consiste de um oráculo
(uma caixa preta que implementa uma função com saída binária) e
o objetivo do algoritmo é extrair alguma informação dessa caixa preta
fazendo o mínimo de consultas possíveis ao oráculo.
No algoritmo de Grover, por exemplo,
dado um banco de dados não ordenado,
o oráculo marca um elemento de $N$ e o algoritmo é capaz de
encontrar o elemento marcado em $O(\sqrt N)$ consultas ao oráculo;
enquanto um computador clássico precisaria de $O(N)$ chamadas ao oráculo.

Um outro tipo de algoritmo quântico bastante estudado são os algoritmos
baseados em passeios quânticos \cite{aharonov1993quantum, farhi1998quantum}.
De fato, há uma equivalência entre passeios quânticos
e qualquer algoritmo por um computador quântico.
\cite{lovett2010universal}.
Além disso, passeios quânticos permitem a implementação de algoritmos em laboratórios
sem a necessidade de um computador quântico \cite{portugal2013quantum}.
Sendo assim, o algoritmo de Grover também pode ser interpretado como
um passeio quântico em grafos.

Shenvi, Kempe e Whaley mostraram que o algoritmo de Grover pode ser interpretado
aproximadamente como um passeio quântico num hipercubo com $N$ vértices
\cite{shenvi2003quantum};
impulsionando algoritmos de busca em outros tipos de grafos.
De fato, mostrou-se que o algoritmo de Grover pode ser interpretado
\emph{exatamente} como um passeio quântico no grafo completo com $N$ vértices
e laços em todos os vértices \cite{ambainis2004coins,wong2015grover}.
Um algoritmo de busca em látices 2-dimensionais quadrados (malha) com $N$ vértices
foi apresentado com complexidade de $O(\sqrt N \log N)$ \cite{ambainis2004coins} e
posteriormente aprimorado para $O(\sqrt{N \log N})$ \cite{tulsi2008faster}.
O algoritmo de busca em látices ``colméia'' com $N$ vértices
também pode ser feita em $O(\sqrt{N \log N})$ \cite{abal2010spatial}.
Um último exemplo (e o de mais interesse para este trabalho)
é a busca realizada no grafo bipartido completo em $O(\sqrt{N})$
\cite{rhodes2019quantum}.

Generalizações do algoritmo de Grover marcam $k$ elementos e o
algoritmo é capaz de encontrar um elemento marcado em $O(\sqrt{N/k})$ passos
\cite{boyer1998tight}.
Entretanto, a generalização de busca para $k$ elementos marcados em
outros grafos não é trivial:
existem algumas escolhas de elementos marcados para as quais
o algoritmo quântico não apresenta ganho em comparação com os algoritmos clássicos
-- chamadas de configurações excepcionais.
Exemplos de casos excepcionais são a diagonal na malha \cite{ambainis2008quantum} e
qualquer quantidade de elementos marcados em grafos cíclicos
\cite{wong2017exceptional}.
Nahimovs e Rivosh encontraram outras configurações excepcionais na malha
\cite{nahimovs2015exceptional}.
Bezerra \textit{et al.} estendem para vários elementos marcados
uma técnica usada para analisar algoritmos de busca
com $N \to \infty$ e apenas um elemento marcado 
\cite{PhysRevA.103.062202,shenvi2003quantum,ambainis2004coins,portugal2013quantum}.
O trabalho não foca em casos excepcionais e a técnica exige que algumas pré-condições sejam verdadeiras; porém,
conjecturaram que não há configurações excepcionais no hipercubo,
encontrando um elemento marcado em $O(\sqrt{N/k})$
\cite{PhysRevA.103.062202}.
Já a referência \cite{rhodes2019quantum} considera a busca no
grafo bipartido completo com $k \ll N$ elementos marcados,
nenhuma configuração excepcional foi encontrada.

O problema de contagem surge naturalmente a partir do problema de busca:
ao invés do objetivo ser encontrar um dos $k$ elementos marcados,
o foco é descobrir o valor de $k$.
Esse problema foi atacado em \cite{brassard2002quantum,kaye2007introduction};
onde o algoritmo de estimativa de fase é utilizado (como uma subrotina)
junto com o operador que descreve o algoritmo de Grover
para obter uma estimativa de $k$.
Essa estimativa de $k$ pode ser utilizada para determinar a quantidade
de iterações necessárias para o algoritmo de busca \cite{nielsen2002quantum}.

Surpreendentemente, não foram encontrados trabalhos que utilizem
o algoritmo de estimativa de fase junto com o operador linear que
descreve o passeio quântico em grafos não completos para
estimar o valor de $k$ nesses grafos.
Portanto, neste trabalho,
propõe-se utilizar esse método para estimar a quantidade $k$ de vértices
marcados no grafo bipartido completo utilizando o operador linear descrito pelo
passeio quântico discreto com moeda nesse grafo -- o mesmo utilizado em \cite{rhodes2019quantum}.
Em particular, foca-se no caso em que ambos os conjuntos disjuntos do grafo
possuem a mesma quantidade de vértices marcados e não marcados.

A estrutura deste trabalho é descrita a seguir.
Capítulo \ref{cap:referencial-teorico} revisa conceitos necessários ao longo do trabalho,
como Álgebra Linear, Postulados da Mecânica Quântica e Teoria dos Grafos.
Capítulo \ref{cap:alg-cont} apresenta o conteúdo necessário para
o entendimento do algoritmo quântico de contagem, além do próprio algoritmo.
Capítulo \ref{cap:alg-cont-grafo-bip-compl} foca na explicação de passeios quânticos em grafos regulares e
na junção do algoritmo quântico de contagem com o passeio quântico de busca no grafo bipartido completo
(as contribuições originais desta dissertação são encontradas nesse capítulo).
Capítulo \ref{cap:conclusao} apresenta as conclusões e considerações finais.
\chapter{Referencial Teórico} \label{cap:referencial-teorico}
Esse Capítulo dedica-se a introduzir conceitos que serão necessários ao longo do documento.
Espera-se que o leitor tenha conhecimento prévio de Álgebra Linear e números complexos.
Seção \ref{sec:comp-quantica} resume os conceitos necessários pra computação quântica e
entendimento dos algoritmos apresentados posteriormente.
Seção \ref{sec:teoria-dos-grafos} aborda os conceitos de Teoria dos Grafos necessários.

\section{Computação Quântica} \label{sec:comp-quantica}

Seção \ref{sec:alg-lin} revisa conceitos de Álgebra Linear utilizando a notação de Dirac;
Seção \ref{sec:postulados} aborda como os conceitos de Álgebra Linear são utilizados
para descrever a Mecânica Quântica e seus postulados;
Seção \ref{sec:circuitos-quanticos} introduz a representação de circuitos
utilizada na implementação ou explicação de algoritmos quânticos.

\subsection{Álgebra Linear} \label{sec:alg-lin}

Em Computação Quântica, trabalha-se normalmente com número complexos.
Ressalta-se aqui algumas definições básicas de números complexos.
\begin{definition}
    A unidade imaginária é representada por $\ii = \sqrt{-1}$.
\end{definition}
\begin{definition}
    Define-se um número complexo $z = a + \ii b$, onde $a, b \in \R$,
    onde $a$ é chamada de parte real e $b$ de parte imaginária.
    Denota-se a parte real de $z$ por $a = \Re(z)$.
\end{definition}
\begin{definition}
    O complexo conjugado de $z \in \mathbb C$ é dado por
    \begin{align}
        z^* = \pr{a + \ii b}^* = a - \ii b.
    \end{align}
\end{definition}

O livro do Axler é uma referência consolidada em Álgebra Linear
\cite{axler2014linear}.
Seja $N \in \mathbb{N}$ tal que $N \geq 1$.
Em Álgebra Linear, normalmente $\vec v$ denota um vetor e
representa uma lista de $N$ elementos complexos (ou entradas complexas);
ou seja, $\vec v \in \mathbb{C}^N$.
Esses $N$ elementos podem ser explicitados através da representação
\begin{align}
    \vec v = \matrx{v_0 \\ v_1 \\ \vdots \\ v_{N-1}},
\end{align}
onde $v_0, v_1, \ldots, v_{N-1} \in \mathbb{C}$.
Ao longo do restante desse trabalho, como usual na Mecânica e Computação Quântica,
utiliza-se a notação de Dirac para representação de vetores.
Então, o mesmo vetor $\vec v$ é representado como $\ket v$ --
leia-se \emph{ket v}.

Toda a análise em Álgebra Linear é feita num espaço denominado espaço vetorial.
Para definir espaço vetorial,
é necessário definir \emph{adição} e \emph{multiplicação por escalar} num conjunto $V$.
\begin{definition}
    Seja $V$ um conjunto de vetores em $\mathbb{C}^N$.
    Para todos $\ket u, \ket v \in V$, a adição é definida como uma função tal que
    $\ket u + \ket v \in V$.
    Para este trabalho, utiliza-se a adição
    \begin{align}
        \ket u + \ket v = \matrx{u_0 \\ u_1 \\ \vdots \\ u_{N-1}} +
            \matrx{v_0 \\ v_1 \\ \vdots \\ v_{N-1}}
        = \matrx{u_0 + v_0 \\ u_1 + v_1 \\ \vdots \\ u_{N-1} + v_{N-1}}.
    \end{align}
\end{definition}

\begin{definition}
    Seja $V$ um conjunto de vetores em $\mathbb{C}^N$.
    Para todos $\ket v \in V$ e $c \in \mathbb{C}$,
    a multiplicação por escalar é uma função tal que
    $c \ket v \in V$.
    Para este trabalho, utiliza-se a multiplicação por escalar
    \begin{align}
        c \ket v = c \matrx{v_0 \\ v_1 \\ \vdots \\ v_{N-1}}
        = \matrx{c v_0 \\ c v_1 \\ \vdots \\ c v_{N-1}} .
    \end{align}
\end{definition}

\begin{definition}
    Sejam $V$ um conjunto de vetores em $\mathbb{C}^N$.
    Um espaço vetorial consiste do conjunto $V$
    munido de adição e multiplicação por escalar definidas em $V$
    de tal modo que as seguintes propriedades sejam verdadeiras;
    para todos $\ket t, \ket u, \ket v \in V$ e $c, c' \in \mathbb{C}$.
    \begin{itemize}
        \item Comutatividade: $\ket u + \ket v = \ket v + \ket u$;
        \item Associatividade: $\pr{\ket t + \ket u} + \ket v = \ket t + \pr{\ket u + \ket v}$;
        \item Identidade aditiva: existe um vetor nulo $\vec 0 \in V$ tal que
            $\vec 0 + \ket v = \ket v$. Para este trabalho,
            \footnote{Para o vetor nulo, utiliza-se $\vec 0$ ao invés de $\ket 0$,
                pois um significado distinto é normalmente atribuído para $\ket 0$.}
            \begin{align}
                \vec 0 = \matrx{0 \\ 0 \\ \vdots \\ 0};
            \end{align}
        \item Inverso aditivo: $\exists \ket{v'} \in V$ tal que $\ket{v'} = - \ket v$ e
            $\ket{v'} + \ket v = -\ket v + \ket v = \vec 0$;
        \item Identidade multiplicativa: $1 \in \mathbb{C}$ tal que $1\ket v = \ket v$;
        \item Propriedades distributivas:
        \begin{itemize}[label=\textperiodcentered]
            \item $c \pr{\ket u + \ket v} = c\ket u + c \ket v$;
            \item $\pr{c + c'} \ket v = c\ket v + c'\ket v$.
        \end{itemize}
    \end{itemize}
\end{definition}

A definição de subespaço vetorial também é relevante para este trabalho.
\begin{definition}
    Seja $V'$ um subconjunto (não necessariamente próprio) de $V$.
    $V'$ é dito um subespaço de $V$ se $V'$ também for um espaço vetorial
    usando a mesma adição e multiplicação por escalar definidas para $V$.
\end{definition}
Por exemplo,
\begin{align}
    V' = \left\{ \matrx{v_0 \\ v_ 1 \\ 0} \right\}
\end{align}
para todo $v_0, v_1 \in \C$ é um subespaço vetorial de $\mathbb{C}^3$,
ou simplesmente um subespaço.

Deseja-se obter maneiras simplificadas para descrever um espaço vetorial.
Para tanto, a definição de conjunto gerador é útil.
\begin{definition}
    \label{def:span}
    Seja $V$ um espaço vetorial e $S$ um subconjunto de vetores de $V$.
    $S$ é um conjunto gerador de $V$ se e somente se $\forall \ket v \in V$,
    $\ket v$ pode ser escrito como uma combinação linear dos vetores em $S$.
    Em outras palavras,
    \begin{align}
        \textnormal{span}\pr{S} = V \iff \forall \ket v \in V,
            \ket v = \sum_{i = 1}^{\card S} \alpha_i \ket{S_i},
    \end{align}
    onde $\alpha_i \in \mathbb{C}$, $\ket{S_i} \in S$, e
    $\alpha_i$ é denominada amplitude de $\ket{S_i}$.
\end{definition}

Um exemplo de um conjunto gerador para $\mathbb{C}^2$ é o conjunto $\set{\ket 0 , \ket 1}$,
onde
\begin{align}
    \ket 0 = \matrx{1 \\ 0} \quad \textnormal{e} \quad \ket 1 = \matrx{0 \\ 1}.
\end{align}
Note que qualquer vetor $\ket v \in \mathbb{C}^2$ pode ser escrito como
\begin{align}
    \ket{v} = \matrx{v_0 \\ v_1} = v_0 \matrx{1 \\ 0} + v_1 \matrx{0 \\ 1} =
        v_0 \ket 0 + v_1 \ket 1 = \sum_{i = 0}^1 v_i \ket{i},
\end{align}
onde $v_0, v_1 \in \C$.

Na definição \ref{def:span} não restringiu-se a cardinalidade de $S$.
Logo, é possível que um vetor tenha múltiplas representações dependendo de $S$.
Por exemplo, se $S = \set{\ket 0, \ket 1, \ket +}$, onde
\begin{align}
    \ket + = \frac{1}{\sqrt 2} \matrx{1 \\ 1};
\end{align}
existem no mínimo duas representações possíveis para $\vec 0$:
\begin{align}
    0 \ket{0} + 0 \ket{1} + 0 \ket + = \vec 0 =
    \frac{1}{\sqrt 2} \ket 0 + \frac{1}{\sqrt 2} \ket 1 - \ket + .
\end{align}
De modo geral, é desejável que $\card S$ tenha o menor valor possível
para simplificar essas representações.
Esses pontos motivam as definições de vetores  \emph{linearmente independentes} e \emph{base}.

\begin{definition}
    Um conjunto $S \in V$ de vetores é dito linearmente independente se
    a única escolha de $\alpha_1, \ldots, \alpha_{\card S} \in \C$ tais que
    \begin{align}
        \sum_{i = 1}^{\card S} \alpha_i \ket{S_i} = \vec 0
    \end{align}
    é $\alpha_1, \ldots, \alpha_{\card S} = 0$,
    onde $\ket{S_i} \in S$.
\end{definition}

\begin{definition}
    Um conjunto $S \in V$ de vetores é uma base de $V$ se e somente se
    $S$ é linearmente independente e $\textnormal{span}\pr{S} = V$.
    Além disso, $V = \mathbb{C}^N \implies \card S = N$.
\end{definition}

Nesse âmbito, $\set{\ket 0, \ket 1, \ket +}$ não são linearmente independentes
(\textit{i.e.} são linearmente dependentes), e tanto $\set{\ket 0, \ket 1}$ quanto
$\set{\ket 0, \ket +}$ são bases de $\mathbb{C}^2$.
Como existe mais de uma base para um espaço vetorial,
é desejável ter um modo sistemático para conversão entre bases (ou vetores)
de um mesmo espaço vetorial ou até mesmo
mapear vetores de um espaço vetorial para outro distinto;
o que motiva a definição de \emph{transformações lineares} e \emph{operadores lineares}.
\begin{definition}
    Sejam $V$ e $W$ espaços vetoriais.
    Uma transformação linear é uma função $T : V \to W$
    que possui as propriedades de aditividade e homogeneidade.
    Ou seja, para $\ket{v_i} \in V$ e $\alpha_i \in \mathbb C$,
    \begin{align}
        T \sum_i \alpha_i \ket{v_i} = T\pr{\sum_i \alpha_i \ket{v_i}}
        = \sum_i \alpha_i T \pr{\ket{v_i}} = \sum_i \alpha_i T \ket{v_i} .
    \end{align}
\end{definition}
\begin{definition}
    Seja $V$ um espaço vetorial e uma transformação linear $T: V \to V$,
    por ter imagem e domínio no mesmo espaço vetorial, chama-se $T$ de um
    operador linear.
\end{definition}
Normalmente, omite-se os parênteses.
Desse modo, sendo $V_1$, $V_2$ e $V_3$ espaços vetoriais e
$T_1 : V_1 \to V_2$ e $T_2: V_2 \to V_3$ transformações lineares;
denota-se a composição de funções
$T_2\pr{T_1\pr{\ket{v}}}$ simplesmente por $T_2 T_1 \ket v$,
onde $\ket v \in V_1$.
Um operador linear de interesse é o operador identidade.
\begin{definition}
    Seja $V$ um espaço vetorial. O operador identidade $I_V: V \to V$
    é tal que $\forall \ket v \in V$,
    \begin{align}
        I_V \ket{v} = \ket{v}.
    \end{align}
    O operador identidade é representado simplesmente por $I$
    quando não há ambiguidade no contexto.
\end{definition}
Para interpretar uma transformação linear como mudança de bases,
considere os espaços vetoriais $V$ e $W$ espaços vetoriais cujas bases são
$\set{\ket{V_j}}$, $\set{\ket{W_i}}$, respectivamente;
e seja $T: V \to W$ uma transformação linear.
Sendo assim, para $\ket{w_j} \in W$ existem valores $T_{i,j} \in \C$ tais que
\begin{align}
    T \ket {V_j} = \ket{w_j} = \sum_i T_{i,j} \ket{W_i}.
\end{align}
Esse resultado sugere que $T$ pode ser representado como uma matriz
cujas entradas são $T_{i,j}$:
\begin{align}
    T = \matrx{
            T_{1, 1} & T_{1, 2} & \cdots & T_{1, \card V} \\
            T_{2, 1} & T_{2, 2} & \cdots & T_{2, \card V} \\
            \vdots & \vdots & \ddots & \vdots \\
            T_{\card W, 1} & T_{\card W, 2} & \cdots & T_{\card W, \card V}
        },
\end{align}
onde $|V| = \card{\set{\ket{V_j}}}$ e $|W| = \card{\set{\ket{W_i}}}$.
Para representar $T$ utilizando notação de Dirac,
é necessário saber como calcular os valores $T_{i,j}$.
Para isso, introduz-se o conceito de \emph{produto interno}.

\begin{figure}[hbt]
    \centering
    \caption{Projeção e comprimento de vetores.}
    \begin{tikzpicture}
    \pgfmathsetmacro\mult{1.5}
    
    \draw[->] (0 , 0) -- (1.1*\mult , 0) node[right] {$\ket{0}$};
    \draw[->] (0, 0) -- (0, 1.1*\mult ) node[above] {$\ket{1}$};

    \draw[->] (0, 0) -- ({1/sqrt(2)*\mult}, {1/sqrt(2)*\mult})
        node[right] {$\ket{+}$};
        
    \draw[dashed,gray] ({1/sqrt(2)*\mult}, {1/sqrt(2)*\mult}) -- ({1/sqrt(2)*\mult}, 0)
        node[below] {$\frac{1}{\sqrt{2}}$};
    \draw[dashed,gray] ({1/sqrt(2)*\mult}, {1/sqrt(2)*\mult}) -- (0, {1/sqrt(2)*\mult})
        node[left] {$\frac{1}{\sqrt{2}}$};
        
    \draw[->] (0, 0) -- ({1*\mult}, {-1*\mult})
        node[right] {$\ket 1 - \ket 0$};
    \draw [decorate, decoration={brace, amplitude=10pt, mirror}, gray]
        (-0.025*\mult, -0.025*\mult) -- ({(1-0.03)*\mult}, {(-1-0.03)*\mult})
        node[midway,yshift=-\mult*10, xshift=-\mult*10]{$\sqrt 2$};
\end{tikzpicture}
    \legend{Fonte: Produzido pelo autor.}
    \label{fig:projecao}
\end{figure}

Há mais motivos para formalizar produto interno.
Considere os vetores $\ket 0$, $\ket 1$ e $\ket +$ representado no plano dos reais --
Figura \ref{fig:projecao} (Fig. \ref{fig:projecao})  --
observa-se que $\ket 0$ e $\ket 1$ são vetores perpendiculares;
logo, a projeção de $\ket 0$ em $\ket 1$ é 0 e vice-versa.
Entretanto, o mesmo não acontece com a projeção de $\ket +$ em $\ket 0$ ou $\ket 1$,
que é igual a $1/\sqrt 2$.
Outro valor escalar que é relavante é o comprimento dos vetores.
Por exemplo, $\ket 0$ tem comprimento 1, mas $\ket 0 - \ket 1$ possui comprimento $\sqrt 2$.
Os cálculos dessas projeções e comprimentos foram obtidos facilmente,
mas definir um modo sistemático e geral para calcular esses valores é mais útil.
Isso pode ser feito utilizando-se \emph{produto interno} e \emph{norma}.

\begin{definition} \label{def:prod-interno}
    Seja $V$ um espaço vetorial.
    O produto interno é uma função $\pr{\cdot, \cdot}: V \times V \to \C$
    tal que $\forall \ket{u}, \ket{v_i} \in V$ e $\alpha_i \in \mathbb{C}$
    as seguintes propriedades são verdadeiras.
    \begin{itemize}
        \item Positividade: $\pr{\ket u, \ket u} \geq 0$ com
            igualdade se e somente se $\ket u = \vec 0$;
        \item  Aditividade e homogeneidade no segundo argumento:
            \begin{align}
                \pr{\ket u, \sum_i \alpha_i \ket{v_i}} =
                    \sum_i \alpha_i \pr{\ket u, \ket{v_i}};
            \end{align}
        \item Simetria Hermitiana: $\pr{\ket u, \ket{v_i}} = \pr{\ket{v_i}, \ket u}^*$.
    \end{itemize}
\end{definition}

Todo produto interno possui as propriedades de aditividade, e de homogeneidade conjugada
no primeiro argumento:
\begin{align}
    \pr{\sum_i \alpha_i \ket{v_i}, \ket{u}} = \pr{\ket{u}, \sum_i \alpha_i \ket{v_i}}^*
    = \sum_i \alpha_i^* \pr{\ket u, \ket {v_i}}^*
    = \sum_i \alpha_i^* \pr{\ket {v_i}, \ket u}.
\end{align}

\begin{definition}
    Se $V$ é um espaço vetorial munido de um produto interno,
    então é denominado \emph{espaço de Hilbert} e denotado por $\hilb$.
\end{definition}
Ao longo deste trabalho,
utiliza-se $\hilb^N$ para se referir ao espaço vetorial $\mathbb{C}^N$
munido de um produto interno.

\begin{definition}
    A norma de um vetor $\ket v \in \hilb$ é dada por
    \begin{align}
        \card{\card{\ket v}} = \sqrt{\pr{\ket v, \ket v}}.
    \end{align}
    Um vetor $\ket v$ é dito unitário se $\card{\card{\ket v}} = 1$.
\end{definition}

Um produto interno em $\mathbb{C}^N$ é o produto escalar.
\begin{definition}
    Sejam $\ket u, \ket v \in \mathbb{C}^N$.
    O produto escalar é definido por
    \begin{align}
        \pr{\ket u, \ket v} = \sum_{i = 0}^{N - 1} u_i^* v_i
        = \matrx{u_0^*, u_1^*, \ldots, u_{N-1}^*}
            \matrx{v_0 \\ v_1 \\ \vdots \\ v_{N-1}} .
    \end{align}
\end{definition}

Note que o produto escalar é equivalente a uma multiplicação de matrizes,
mas antes de realizar essa multiplicação, tomou-se o transposto conjugado de $\ket u$.
O vetor transposto conjugado também é chamado de vetor dual.
Observe também que o vetor dual pode ser interpretado como uma transformação linear
$\C^N \to \C$; de fato, ele é definido de tal forma.
\begin{definition} \label{def:vetor-dual}
    Para todo $\ket{u}, \ket{v} \in \hilb^N$,
    o vetor dual (leia-se \emph{bra u}) $\bra{u} \equiv \ket{u}^\dagger$ é
    uma transformação linear $\C^N \to \C$ tal que
    \begin{align}
        \braket{u | v} = \bra  u \ket v = \pr{\ket u, \ket v}.
    \end{align}
\end{definition}

Em posse da definição de produto interno e de norma,
é possível definir ortogonalidade e ortonormalidade,
que se relacionam com a perpendicularidade argumentada previamente.
\begin{definition}
    Sejam $\ket u, \ket v \in \hilb$. Esses vetores são ditos ortogonais se $\braket{u | v} = 0$.
\end{definition}
\begin{definition}
    Sejam $\ket u, \ket v \in \hilb$. Esses vetores são ditos ortonormais se
    $\braket{u | v} = 0$ e $\card{\card{\ket u}} = \card{\card{\ket v}} = 1$.
\end{definition}

Recorrendo novamente à Fig. \ref{fig:projecao} têm-se uma interpretação
geométrica de vetores ortogonais ($\ket +$ e $\ket 0 - \ket 1$) e
vetores ortonormais ($\ket 0$ e $\ket 1$).
Essas definições estendem-se facilmente para um conjunto de vetores.
O conjunto $\set{\ket{V_i}} \in V$ é dito ortonormal se
$\braket{V_i|V_j} = 0$ e $\card{\card{\ket{V_i}}} = 1$ para todo $i \neq j$.
Além disso, se o conjunto for uma base de $V$, diz-se que $\set{\ket{V_i}}$ é uma base ortonormal.
Isso motiva a definição do delta de Kronecker.
\begin{definition}
    Seja $\set{\ket{V_i}}$ um conjunto de vetores ortonormais indexados por $i$.
    O delta de Kronecker é uma função de duas variáveis tal que
    \begin{align}
        \delta_{i, j} = \braket{V_i | V_j} = \begin{cases}
            1 \quad \textnormal{se } i = j; \\
            0 \quad \textnormal{se } i \neq j.
        \end{cases}
    \end{align}
\end{definition}

Utilizando bases ortonormais, delta de Kronecker, e as propriedades de produto interno,
é possível obter uma representação matricial para o vetor dual.
Sejam dois vetores $\ket v = \sum_{i} v_i \ket{S_i}$ e $\ket u = \sum_{i} u_i \ket{S_i}$
onde $v_i, u_i, \in \mathbb{C}$,
\begin{align}
    \braket{u|v} &= \pr{\sum_i u_i \ket{S_i} , \sum_j v_j \ket{S_j}}
    \\
    &= \sum_{i,j} u_i^* v_j \braket{S_i | S_j}
    \\
    &= \sum_{i,j} u_i^* v_j \delta_{i,j}
    \\
    &= \sum_i u_i^* v_i .
\end{align}
Ou seja, se $\ket u$ e $\ket v$ forem representados com a mesma base,
o vetor dual $\bra u$ é o transposto conjugado de $\ket u$.

Retornando ao tópico de transformações lineares,
é possível representar uma transformação linear entre espaços de Hilbert
utilizando notação de Dirac.
Essa transformação é representada através de \emph{produtos externos}.
\begin{definition}
    Sejam $V$ e $W$ espaços de Hilbert, e $\ket v, \ket{v'} \in V$ e $\ket{w} \in W$.
    Utiliza-se o produto externo $\ket w \bra v$ para
    definir uma transformação linear $\ket w \bra v : V \to W$ cuja ação é dada por
    \begin{align}
        \pr{\ket{w}\bra{v}} \ket{v'} = \ket{w} \bra{v} \ket{v'} = \ket{w} \braket{v | v'}
        = \braket{v | v'} \ket{w} .
    \end{align}
\end{definition}

Essa representação em produto externo sugere que uma transformação linear $T: V \to W$
descreve uma relação entre bases ortonormais $\set{\ket{V_j}}$ de $V$ e $\set{\ket{W_i}}$ de $W$.
Para dar continuidade, a \emph{relação de completude} faz-se necessária
\begin{definition}
    Seja $\set{\ket{V_i}}$ uma base ortonormal de Hilbert $V$.
    A relação de completude assegura que
    \begin{align}
        \sum_i \ket{V_i} \bra{V_i} = I.
    \end{align}
\end{definition}
Note que isso é verdade já que $\forall \ket v \in V$,
\begin{align}
    \ket v &= I \ket v = I \sum_{i'} v_{i'} \ket{V_{i'}}
    \\
    &= \sum_i \ket{V_i}\bra{V_i} \sum_{i'} v_{i'} \ket{V_{i'}}
    \\
    &= \sum_{i,i'} \ket{V_i} v_{i'} \delta_{i,i'}
    \\
    &= \sum_{i'} v_{i'} \ket{V_{i'}}
    \\
    &= \ket v,
\end{align}
onde $v_{i'} \in \C$.
Analogamente, considere o vetor dual $\ket v$,
então $\forall \ket{v'} \in V$,
\begin{align}
    \braket{v | v'} &= \bra{v} I \ket{v'}
    \\
    &= \bra{v} \pr{ \sum_i\ket{V_i}\bra{V_i} } \sum_{i'} v_{i'} \ket{V_{i'}}
    \\
    &= \bra{v} \sum_{i,i'} \ket{V_i} v_{i'} \delta_{i,i'}
    \\
    &= \bra{v} \ket{v'} = \braket{v | v'}.
\end{align}

Sendo assim, denote por $\set{\ket{V_j}}$ e $\set{\ket{W_i}}$
bases ortonormais dos espaços de Hilbert $V$ e $W$, respectivamente;
e $T: V \to W$ uma transformação linear.
Então,
\begin{align}
    T &= I_W T I_V
    \\
    &= \sum_i \ket{W_i} \bra{W_i} T \sum_j \ket{V_j} \bra{V_j}
    \\
    &= \sum_{i,j} \ket{W_i} \bra{W_i} T \ket{V_j} \bra{V_j}
    \\
    &= \sum_{i,j} \bra{W_i} T \ket{V_j} \ket{W_i} \bra{V_j} .
    \label{eq:outer-prod-T}
\end{align}
Note que $\bra{W_i} T \ket{V_j} \in \mathbb C$,
já que as transformações lineares
\begin{align}
     T: V \to W \textnormal{ e } \bra{W_i}: W \to \C
    \implies \bra{W_i} T : V \to \C .
\end{align}

A Equação \ref{eq:outer-prod-T} (Eq. \ref{eq:outer-prod-T}) sugere que
é possível obter uma representação matricial para $T$
com respeito à base de entrada $\set{\ket{V_j}}$ e à base de saída $\set{\ket{W_i}}$.
Ordene as bases ortonormais pelos índices $i$ e $j$,
e denote $T_{i,j} = \bra{W_i} T \ket{V_j}$.
A matriz de $T$ é dada por
\begin{align}
    T &= T_{1,1} \ket{W_1} \bra{V_1} +
        \sum_{\substack{i, j \\ (i, j) \neq (1, 1)}} T_{i,j} \ket{W_i} \bra{V_j}
    \\
    &= T_{1, 1} \matrx{1 \\ 0 \\ \vdots \\ 0} \matrx{1 & 0 & \cdots & 0} +
        \sum_{\substack{i, j \\ (i, j) \neq (1, 1)}} T_{i,j} \ket{W_i} \bra{V_j}
    \\
    &= \matrx{
            T_{1, 1} & 0 & \cdots & 0 \\
            0 & 0 & \cdots & 0 \\
            \vdots & \vdots & \ddots & \vdots \\
            0 & 0 & \cdots & 0
        } +
        \sum_{\substack{i, j \\ (i, j) \neq (1, 1)}} T_{i,j} \ket{W_i} \bra{V_j}
    \\
    &= \matrx{
            T_{1, 1} & T_{1, 2} & \cdots & T_{1, \card V} \\
            T_{2, 1} & T_{2, 2} & \cdots & T_{2, \card V} \\
            \vdots & \vdots & \ddots & \vdots \\
            T_{\card W, 1} & T_{\card W, 2} & \cdots & T_{\card W, \card V}
        },
\end{align}
onde $(i, j)$ representam tuplas;
$\card V = \card{\set{\ket{V_i}}}$; $\card W = \card{\set{\ket{W_j}}}$;
$1 \leq i \leq \card V$; $1 \leq j \leq \card W$; e
os vetores linha $\matrx{b_1 & \cdots & b_N }$ com
$b_i = 1$ e $b_{i'} = 0$ caso $i' \neq i$
indicam que deve-se utilizar o dual de $\ket{V_i}$
(análogo para a relação entre vetores coluna com a utilização dos vetores $\ket{W_j}$).

Do mesmo modo que existem funções invertíveis,
deseja-se saber se um operador linear $T$ é invertível ou não.
Uma transformação linear $T: V \to W$ é invertível se e somente se
for injetiva e sobrejetiva \cite{axler2014linear}.
De particular interesse para esse trabalho,
são as transformações lineares invertíveis com imagem e domínio iguais,
\textit{e.g.} $T: V \to V$.
Se $T$ for invertível, então existe uma transformação linear $T^{-1}$ tal que
\begin{align}
    T T^{-1} = T^{-1} T = I.
\end{align}
Para dar continuidade, é necessário definir transformações lineares adjuntas,
análogo ao que foi feito na definição \ref{def:vetor-dual} (def. \ref{def:vetor-dual}).
Essa definição é consequência do Teorema de representação de Riesz
\cite{axler2014linear}.

\begin{definition}
    Sejam $V, W$ espaços de Hilbert;
    denotando o produto interno de $W$ por $(\cdot, \cdot)$ --
    def. \ref{def:prod-interno} -- e uma transformação linear $T: V \to W$.
    A transformação linear adjunta $T^\dagger: W \to V$ de $T: V \to W$ é tal que
    $\forall v \in V$ e $\forall \ket w \in W$,
    \begin{align}
        \pr{\ket w, T \ket v} = \pr{T^\dagger \ket w, \ket v}.
    \end{align}
\end{definition}

Sejam $V$, $W$ e $W'$ espaços de Hilbert;
$T, T_1, T_2 : V \to W$, $T_1 : W' \to W$ e $T_2 : V \to W'$ transformações lineares
tais que $T = L_1 + L_2$, $T = T_1 T_2$ e
$T = T_{i,j} \ket{W_i}\bra{V_j}$ para $T_{i,j} \in \C$ e
bases ortonormais $\set{\ket{V_j}}$, $\set{\ket{W_i}}$ de $V$, $W$, respectivamente.
Usando as propriedades de produtos internos, e a definição de transformação linear adjunta,
conclui-se que as seguintes propriedades são verdadeiras.
\begin{itemize}
    \item $\pr{L_1 + L_2}^\dagger =
        L_1^\dagger + L_2^\dagger$:
    \begin{align}
        \pr{\ket w, T \ket v}
        &= \pr{\ket w, L_1 \ket v} + \pr{\ket w, L_2 \ket v}
        \\
        &= \pr{L_1^\dagger \ket w, \ket v} +
            \pr{L_2^\dagger \ket w, \ket v}
        \\
        &= \pr{\pr{L_1^\dagger + L_2^\dagger} \ket w, \ket v}
        \\
        &= \pr{T^\dagger \ket w, \ket v};
    \end{align}
    \item $\pr{T_1T_2}^\dagger = T_2^\dagger T_1^\dagger$:
    \begin{align}
        \pr{\ket w, T \ket v} &= \pr{\ket w, T_1 T_2 \ket v}
        = \pr{T_1^\dagger \ket w, T_2 \ket v}
        \\
        &= \pr{T_2^\dagger T_1^\dagger \ket w, \ket v}
        = \pr{T^\dagger \ket w, \ket v};
    \end{align}
    \item Notando que $\ket v$ pode ser interpretado como uma transformação linear de $\C$ para $V$ e
    relambrando que $\bra v \equiv \ket v ^\dagger$,
    obtém-se que $\pr{T \ket v}^\dagger = \bra v T^\dagger$:
    \begin{align}
        \pr{\ket w 1, \pr{T \ket v} 1} = \pr{T^\dagger \ket w 1, \ket v 1}
        = \pr{\bra v T^\dagger \ket w 1, 1}  = \pr{\pr{T \ket v}^\dagger \ket w 1, 1};
    \end{align}
    \item $\pr{T^\dagger}^\dagger = T$:
    \begin{align}
        \pr{\ket w, T \ket v} &= \pr{T^\dagger \ket w, \ket v}
        = \pr{\ket v, T^\dagger \ket w}^*
        \\
        &= \pr{\pr{T^\dagger}^\dagger \ket v,  \ket w}^*
        = \pr{\ket w, \pr{T^\dagger}^\dagger \ket v} .
    \end{align}
    \item $T^\dagger = \sum_{i,j} T_{i,j}^* \ket{V_j} \bra{V_i}$ --
    \textit{i.e.} representação de $T^\dagger$ em notação de Dirac:
    \begin{align}
        \pr{\ket w, T \ket v} &= \pr{\ket w, \sum_{i,j} T_{i,j} \ket{V_i}\bra{V_j} \ket v}
        = \sum_{i,j} T_{i,j} \pr{\ket w, \ket{V_i}\bra{V_j} \ket v}
        \\
        &= \sum_{i,j} T_{i,j} \pr{\bra{V_j}^\dagger \ket{V_i}^\dagger \ket w, \ket v}
        = \sum_{i,j} T_{i,j} \pr{\ket{V_j} \bra{V_i} \ket w, \ket v}
        \\
        &= \pr{\sum_{i,j} T_{i,j}^* \ket{V_j} \bra{V_i} \ket w, \ket v}
        = \pr{T^\dagger \ket w, \ket v};
    \end{align}
    \item $I^\dagger = I$: segue da relação de completude e da propriedade anterior.
\end{itemize}

Essas propriedades permitem manipular transformações lineares adjuntas
sem ter que recorrer à definição frequentemente.

A definição de adjunto leva à definição de outros tipos de operadores,
\textit{e.g.} Hermitiano, projetores, normais e unitários.
Seja $V$ um espaço de Hilbert com dimensão $N$, então,
\begin{definition}
    Um operador $T: V \to V$ é Hermitiano se $T = T^\dagger$.
\end{definition}
\begin{definition}
    Seja $W$ um subespaço de $V$ com bases ortonormais
    $\set{\ket{W_i}} \subseteq \set{\ket{V_j}}$, respectivamente.
    Define-se um projetor de $V$ em $W$ por
    \begin{align}
        P = \sum_i \ket{W_i} \bra{W_i}.
    \end{align}
\end{definition}
Um projetor $P$ é Hermitiano já que
\begin{align}
    P^\dagger = \pr{\sum_i \ket{W_i} \bra{W_i}}^\dagger
    = \sum_i \bra{W_i}^\dagger \ket{W_i}^\dagger
    = \sum_i \ket{W_i} \bra{W_i} = P,
\end{align}
e $P^2 = P$ já que
\begin{align}
    P^2 = \pr{\sum_i \ket{W_i} \bra{W_i}} \pr{\sum_j \ket{W_j} \bra{W_j}}
    = \sum_{i,j} \ket{W_i} \delta_{i,j} \bra{W_j}
    = \sum_i \ket{W_i} \bra{W_i} = P.
\end{align}

\begin{definition}
    Um operador $T: V \to V$ é normal se $T^\dagger T = T T^\dagger$.
\end{definition}

Note que operadores Hermitianos são normais.
Operadores normais têm uma representação especial dada pelo Teorema Espectral.
\begin{theorem} \label{teo:espectral}
    Teorema Espectral:
    Um operador normal $T: V \to V$ é diagonalizável.
    Isto é, existem valores $\lambda_i \in \C$ e
    uma base ortonormal $\set{\ket{\lambda_i}}$ de $T$ tal que
    \begin{align}
        T = \sum_i \lambda_i \ket{\lambda_i}\bra{\lambda_i} =
            \matrx{
                \lambda_1 & 0 & \cdots & 0 \\
                0 & \lambda_2 & \cdots & 0 \\
                \vdots & \vdots & \ddots & \vdots \\
                0 & 0 & \cdots & \lambda_N
            } .
    \end{align}
\end{theorem}

Devido ao nome do Teorema, a base $\set{\ket{\lambda_i}}$ é comumente chamada de
\emph{decomposição espectral} de $T$.
Note que $\set{\lambda_i}$ permite calcular a saída de $T$ de modo simples,
já que
\begin{align}
    T \ket{\lambda_j} = \sum_i \lambda_i \ket{\lambda_i} \braket{\lambda_i | \lambda_j}
    = \sum_i \lambda_i \ket{\lambda_i} \delta_{i,j} = \lambda_j \ket{\lambda_j}
\end{align}
e para qualquer vetor $\ket v$ de $V$, $\ket v = \sum_j v_j \ket{\lambda_j}$ com $v_j \in \C$,
\begin{align}
    T \ket v = T \sum_j v_j \ket{\lambda_j} = \sum_j v_j \lambda_j \ket{\lambda_j}.
\end{align}
Por proporcionarem tal simplicidade,
atribui-se nomes específicos para esses valores e vetores.
\begin{definition}
    Para uma transformação linear $T: V \to V$,
    seja um vetor $\ket\lambda \in V$ e um valor $\lambda \in \C$ tais
    \begin{align}
        T \ket\lambda = \lambda \ket\lambda.
    \end{align}
    Então, diz-se $\lambda$ é um \emph{autovalor} de $T$ e que
    $\ket\lambda$ é um \emph{autovetor} de $T$ associado ao autovalor $\lambda$.
    Alternativamente, diz-se que $\ket\lambda$ é um $\lambda$-autovetor de $T$.
\end{definition}

Um tipo especial de operador normal são os operadores unitários.
\begin{definition}
    Um operador $U : V \to V$ é unitário se $U^\dagger U = U U^\dagger = I$.
\end{definition}

Da definição segue que $U$ é invertível e sua inversa é $U^\dagger$.
Operadores unitários têm a propriedade de preservar o produto interno entre dois vetores,
consequentemente a norma se preserva.
Sejam $\ket u$, $\ket v \in V$, então
\begin{align}
    \pr{U\ket u, U \ket v} = \bra u U^\dagger U \ket v = \braket{u | v} .
\end{align}
Já que o produto interno é preservado,
ao tomar uma base $\set{V_i}$ de $V$ e denotando $\ket{U_i} = U \ket{V_i}$,
conclui-se que $U = \sum_i \ket{U_i}\bra{V_i}$.
Note que esse resultado é consistente com a eq. \ref{eq:outer-prod-T} e que
$\braket{V_i | V_j} = \braket{U_i | U_j}$.
Uma consequência da preservação de norma é que os autovalores de $U$
também têm norma 1, \textit{i.e.}
se $\ket{\lambda_j}$ é um $\lambda_j$-autovetor de $U$,
então $\lambda_j = e^{\ii\theta_j} = \exp{\ii\theta_j}$ onde
$e$ é o número de Euler e $\theta_j \in \mathbb{R}$.
Note que
\begin{align}
    U \ket{\lambda_j} = e^{\ii\theta_j}\ket{\lambda_j}
    \implies \card{\card{e^{\ii\theta_j}\ket{\lambda_j}}}^2
    = \bra{\lambda_j} e^{-\ii\theta_j} e^{\ii\theta_j} \ket{\lambda_j}
    = \braket{\lambda_j | \lambda_j} = 1,
\end{align}
conforme desejado.

Sejam $V_1$ e $V_2$ espaços de Hilbert com dimensões $N_1$ e $N_2$, respectivamente.
Analisa-se o problema de utilizar $V_1$ e $V_2$ para representar um espaço vetorial de maior dimensão.
Mais especificamente, o espaço de Hilbert $V = V_1 \otimes V_2$ (leia-se $V_1$ tensor $V_2$)
com dimensão $N_1 N_2$.
O símbolo $\otimes$ indica produto tensorial, cuja definição é dada a seguir.
\begin{definition}
    Sejam $V_1$ e $V_2$ espaços vetoriais, $\ket{v_1}, \ket{v_1'} \in V_1$,
    $\ket{v_2}, \ket{v_2'} \in V_2$ e $c \in \C$.
    Então um produto tensorial $V_1 \otimes V_2$ satisfaz as seguintes propriedades.
    \begin{itemize}
        \item $c \pr{\ket{v_1} \otimes \ket{v_2}} = \pr{c \ket{v_1}} \otimes \ket{v_2}
            = \ket{v_1} \otimes \pr{c \ket{v_2}}$;
        \item $\pr{\ket{v_1} + \ket{v_1'}} \otimes \ket{v_2} =
            \pr{\ket{v_1} \otimes \ket{v_2}} + \pr{\ket{v_1'} \otimes \ket{v_2}}$;
        \item $\ket{v_1} \otimes \pr{\ket{v_2} + \ket{v_2'}} =
            \pr{\ket{v_1} \otimes \ket{v_2}} + \pr{\ket{v_1} \otimes \ket{v_2'}}$.
    \end{itemize}
\end{definition}

Em outras palavras, o produtor tensorial mantém a linearidade em ambos os subespaços separados e
``estende'' a linearidade para $V_1 \otimes V_2$.
Note também que
\begin{align}
    \pr{\ket{v_1} + \ket{v_1'}} \otimes \pr{\ket{v_2} + \ket{v_2'}} =
        \ket{v_1} \otimes \ket{v_2} + \ket{v_1'} \otimes \ket{v_2} +
        \ket{v_1} \otimes \ket{v_2} + \ket{v_1} \otimes \ket{v_2'}
\end{align}
é a combinação linear de quatro vetores em $V_1 \otimes V_2$.
Seguindo esse mesmo raciocínio,
para bases ortonormais $\set{V_{1,i}}$ de $V_1$ e $\set{V_{2,j}}$ de $V_2$,
qualquer vetor $\ket v \in V$ pode ser escrito como uma combinação linear
\begin{align}
    \pr{\sum_i v_{1,i} \ket{V_{1, i}}} \otimes \pr{\sum_j v_{2,j} \ket{V_{2, j}}}
    = \sum_{i,j} v_{1,i} v_{2, j} \ket{V_{1,i}} \otimes \ket{V_{2,j}},
\end{align}
onde $v_{1,i}, v_{2,j} \in \C$.
Perceba que isso resulta numa base $N_1 N_2$-dimensional de $V$.
A representação matricial pode ajudar no entendimento.
Considerando $\ket{v_1} \otimes \ket{v_2} \in V$, então, para $v_{1,i}, v_{2, j} \in \C$,
\begin{align}
    \ket{v_1} \otimes \ket{v_2} = \pr{\sum_i v_{1,i}} \otimes \ket{v_2} =
    \matrx{v_{1, 1} \ket{v_2} \\ \vdots \\ v_{1, N_1} \ket{v_2}}
    = \matrx{v_{1, 1} v_{2, 1} \\ \vdots \\ v_{1, 1} v_{2, N_2} \\ v_{1, 2} v_{2, 1} \\
        \vdots \\ v_{1, 2} v_{2, N_2} \\ \vdots \\ v_{1, N_1} v_{2, N_2}}.
\end{align}
Para simplificar a notação,
é comum representar $\ket{v_1} \otimes \ket{v_2} = \ket{v_1} \ket{v_2} = \ket{v_1, v_2} = \ket{v_1 v_2}$.

É interessante também estender os conceitos vistos até então em $V_1$ e $V_2$ para o espaço $V$ --
\textit{e.g.} transformações lineares e produtos internos --
de modo que a linearidade seja mantida.

Sejam $T_1 : V_1 \to V_1'$ e $T_2 : V_2 \to V_2'$ transformações lineares, onde
$V_1'$ e $V_2'$ possuem dimensão $M_1$ e $M_2$, respectivamente.
Então, define-se a transformação linear $T = T_1 \otimes T_2 : V_1 \otimes V_2 \to V_1' \otimes V_2'$
conforme a equação
\begin{align}
    \pr{T_1 \otimes T_2} \pr{\ket{v_1} \otimes \ket{v_2}} = \pr{T_1 \ket{v_1}} \otimes \pr{T_2 \otimes \ket{v_2}}.
\end{align}
A representação matricial de $T_1 \otimes T_2$ é análoga à de vetores e é dada por
\begin{align}
    T_1 \otimes T_2 &= \matrx{
        T_{1, 1,1} T_2 & T_{1, 1,2} T_2 & \cdots & T_{1, 1,N_1} T_2 \\
        T_{1, 2,1} T_2 & T_{1, 2,2} T_2 & \cdots & T_{1, 2,N_1} T_2 \\
        \vdots & \vdots & \ddots & \vdots \\
        T_{1, M_1,1} T_2 & T_{1, M_1,2} T_2 & \cdots & T_{1, M_1,N_1} T_2
    },
\end{align}
onde $T_{1,i,j}$ são as entradas de $T_1$.
Note que nem toda transformação linear $T$ pode ser representada usando
um único produto tensorial.
Por exemplo, seja
\begin{align}
    \CNOT = \matrx{
        1 & 0 & 0 & 0 \\
        0 & 1 & 0 & 0 \\
        0 & 0 & 0 & 1 \\
        0 & 0 & 1 & 0
    }
\end{align}
e deseja-se encontrar $A, B: \hilb^2 \to \hilb^2$ tal que $\CNOT = A \otimes B$.
Isso implicaria que
\begin{align}
    \CNOT = \matrx{a_{1,1} B & a_{1,2} B \\ a_{2,1} B & a_{2,2} B}
    = \matrx{
        a_{1,1} b_{1,1} & a_{1,1} b_{1,2} & a_{1,2} b_{1,1} & a_{1,2} b_{1,2} \\
        a_{1,1} b_{2,1} & a_{1,1} b_{2,2} & a_{1,2} b_{2,1} & a_{1,2} b_{2,2} \\
        a_{2,1} b_{1,1} & a_{2,1} b_{1,2} & a_{2,2} b_{1,1} & a_{2,2} b_{1,2} \\
        a_{2,1} b_{2,1} & a_{2,1} b_{2,2} & a_{2,2} b_{2,1} & a_{2,2} b_{2,2}
    },
\end{align}
onde $a_{i,j}, b_{i,j} \in \C$ são respectivamente os elementos de $A$ e $B$
para $i,j \in \set{1,2}$.
Essa equação implica que $a_{1,1}b_{1,1} = a_{2,2}b_{1,2} = 1$ e
$a_{2,2}b_{1,1} = 0$, o que gera a seguinte contradição:
como $a_{2,2}b_{1,1} = 0$, $a_{2,2} = 0$ ou $b_{1,1} = 0$;
se $a_{2,2} = 0$, então $a_{2,2}b_{1,2} \neq 1$.
se $b_{1,1} = 0$, então $a_{1,1}b_{1,1} \neq 1$.
Então conlui-se que $\CNOT \neq A \otimes B$ para qualquer $A, B \in \hilb^2$.

O produto interno em $V$ segue de uma maneira muito similar e
utilizando os produtos internos que já foram definidos para $V_1$ e $V_2$.
Sendo $\ket v = \sum_i v_i \ket{v_{1, i}} \otimes \ket{v_{2, i}} \in V$ e
$\ket{v'} = \sum_j v_j' \ket{v_{1, j}'} \otimes \ket{v_{2, j}'} \in V$
onde $v_i, v_j' \in \C$, o produto interno $\braket{v | v'}$ é dado por
\begin{align}
    \pr{\sum_i v_i \ket{v_{1, i}} \otimes \ket{v_{2, i}}, \sum_j v_j' \ket{v_{1, j}'} \otimes \ket{v_{2, j}'}}
    = \sum_{i,j} v_i^* v_j' \braket{v_{1, i} | v_{1, j}'} \braket{v_{2, i} | v_{2, j}'} .
\end{align}
A partir da qual seguem as definições de transformação linear adjunta, operadores Hermitianos e operadores unitários.
Em particular, vale salientar que para uma transformação linear
$T = T_1 \otimes T_2$ com domínio em $V_1 \otimes V_2$,
\begin{align}
    T^\dagger = \pr{T_1 \otimes T_2}^\dagger = T_1^\dagger \otimes T_2^\dagger .
\end{align}
Além disso, se $T_1$ e $T_2$ forem operadores unitários,
então $T$ também é unitário.
Uma notação útil utilizada no trabalho visa compactar múltiplos produtos tensoriais:
sendo $\ket v, \ket {v_i}$ vetores de um espaço de Hilbert e $n \in \mathbb{N}$,
\begin{align}
    \bigotimes_{i=0}^{n-1} \ket{v_i} = \ket{v_0} \otimes \ket{v_1} \otimes \cdots \otimes \ket{v_{n-1}};
\end{align}
analogamente,
\begin{align}
    \ket{v}^{\otimes n} = \bigotimes_{i = 0}^{n-1} \ket{v}.
\end{align}
O mesmo se aplica a transformações lineares.

Uma última definição que será necessária é a do operador $\oplus$,
utilizado para representar soma direta.
\begin{definition}
    Sejam $V_1$, $V_2$ e $V$ espaços de Hilbert com o mesmo produto interno e
    com dimensões $N_1$, $N_2$, $N = N_1 + N_2$, respectivamente.
    É possível descrever $V$ em termos da soma direta ($\oplus$) de $V_1$ e $V_2$.
    Para todo $\ket v \in V$ existem $\ket{v_1} \in V_1$ e $\ket{v_2} \in V_2$
    tais que
    \begin{align}
        \ket v = \ket{v_1} \oplus \ket{v_2} = \matrx{\ket{v_1} \\ \ket{v_2}}
        = \matrx{v_{1,1} \\ \vdots \\ v_{1, N_1} \\ v_{2, 1} \\ \vdots \\ v_{2, N_2}},
    \end{align}
    onde $v_{1,1}, \cdots v_{1, N_1} \in \C$ são os elementos de $\ket{v_1}$ e
    $v_{2,1}, \cdots v_{2, N_2} \in \C$ são os elementos de $\ket{v_2}$.
\end{definition}
\subsection{Mecânica Quântica -- Postulados} \label{sec:postulados}
Nesta Seção são listados os postulados da Mecânica Quântica.
Esses postulados descrevem como os conceitos da Seção \ref{sec:alg-lin} são utilizados
para descrever a Mecânica Quântica.
Um computador quântico precisa agir de acordo com as regras definidas por esses postulados.

O primeiro postulado diz respeito ao escopo onde a Mecânica Quântica atua.
\begin{definition}
    Postulado do Espaço dos estados:
    Um sistema físico isolado é descrito por um espaço de Hilbert e
    o estado em que esse sistema se encontra é descrito por um vetor unitário nesse espaço de Hilbert.
\end{definition}
Como mencionado na seção \ref{sec:alg-lin},
ao longo deste trabalho utiliza-se espaços de Hilbert com dimensão $N$.
Esse espaço de estados é representado por $\hilb^N$ e vetores que pertecem a esse estado
$\ket v \in \hilb^N$.
Vetores $\ket v$ podem ser descritos unicamente como uma combinação linear
$\sum_i v_i \ket{V_i}$ dados valores $v_i \in \C$ (denominados de amplitudes) e
uma base $\set{\ket{V_i}}$ de $V$.
Normalmente utiliza-se uma base ortonormal.
Quando um estado $\ket{v}$ é representado como uma combinação linear de outros estados --
\textit{e.g.} $\ket{v} = v_0 \ket{V_0} + v_1 \ket{V_1}$ para $v_0, v_1 \neq 0$ --
diz-se que $\ket{v}$ é uma \emph{sobreposição} dos estados $\ket{V_0}$ e $\ket{V_1}$.

Para realizar computação, é interessante que os estados mudem e que
haja algum modo de descrever essa evolução.
\begin{definition}
    Postulado da Evolução: A evolução de um sistema quântico isolado é descrita
    por um operador unitário $U$ e dependendo do tempo.
    Seja $\ket{\psi_0}$ o estado correspondente ao instante de tempo $t_0$,
    o estado no instante de tempo seguinte $t_1$ é dado por
    $\ket{\psi_1} = U \ket{\psi_0}$.
\end{definition}
Esse postulado permite que a norma dos estados seja mantida
(ou seja, trabalha-se apenas com estados unitários).
Normalmente, para facilitar a análise,
decompõ-se $U$ em diversos operadores unitários -- e.g $U = U_1 U_0$ --
e a análise é feita passo a passo: sendo o estado inicial $\ket{\psi_0}$,
o estado no passo 1 é $\ket{\psi_1} = U_0 \ket{\psi_0}$ e
o estado no passo 2 é $\ket{\psi_2} = U_1 \ket{\psi_1}$;
sendo $\ket{\psi_f}$ o estado final, nesse exemplo tem-se que
$\ket{\psi_f} = U\ket{\psi_0} = \ket{\psi_2}$.

Para realizar computação, é necessário obter algum resultado,
mas infelizmente, obter explicitamente o estado final $\ket{\psi_f}$ não é fácil.
Inclusive, é comum $\ket{\psi_f}$ ser uma sobreposição de outros estados -- \textit{e.g.} sobreposição de $\set{\ket{V_i}}$ --
cuja obtenção é mais fácil.
\begin{definition}
   Postulado da Medição: Medições de um estado são descritas por
   um conjunto de \emph{operadores de medida} $\set{M_i}$ e os
   índices $i$ referem-se aos possíveis resultados.
   Dado um vetor $\ket{\psi}$, a probabilidade de obter $i$ após a medição
   é dada por
   \begin{align}
       \prob{i} = \bra{\psi}M_i^\dagger M_i \ket{\psi}
   \end{align}
   e o estado correspondente $\ket{\psi_i}$ obtido após a medida é dado por
   \begin{align}
       \ket{\psi_i} = \frac{M_i \ket{\psi}}{\sqrt{\prob{i}}}.
   \end{align}
   Como a soma das probabilidades precisa ser igual a 1,
   \begin{align}
       1 = \sum_i \prob{i} = \sum_i \bra{\psi}M_i^\dagger M_i \ket{\psi}
       = \braket{\psi | \psi},
   \end{align}
   conclui-se que que os operadores de medida satisfazem a relação de completude:
   \begin{align}
       \sum_i M_i^\dagger M_i = I.
   \end{align}
\end{definition}

Note que o fato de $\ket{\psi}$ precisar ser unitário está
intrinsecamente relacionado com a soma das probabilidades resultar em 1.
Além disso, suponha que $\ket\psi = \sum_i v_i \ket{V_i}$ e
note que $\set{M_i} = \set{\ket{V_i}\bra{V_i}}$ é um conjunto de operadores de medida
e que a relação de completude é respeitada:
\begin{align}
    \sum_i M_i^\dagger M_i &= \sum_i \pr{\ket{V_i}\bra{V_i}}^\dagger
        \pr{\ket{V_i}\bra{V_i}}
    = \sum_i \pr{\ket{V_i}\bra{V_i}}\pr{\ket{V_i}\bra{V_i}}
    \\
    &= \sum_i \ket{V_i} 1 \bra{V_i}
    = I .
\end{align}
Espera-se que as amplitudes de $\ket{V_i}$ influenciem
na probabilidade de obtenção de cada estado.
De fato,
\begin{align}
    \prob{i} &= \sum_j v_j^* \bra{V_j} M_i^\dagger M_i \sum_k v_k \ket{V_k}
    \\
    &= \sum_j v_j^* \bra{V_j} \ket{V_i} \bra{V_i} \sum_k v_k \ket{V_k}
    \\
    &= \sum_j v_j^* \delta_{j,i} \sum_k v_k \delta{i, k}
    \\
    &= v_i^* v_i = \card{v_i}^2.
\end{align}
Logo, o estado obtido após a medida é
\begin{align}
    \ket{\psi_i} &= \frac{\ket{V_i}\bra{V_i} \sum_j{v_j}\ket{V_j}}{\card{v_i}}
    \\
    &= \frac{\ket{V_i} \sum_j{v_j} \delta_{i,j}}{\card{v_i}}
    \\
    &= \frac{v_i \ket{V_i}}{\card{v_i}} .
\end{align}
Observe que esses operadores de medida foram definidos de forma a
estarem diretamente relacionados com o estado obtido após a medição --
\textit{i.e.} $M_i \implies \ket{V_i}$.
Além disso, $v_i/\card{v_i} = \exp\pr{\ii\theta_i}$ para algum
$\theta_i \in \mathbb{R}$.
Diz-se que $\exp\pr{\ii\theta_i}$ é a fase de $\ket{V_i}$ e
dependendo do contexto, essa fase pode ser ignorada após a medida já que
para dois vetores $\ket\psi$ e $e^{\ii\theta}\ket\psi$,
$e^{-\ii\theta} \bra\psi M_i^\dagger M_i e^{\ii\theta} \ket{\psi} =
\bra\psi M_i^\dagger M_i \ket{\psi}$.

Considerando um exemplo mais concreto, considere a medição de
$\ket +$ em relação aos operadores de medida
$M_0 = \ket 0 \bra 0$ e $M_1 = \ket 1 \bra 1$:
\begin{align}
    \prob{0} = \prob{1} = \pr{\frac{1}{\sqrt 2}}^2 = \frac{1}{2},
\end{align}
e os possíveis estados após a medição são
\begin{align}
    \ket{\psi_0} = \frac{\frac{1}{\sqrt 2} \ket{0}}{\card{1/\sqrt{2}}}
    = \ket{0}
    \quad \textnormal{ e }
    \quad \ket{\psi_1} = \ket 1.
\end{align}

Por último, resta analisar como sistemas quânticos compostos se comportam --
\textit{i.e.} dois ou mais sistemas distintos evoluindo concomitantemente.
Para isso, utiliza-se a notação tensorial.
\begin{definition}
   Postulado dos sistemas compostos:
   Sejam $V_1, V_2$ espaços de Hilbert.
   A composição desses sistemas é dada por $V = V_1 \otimes V_2$.
   Sendo assim, se $\ket{v_1} \in V_1$ e $\ket{v_2} \in V_2$,
   a composição desses estados
   pode ser representada como $\ket v = \ket{v_1} \otimes \ket{v_2}$,
   onde $\ket v \in V$.
   Estados $\ket v \in V$ que \emph{não} podem ser descritos dessa forma
   são chamados de estados emaranhados --
   \textit{i.e.} $\ket v \neq \ket{v_1} \otimes \ket{v_2}$ para qualquer $\ket{v_1} \in V_1$ e
   $\ket{v_2} \in V_2$.
\end{definition}

A evolução do sistema composto $V$ também é dada por operadores unitários,
o que inclui $U_1 \otimes U_2$ desde que
$U_1$ seja unitário em $V_1$ e $U_2$ unitário em $V_2$.
Para este trabalho,
é útil analisar o que acontece ao realizar uma medição num subsistema de $V$.
Considere que deseja-se fazer uma medição no primeiro subsistema do estado
$\ket v = \sum_{i} c_i \ket{v_i} \otimes \ket{w_i}$
onde $c_i \in \C$, $\ket{v_i} \in V_1$ e $\ket{w_i} \in V_2$.
Como não deseja-se medir o segunda subsistema,
considere os operadores de medida $\set{M_k} = \set{\ket{v_k}\bra{v_k} \otimes I}$.
Então, a probabilidade de obter o resultado $k$ é:
\begin{align}
    \prob{k} &= \bra{v} M_k^\dagger M_k \ket{v}
    \\
    &= \sum_{i} c_i^* \bra{v_i} \otimes \bra{w_i}
        \pr{\ket{v_k}\bra{v_k} \otimes I}^\dagger
        \pr{\ket{v_k}\bra{v_k} \otimes I}
        \sum_{j} c_{j} \ket{v_j} \otimes \ket{w_j}
    \\
    &= \sum_{i,j} c_i^* c_j \bra{v_i} \otimes \bra{w_i}
        \pr{\ket{v_k}\bra{v_k} \otimes I}
        \ket{v_j} \otimes \ket{w_j}
    \\
    &= \sum_{i,j} c_i^* c_j \braket{v_i|v_k} \braket{v_k|v_j} \braket{w_i|w_j}
    \\
    &= \card{c_k}^2;
\end{align}
e o estado após a medida é
\begin{align}
    \frac{M_k \ket{v}}{\sqrt{\prob{k}}} &=
    \frac{1}{\card{c_k}} \pr{\ket{v_k}\bra{v_k} \otimes I }
        \sum_{i} c_i \ket{v_i} \otimes \ket{w_i}
    \\
    &= \frac{1}{\card{c_k}}
        \sum_i \pr{c_i \ket{v_k} \braket{v_k | v_i}} \otimes \ket{w_i}
    \\
    &= \frac{c_k}{\card{c_k}} \ket{v_k} \otimes \ket{w_k} .
\end{align}
Perceba que nessa situação,
os estados do primeiro subsistema estavam diretamente associados a
estados do segundo subsistema e como consequência,
ao realizar a medição do primeiro subsistema,
o segundo também foi afetado.
\subsection{Circuitos Quânticos} \label{sec:circuitos-quanticos}

Nessa seção, introduz-se a notação de circuitos que é comum na
Computação Quântica para descrever ou ilustrar algoritmos.
A representação de circuitos é equivalente à representação de matrizes
e é de sumo interesse para implementação de algoritmos em computadores quânticos,
já que um circuito descreve um algoritmo.

Num computador clássico a menor unidade de informação é representada por um \emph{bit}.
Um bit $b$ pode codificar, por exemplo,
se há ausência (valor 0) ou presença (valor 1) de corrente elétrica.
O bit quântico (\emph{qubit}) $\ket b$ é análogo, mas utiliza vetores:
um paralelo é feito entre $\ket 0$ e 0, e entre $\ket 1$ e 1.
Análogo a um circuito clássico,
um circuito quântico possui um (ou mais) fio,
representando um qubit.
A informação flui da esquerda pra direita e pode ser alterada
de acordo com uma \emph{porta}.

\begin{figure}[htb]
    \centering
    \caption{Exemplo de circuito: porta NOT.}
    \includegraphics[width=0.5\textwidth]{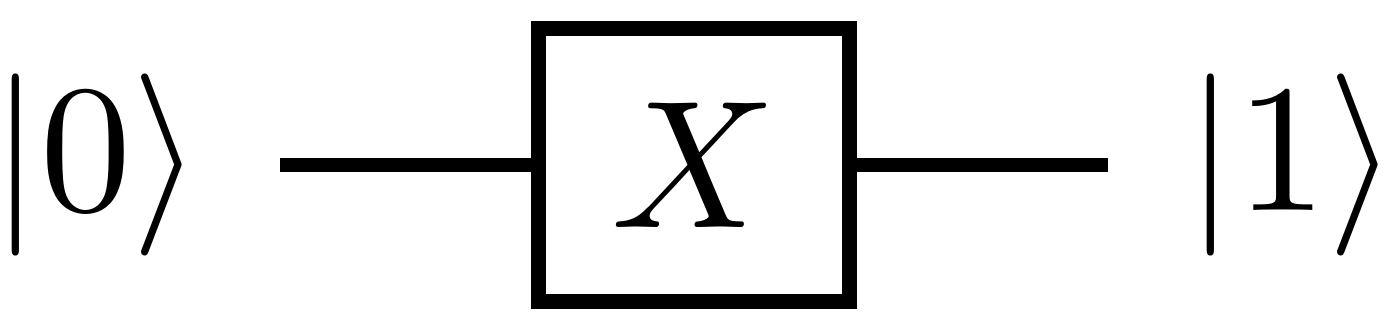}
    \legend{Fonte: Produzido pelo autor.}
    \label{fig:circ-X}
\end{figure}

Por exemplo, o circuito da Fig. \ref{fig:circ-X} possui um qubit.
A entrada do circuito está representada à esquerda do fio ($\ket 0$).
O bloco $X$ no meio do circuito representa uma porta cuja ação no qubit é
descrita por uma matriz unitária (conforme descrito nos postulados).
No caso,
\begin{align}
    X = \matrx{0 & 1 \\ 1 & 0} \implies
    X\ket 0 = \matrx{0 & 1 \\ 1 & 0} \matrx{1 \\ 0}
    = \matrx{0 \\ 1} = \ket 1.
\end{align}
Sendo assim, a ação da porta $X$ no circuito da Fig. \ref{fig:circ-X} é
transformar $\ket 0$ em $\ket 1$, que é a saída do circuito,
representada à direita do fio.
Como $X$ mapeia $\ket 0 \to \ket 1$ e $\ket 1 \to \ket 0$,
invertendo a informação do qubit, $X$ é chamada de porta NOT.
Já que há uma equivalência entre portas e matrizes unitárias,
o longo deste trabalho,
os termos porta e operador unitário são usados como sinônimos.

\begin{figure}[htb]
    \centering
    \caption{Exemplo de circuito com múltiplas portas e estados intermediários.}
    \includegraphics[width=0.5\textwidth]{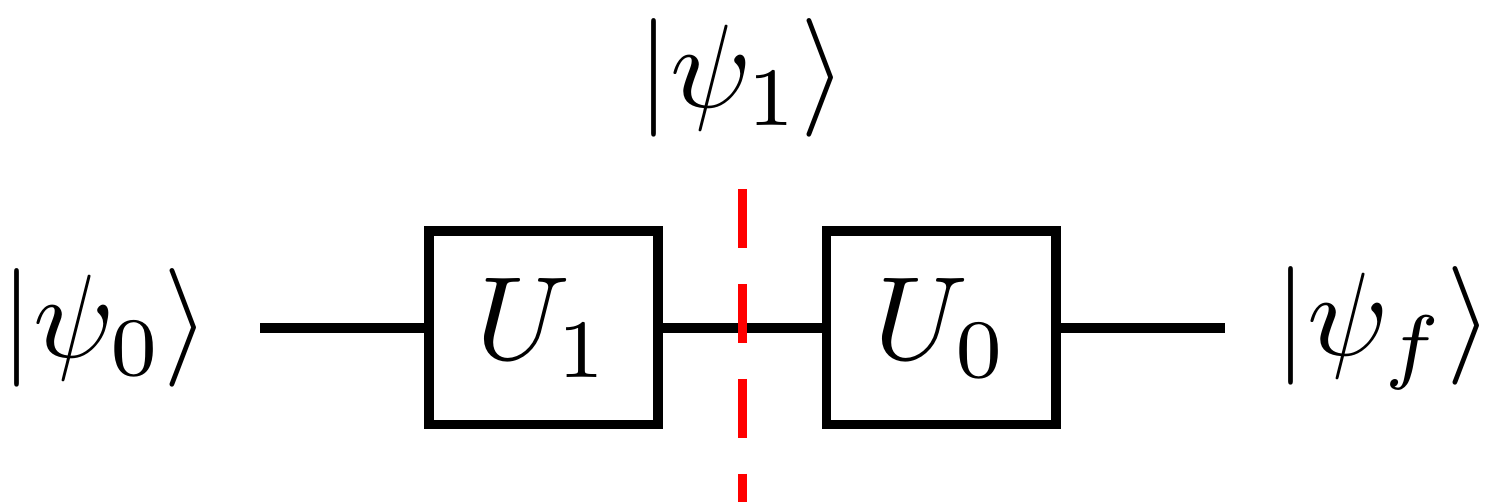}
    \legend{Fonte: Produzido pelo autor.}
    \label{fig:circ-mult-op}
\end{figure}

Um operador unitário $U$ pode ser decomposto como um produto de vários
operadores unitários --
\textit{e.g.} $U = U_0 U_1 \cdots U_n = \prod_{i=1}^n U_i$ onde $n \in \N$.
Essa decomposição pode facilitar a análise e a concepção de circuitos quânticos.
Considere o circuito apresentado na Fig. \ref{fig:circ-mult-op}.
A entrada do circuito é $\ket{\psi_0}$.
Primeiro aplica-se a porta $U_1$, resultado no estado intermediário
$\ket{\psi_1} = U_1 \ket{\psi_0}$
(estados intermediários são representados por linhas tracejadas que
``cortam'' o circuito na vertical para fins didáticos apenas).
Posteriormente, aplica-se a porta $U_0$ em $\ket{\psi_1}$,
resultando no estado final $\ket{\psi_f} = U_0 \ket{\psi_1}$.
Note que $\ket{\psi_f} = U_0 U_1 \ket{\psi_0}$,
portanto, a ordem em que as portas são representadas nos circuitos é
a ordem inversa de sua representação algébrica.

\begin{figure}[htb]
    \centering
    \begin{minipage}[t]{0.45\textwidth}
        \centering
        \caption{Circuito que inverte 2 qubits.}
        \includegraphics[width=\textwidth]{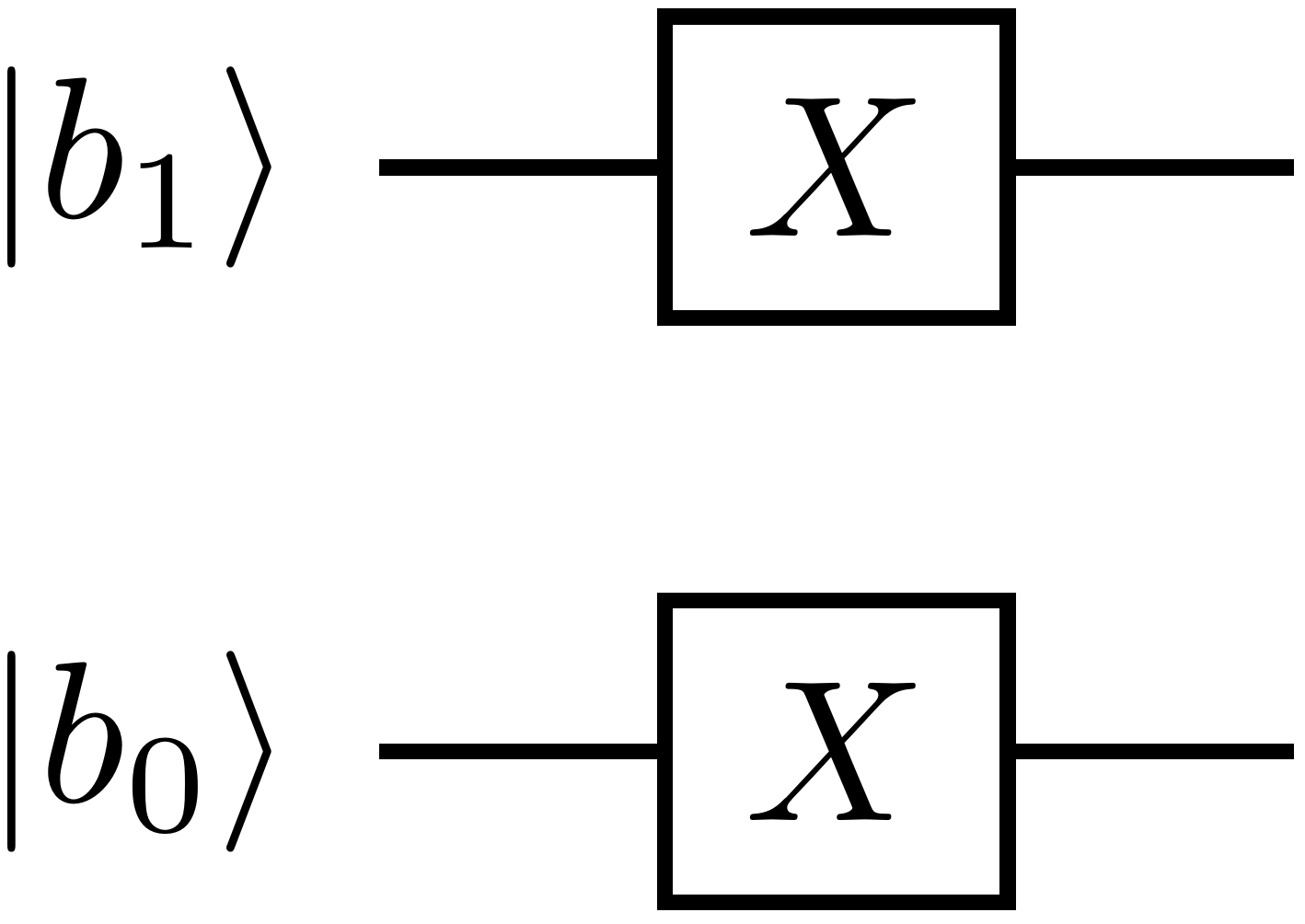}
        \legend{Fonte: Produzido pelo autor.}
        \label{fig:circ-X-2}
    \end{minipage}
    \hfill
    \begin{minipage}[t]{0.45\textwidth}
        \centering
        \caption{Circuito que inverte 2 qubits (compacto).}
        \includegraphics[width=\textwidth]{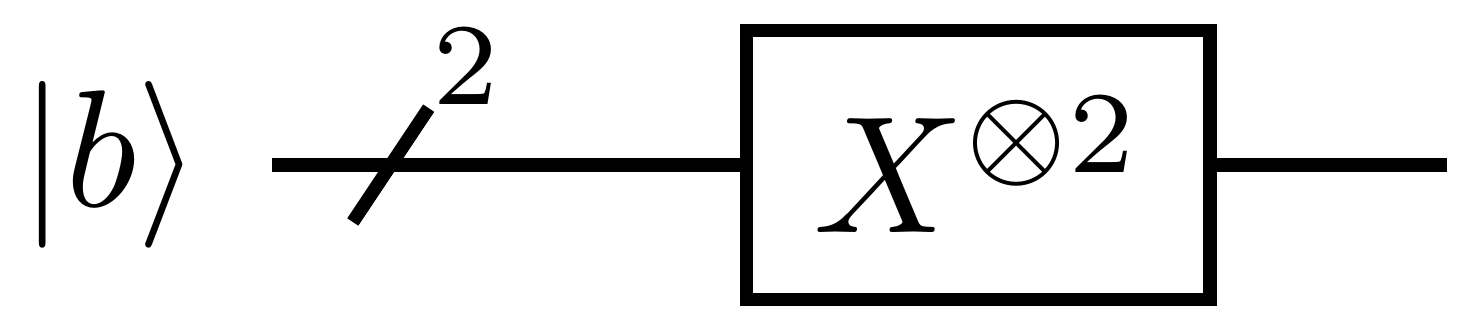}
        \legend{Fonte: Produzido pelo autor.}
        \label{fig:circ-X-2-comp}
    \end{minipage}
\end{figure}

Em computação, normalmente se trabalha com múltiplos bits de
modo a representar um número maior de informação.
Considere um sistema composto de 2 qubits: $\ket{b_1}$ e $\ket{b_0}$.
Pelos postulados, é possível representar o estado composto utilizando
produto tensorial: $\ket b = \ket{b_1} \otimes \ket{b_0}$
(desde que $\ket b$ não seja um estado emaranhado).
O circuito que inverte os dois qubits está representado na Fig. \ref{fig:circ-X-2}.
Usando produto tensorial, é possível representar o mesmo circuito de
forma mais compacta (Fig. \ref{fig:circ-X-2-comp}):
a linha na diagonal com rótulo 2 indica que estão sendo utilizados 2 qubits e
$X^{\otimes 2}$ que aplica-se a porta $X$ em ambos os qubits.
Em ambos os circuitos, a saída foi omitida.

\begin{figure}[htb]
    \centering
    \begin{minipage}[t]{0.45\textwidth}
        \centering
        \caption{Circuito que inverte n qubits.}
        \includegraphics[height=0.2\textheight]{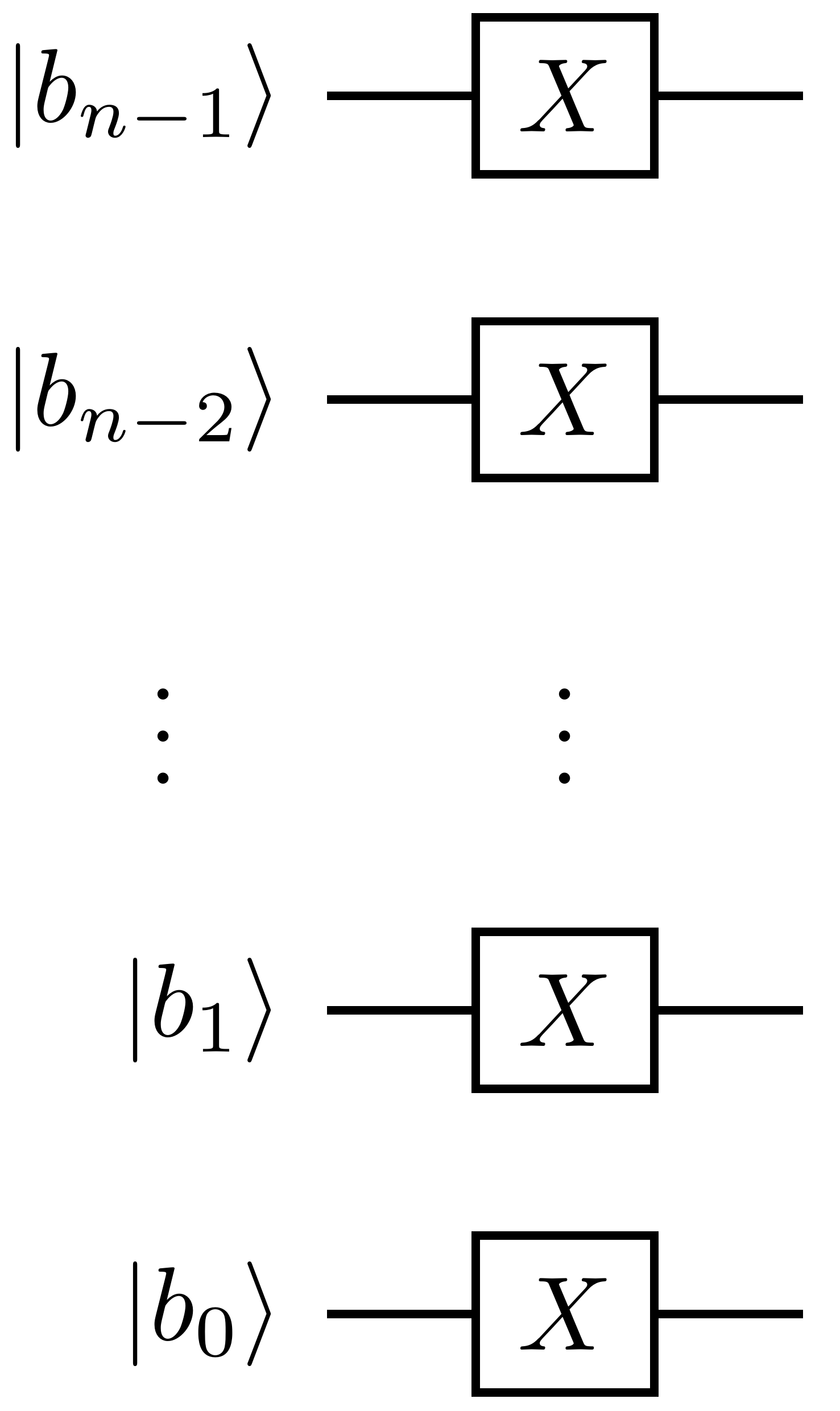}
        \legend{Fonte: Produzido pelo autor.}
        \label{fig:circ-X-n}
    \end{minipage}
    \hfill
    \begin{minipage}[t]{0.45\textwidth}
        \centering
        \caption{Circuito que inverte n qubits (compacto).}
        \includegraphics[width=\textwidth]{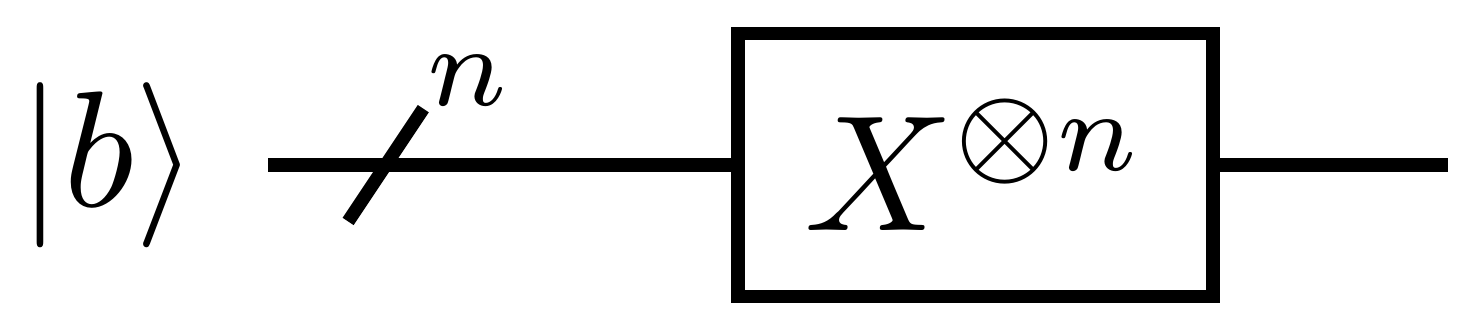}
        \legend{Fonte: Produzido pelo autor.}
        \label{fig:circ-X-n-comp}
    \end{minipage}
\end{figure}

O exemplo anterior pode ser facilmente generalizado para $n$ qubits
($n \in \N$ e $n \geq 1$) e tomando
\begin{align}
    \ket{b} = \ket{b_{n-1}}\ket{b_{n-2}} \cdots \ket{b_1} \ket{b_0}
    = \otimes_{i = 1}^n \ket{b_{n - i}}.
\end{align}
Fig. \ref{fig:circ-X-n} ilustra o circuito que inverte $n$ qubits e
a Fig. \ref{fig:circ-X-n-comp} ilustra o mesmo circuito de maneira compacta
usando produto tensorial.

Faz-se um paralelo dos qubits $\ket b$ com bits.
Ao concatenar vários bits, é possível representar números naturais.
Por exemplo, considere a sequência (ou cadeia) de bits
$b = b_{n-1}b_{n-2}\cdots b_1 b_0$ para $0 \leq i < n$.
O número natural $b_\N$ representado por essa cadeia de bits é
\begin{align}
    b_\N = \sum_{i = 0}^{n-1} b_i \cdot 2^i .
\end{align}
Diz-se que $b$ é a representação binária de um número natural e
$b_\N$ a representação decimal do mesmo número.
Como $b$ e $b_\N$ representam o mesmo número mudando apenas a representação,
utiliza-se um único rótulo para representar ambos -- \textit{e.g.} $b$ --
desde que especificado pelo contexto qual representação está sendo utilizada.
Desse modo, tanto $b$ quanto $\ket b$ são capazes de representes $2^n$ valores e
esses vetores possuem uma relação direta com a
\emph{base canônica} ou \emph{base computacional}.

\begin{definition}
    Seja $N \in \N$ tal que $N \geq 1$.
    Os $N$ vetores da base canônica (ou base computacional) são dados por
    \begin{align}
        \ket i = \matrx{b_0 \\ \vdots \\ b_i \\ \vdots \\ b_{N-1}},
    \end{align}
    onde $b_i = 1$, $b_j = 0$ se $j \neq i$, e $0 \leq i, j \leq N-1$.
    Por exemplo, o conjunto $\set{\ket 0, \ket 1}$ visto anteriormente é a base computacional de $\hilb^2$.
\end{definition}

Normalmente, $N$ é uma potência de 2,
traçando-se um paralelo direto de um qubit com um bit e
um fio de circuito por bit ou qubit (conforme retratado até aqui).
Tomar um valor de $N$ que não é potência de 2, entretanto,
pode facilitar manipulações algébricas dependendo do problema.
Nesse trabalho, ambas as situações são utilizadas (e, inclusive, combinadas).

\begin{figure}[htb]
    \centering
    \caption{Exemplo de circuito atuando em dois espaços de Hilbert
        ($\hilb^N \otimes \hilb^{N'}$).}
    \includegraphics[width=0.4\textwidth]{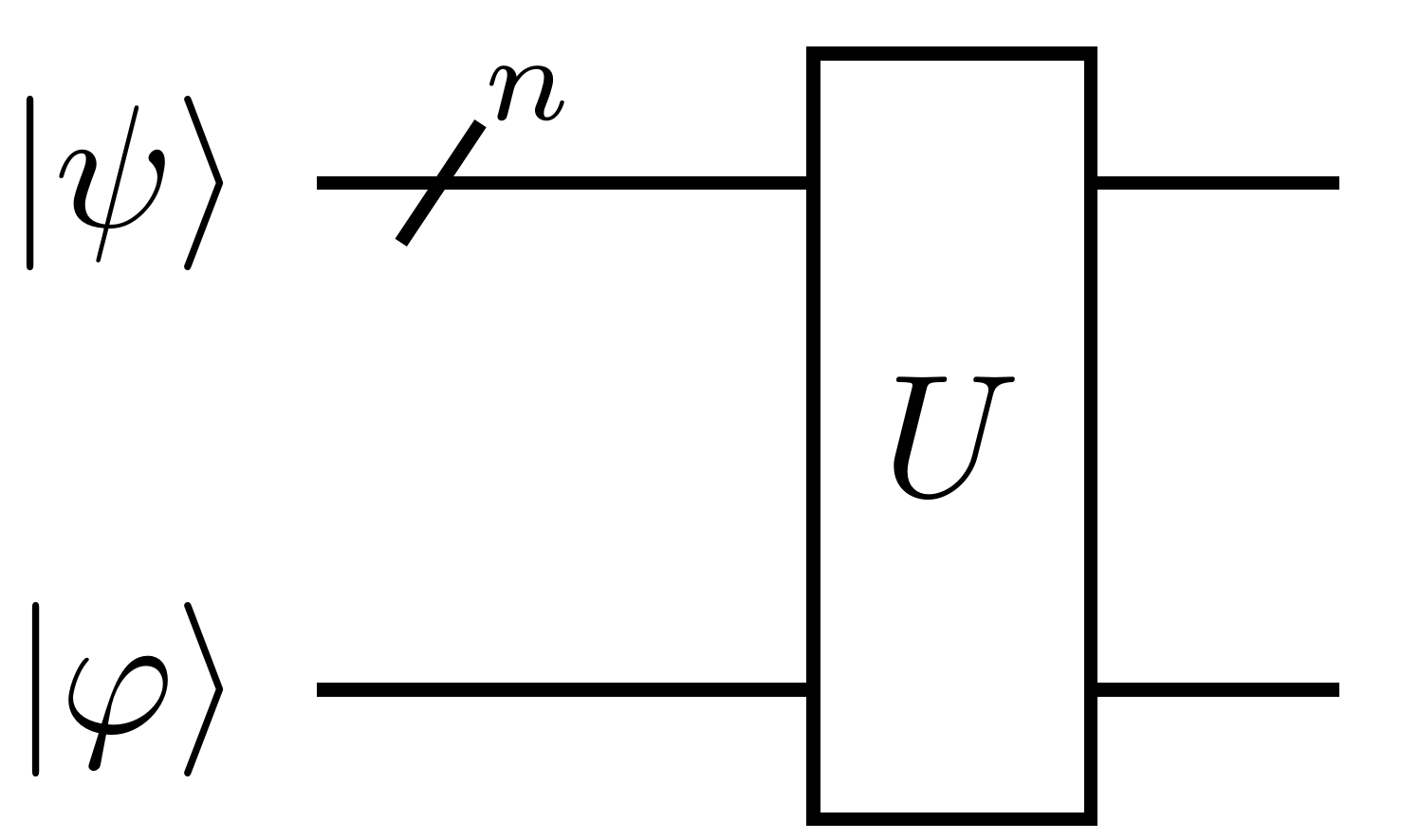}
    \legend{Fonte: Produzido pelo autor.}
    \label{fig:qudit-regist}
\end{figure}

Suponha que $N = 2^n$ mas $N'$ não é uma potência de 2 e
que deseja-se construir um circuito que atua em $\hilb^N \otimes \hilb^{N'}$.
Como $\hilb^{N'}$ pode ser interpretado como uma generalização de qubit --
\textit{i.e.} um qudit \footnote{um qubit com mais de 2 níveis.} \cite{thew2002qudit} --
sua representação num circuito é feita através de um único fio.
Suponha que $\ket\psi \in \hilb^N$ e $\ket\varphi \in \hilb^{N'}$ são vetores; e
$U: \hilb^N\otimes\hilb^{N'} \to \hilb^N\otimes\hilb^{N'}$ um operador linear.
Um circuito cuja ação é definida pela porta $U \otimes U'$ é ilustrado
na Fig. \ref{fig:qudit-regist}.
Nesse circuito, combina-se dois espaços de Hilbert e quando isso acontece,
normalmente cada espaço de Hilbert tem uma função diferente,
recebendo o nome de \emph{registradores}.
No caso da Fig. \ref{fig:qudit-regist},
$\hilb^N$ corresponde ao primeiro registrador e $\hilb^{N'}$ ao segundo.

Vale salientar que é possível representar $\hilb^{N'}$
usando um outro espaço de Hilbert $\hilb^{N'_2}$ sendo $N'_2$ uma potência de 2,
desde que $\hilb^{N'}$ seja um subespaço de $\hilb^{N'_2}$.
Isso é útil principalmente para a implementação de algoritmos
em computadores quânticos, tópico que foge do escopo do trabalho.

Retornando aos circuitos que usam apenas qubits,
introduz-se alguma portas que serão utilizadas ao longo do trabalho.
Especificamente: porta Hadamard, portas $\SWAP$ e portas controladas.
Para o resto da seção, considere que $n \in \N$, $n \leq 1$ e $N = 2^n$.

\begin{definition}
    A porta de Hadamard atua em $\hilb^2$ e é definida pela matriz
    \begin{align}
        H = \frac{1}{\sqrt 2} \matrx{1 & 1 \\ 1 & -1}.
    \end{align}
    A ação desse operador também pode ser entendida simplesmente por
    $\ket 0 \to \ket +$ e $\ket 1 \to \ket -$,
    onde $\ket - = \pr{\ket 0 - \ket 1}/\sqrt{2}$.
\end{definition}
A porta de Hadamard é unitária já que $H^2 = I$.
A porta de Hadamard é utilizada na maioria dos algoritmos,
sendo responsável pelo paralelismo quântico:
gera um estado que é uma sobreposição uniforme de todas as possíveis entradas
do algoritmo.
Note que para $\hilb^2$, tem-se $H\ket 0 = \pr{\ket 0 + \ket 1}/\sqrt{2}$;
já para $\hilb^4$,
\begin{align}
    H^{\otimes 2} \ket{0}^{\otimes 2} &= \frac{\ket 0 + \ket 1}{\sqrt 2}
        \otimes \frac{\ket 0 + \ket 1}{\sqrt 2}
    \\
    &= \frac{\ket{00} + \ket{01} + \ket{10} + \ket{11}}{2}
    \\
    &= \frac{1}{2} \sum_{i = 0}^{3} \ket{i}.
\end{align}
De um modo geral,
\begin{align}
    H^{\otimes n} \ket 0 = \frac{1}{\sqrt{N}} \sum_{i=0}^{N-1} \ket{i}.
\end{align}

\begin{definition}
    A operação SWAP que atua nos qubits indexados por $i$ e $j$ é representada por $\SWAP_{i, j}$.
    A ação da porta $\SWAP_{i,j}$ é trocar a informação representada
    pelo $i$-ésimo qubit com a do $j$-ésima qubit.
    Fig. \ref{fig:swap} ilustra uma porta $\SWAP$.
    \begin{figure}[htb]
    \centering
        \caption{Porta $\SWAP$.}
        \includegraphics[height=0.075\textheight]{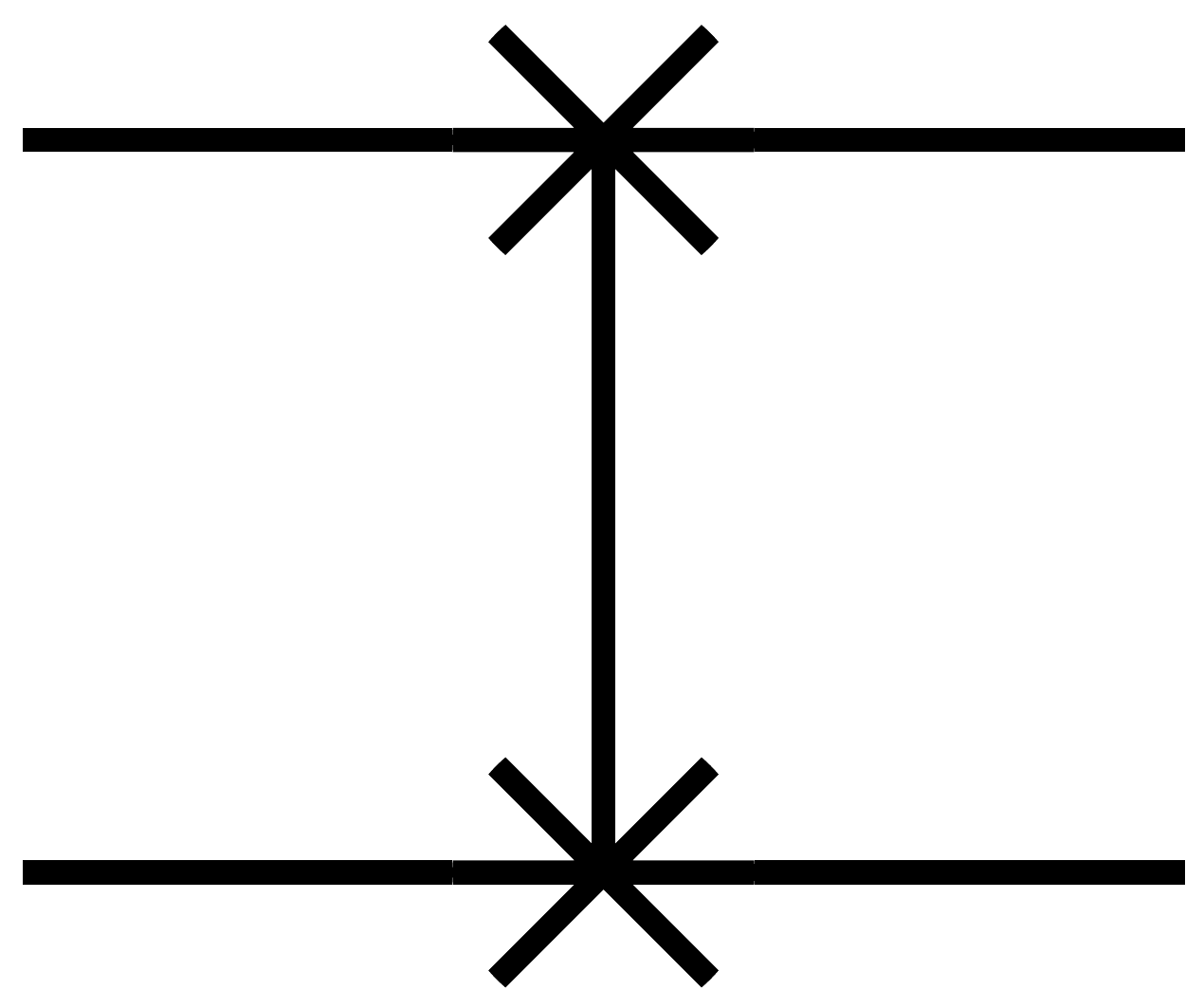}
        \legend{Fonte: Produzido pelo autor.}
        \label{fig:swap}
    \end{figure}
\end{definition}

\begin{figure}[htb]
\centering
    \caption{Exemplo da ação de uma porta $\SWAP$.}
    \includegraphics[height=0.1\textheight]{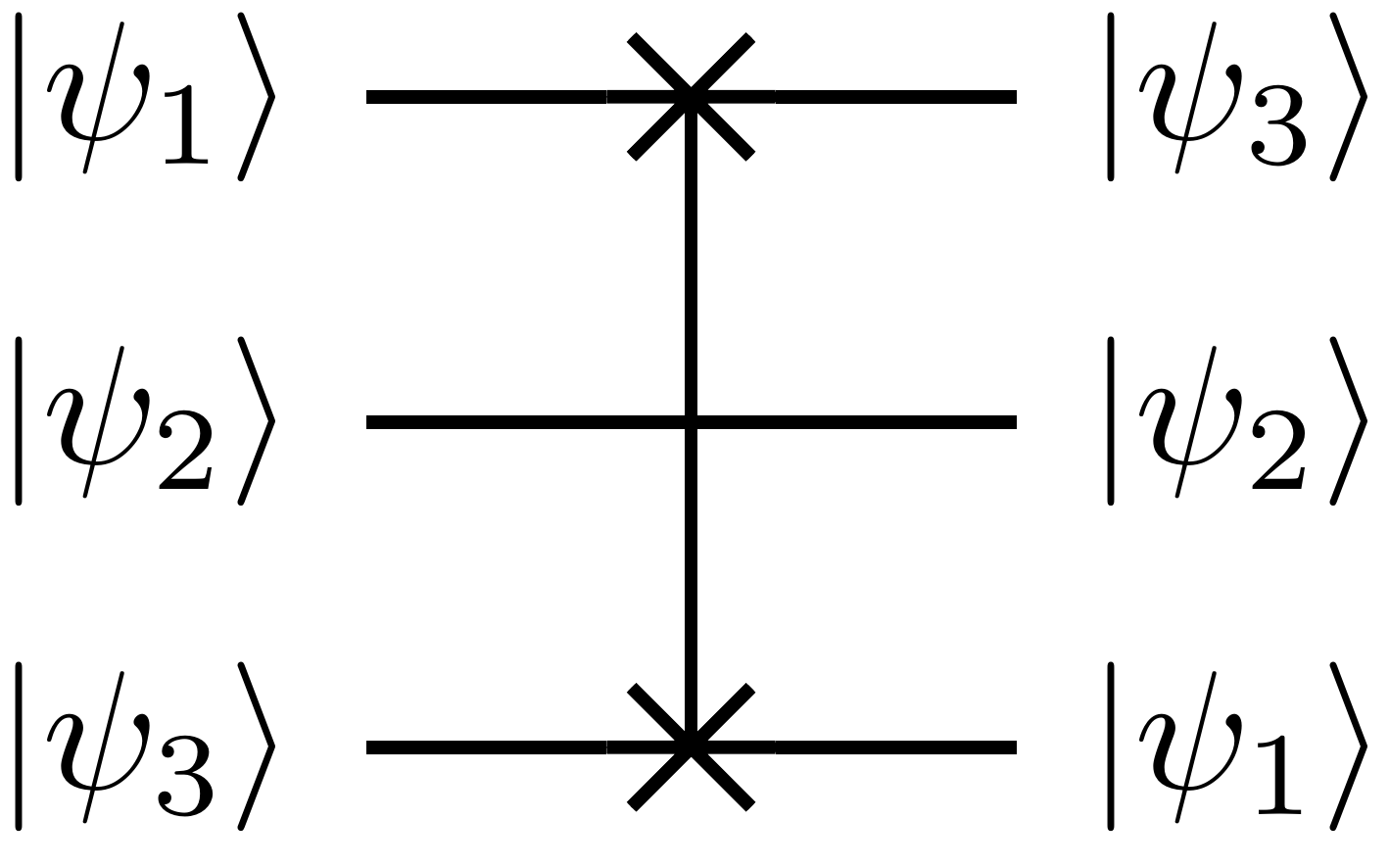}
    \legend{Fonte: Produzido pelo autor.}
    \label{fig:exemplo-swap}
\end{figure}

Por exemplo, suponha que $\ket{\psi_1}, \ket{\psi_2}, \ket{\psi_3} \in\hilb^2$ e
que deseja-se trocar o conteúdo do primeiro e do terceiro qubit
do estado $\ket{\psi_1}\ket{\psi_2}\ket{\psi_3}$.
Sendo assim, o circuito desejado é dado pela porta $\SWAP_{1,3}$ já que
\begin{align}
    \SWAP_{1, 3} \ket{\psi_1}\ket{\psi_2}\ket{\psi_3} =
        \ket{\psi_3}\ket{\psi_2}\ket{\psi_1},
\end{align}
ilustrado na Fig. \ref{fig:exemplo-swap}.

\begin{definition}
    Seja $U$ uma porta unitária que atua em $\hilb^2$.
    A porta $U$-controlada depende de dois qubits:
    um qubit de controle e um qubit alvo.
    Se o qubit de controle for $\ket 0$, o estado permanece inalterado.
    Se o qubit de controle for $\ket 1$, aplica-se a porta $U$
    no qubit alvo.
    A porta $U$-controlada com controle no qubit $c$-ésimo qubit e
    alvo no $a$-ésimo qubit é representada por
    \begin{align}
        \bctrl c a U.
    \end{align}
    Fig. \ref{fig:ctrl-U} ilustra uma porta $U-$controlada.
    
    \begin{figure}[htb]
    \centering
        \caption{Porta $U$-controlada.}
        \includegraphics[height=0.075\textheight]{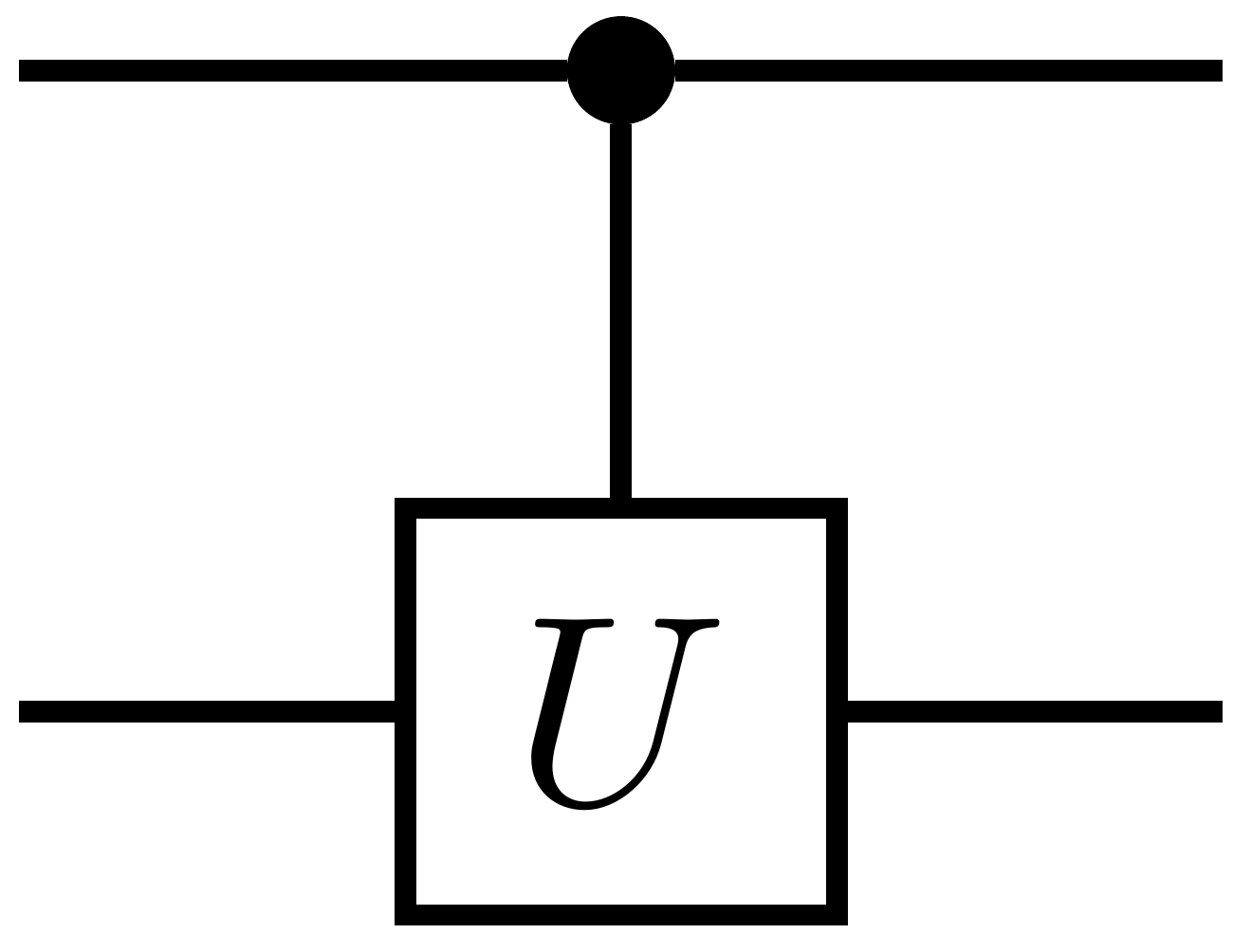}
        \legend{Fonte: Produzido pelo autor.}
        \label{fig:ctrl-U}
    \end{figure}
\end{definition}

\begin{figure}[htb]
\centering
    \caption{Circuito gerador de estados de Bell.}
    \includegraphics[width=0.4\textwidth]{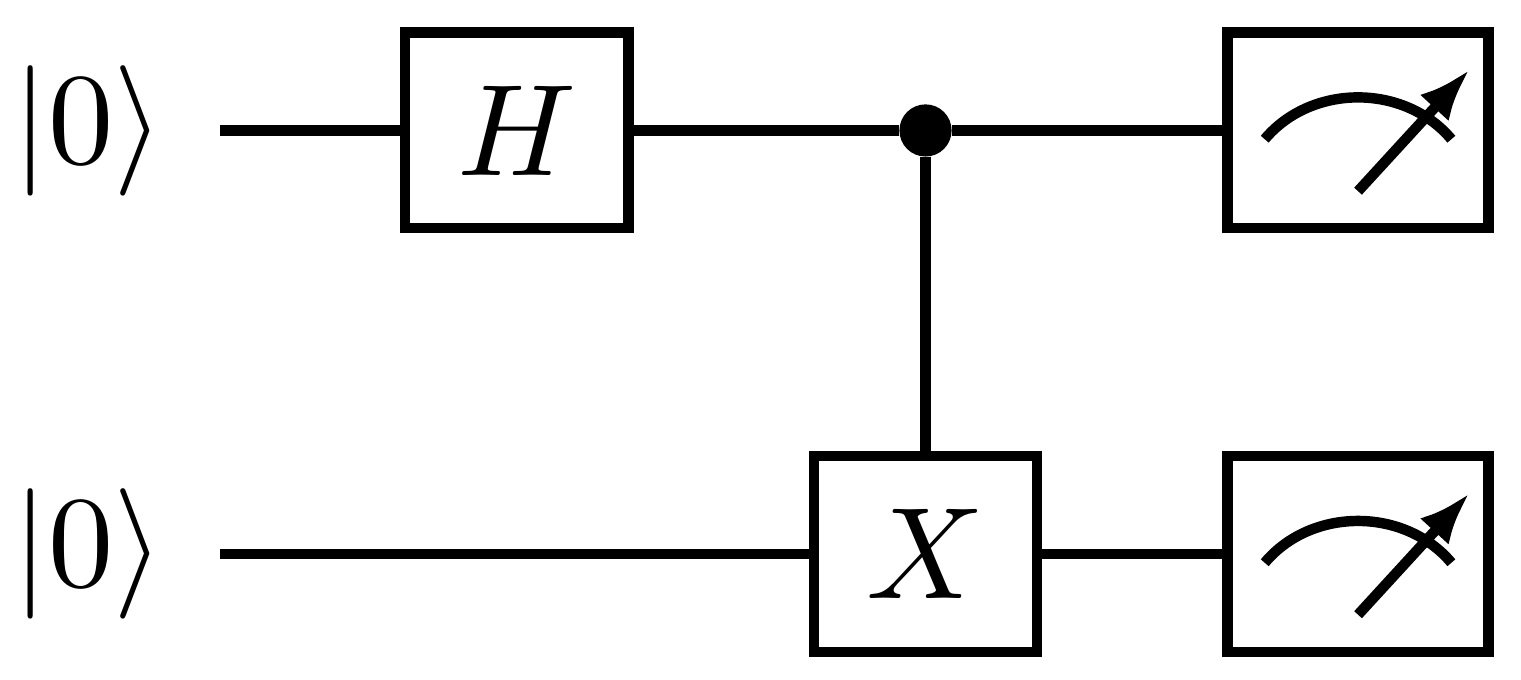}
    \legend{Fonte: Produzido pelo autor.}
    \label{fig:circ-bell}
\end{figure}

Um exemplo de circuito que usa a portas controladas,
é o circuito que gera estados de Bell,
descrito por $\bctrl 1 2 X \pr{H \otimes I}$.
Esse circuito está ilustrado na Fig. \ref{fig:circ-bell}
para entrada $\ket{00}$.

Após executar esse circuito, obtém-se
\begin{align}
    \bctrl 1 2 X \pr{H \otimes I} \ket{00} &=
        \bctrl 1 2 X \frac{1}{\sqrt 2} \pr{\ket {00} + \ket {10}}
    \\
    &= \frac{1}{\sqrt 2} \pr{\ket {00} + \pr{I \otimes X} \ket {10}} 
    \\
    &= \frac{1}{\sqrt 2} \pr{\ket {00} + \ket {11}}.
\end{align}
Os ícones à direita (no final) do circuito indicam que está sendo feita uma
medição na base computacional.
Como resultado desse exemplo,
uma medição na base computacional resulta no estado
$\ket{00}$ ou no estado $\ket{11}$,
ambos com probabilidade de $1/2$.
\section{Teoria dos Grafos} \label{sec:teoria-dos-grafos}
O livro do West é uma referência consolidada em Teoria dos Grafos
\cite{west2001introduction}.
Considere as seguintes definições.

\begin{definition}
    Um grafo $\Gamma(V, E)$ consiste de dois conjuntos finitos:
    um de vértices $V$ e um de arestas $E$.
    A cada vértice é associado um rótulo (normalmente um número natural) e
    cada aresta $a = \pr{v_1, v_2}$ é uma tupla de dois vértices
    $v_1, v_2 \in V$.
\end{definition}
\begin{definition}
    Seja $\Gamma(V, E)$ um grafo.
    Qualquer aresta $(v, v) \in E$ onde $v \in V$ é chamada de laço.
\end{definition}
\begin{definition}
    Seja $\Gamma(V, E)$ um grafo tal que $v_1, v_2 \in V$.
    Quaisquer arestas que possuam o mesmo par de vértices
    são chamadas de arestas múltiplas --
    \textit{e.g.} $(v_1, v_2)$ e $(v_2, v_1)$.
\end{definition}
\begin{definition}
    Um grafo $\Gamma(V, E)$ é um grafo simples se não possui
    laços ou arestas múltiplas.
    Nesse caso, as arestas podem ser representadas como um conjunto de
    vértices ao invés de tuplas ou até mesmo contatenando-se os vértices.
    Por exemplo, se $v_1, v_2 \in V$, a aresta entre esses dois vértices pode
    ser representada equivalentemente como $\set{v_1, v_2}$, $v_1 v_2$ ou $v_2 v_1$.
\end{definition}

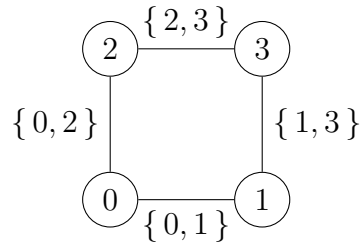
\begin{figure}[htb]
    \centering
    \caption{Exemplo de grafo simples.}
    \begin{tikzpicture}
    \pgfmathsetmacro\mult{2}
    \pgfmathsetmacro\N{2}
    
    \foreach \xx in {1,...,\N}{
        \foreach \yy in {1,...,\N} {
            \pgfmathsetmacro{\x}{int(\xx - 1)}
            \pgfmathsetmacro{\y}{int(\yy - 1)}
            \pgfmathsetmacro{\res}{int(\y*\N + \x)}
            \node[circle,draw] (\res) at (\x*\mult, \y*\mult) {\res};
        }
    }
    
    \path (0) edge node[below] {$\set{0, 1}$} (1);
    \path (1) edge node[right] {$\set{1, 3}$} (3);
    \path (3) edge node[above] {$\set{2, 3}$} (2);
    \path (2) edge node[left] {$\set{0, 2}$} (0);
    
\end{tikzpicture}
    \legend{Fonte: produzido pelo autor.}
    \label{fig:grafo-exemplo}
\end{figure}

Ao longo deste trabalho, foca-se apenas em grafos simples.
Um exemplo de grafo simples está ilustrado na Fig. \ref{fig:grafo-exemplo}.
Esse grafo possui quatro vértices ($V = \set{1, 2, 3, 4}$) e quatro arestas.
Os rótulos dos vértices e arestas podem ser omitidos.

Introduz-se agora alguns conceitos que relacionam vértices e arestas:
incidência, grau e adjacência.

\begin{definition}
    Seja $\Gamma(V, E)$ um grafo simples, $a \in E$ uma aresta tal que
    $a = \set{v_1, v_2}$ e $v_1, v_2 \in V$.
    Então, diz-se que a aresta $a$ é \emph{incidente} nos vértices $v_1$ e $v_2$.
\end{definition}
\begin{definition}
    Seja $\Gamma(V, E)$ um grafo simples e um vértice $v \in V$,
    diz-se que $v$ tem grau $d$ se existem $d$ arestas distintas
    incidentes a $v$.
\end{definition}
\begin{definition}
    Seja $\Gamma(V, E)$ um grafo simples, $v_1, v_2 \in V$ e $\set{v_1, v_2} \in E$.
    Então, diz-se que os vértices $v_1$ e $v_2$ são \emph{adjacentes}.
\end{definition}

Grafos podem ser conexos ou não conexos.
Para definir grafos conexos, é necessário definir passeio e caminho.
\begin{definition}
    Seja $\Gamma(V, E)$ um grafo simples.
    Um passeio em $\Gamma$ é uma lista de vértices (não necessariamente distintos)
    $v_0, \ldots, v_n \in V$ tal que $\set{v_i, v_{i+1}} \in E$
    para $0 \leq i \leq n-1$.
    Ou seja, uma lista de vértices adjacentes.
\end{definition}
Um exemplo de passeio no grafo da Fig. \ref{fig:grafo-exemplo} é
0, 1, 3, 1, 3, 2.
\begin{definition}
    Um caminho num grafo simples $\Gamma(V, E)$ é um passeio onde todos os vértices
    da lista são distintos.
\end{definition}
Um exemplo de caminho no grafo da Fig. \ref{fig:grafo-exemplo} é
0, 1, 3, 2.
\begin{definition}
    Um grafo simples $\Gamma(V, E)$ é dito conexo se para todo par distinto de vértices
    $v_1, v_2 \in V$ existe um caminho de $v_1$ até $v_2$.
\end{definition}
O grafo da Fig. \ref{fig:grafo-exemplo} é um grafo conexo.
Entretanto, ao se retirar duas arestas quaisquer desse grafo,
ele se torna um grafo não-conexo.
Ao longo deste trabalho, pressupõe-se que todos os grafos são conexos.

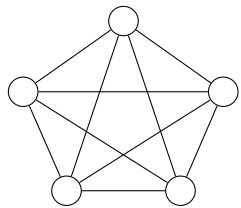
\begin{figure}[htb]
    \centering
    \caption{Grafo completo com cinco vértices.}
    \begin{tikzpicture}
    \pgfmathsetmacro\mult{0.75}
    \pgfmathsetmacro\N{5}
    
    \node[circle,draw] (1) at (0*\mult, 0*\mult) {};
    \node[circle,draw] (2) at (-1.76*\mult, -1.25*\mult) {};
    \node[circle,draw] (3) at (-1*\mult, -3*\mult) {};
    \node[circle,draw] (4) at (1*\mult, -3*\mult) {};
    \node[circle,draw] (5) at (1.75*\mult, -1.25*\mult) {};
    
    \pgfmathsetmacro\n{\N - 1}
    \foreach \x in {1,...,\n}{
        \pgfmathsetmacro\yy{int(\x + 1)}
        \foreach \y in {\yy,...,\N} {
            \path (\x) edge (\y);
        }
    }
\end{tikzpicture}
    \legend{Fonte: produzido pelo autor.}
    \label{fig:grafo-completo}
\end{figure}

Foca-se agora em alguns tipos de grafo que estão relacionados com o trabalho.
\begin{definition}
    Um grafo simples $\Gamma(V, E)$ é dito um grafo completo se
    todos os vértices distintos forem adjacentes entre si.
\end{definition}
Fig. \ref{fig:grafo-completo} ilustra o grafo completo com 5 vértices.

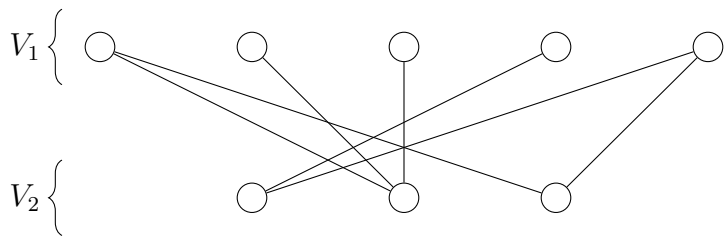
\begin{figure}[htb]
    \centering
    \caption{Exemplo de grafo bipartido.}
    \begin{tikzpicture}
    \pgfmathsetmacro\N{5}
    \pgfmathsetmacro\M{3}
    \pgfmathsetmacro\mult{2}
    
    \foreach \x in {1,...,\N}{
        \node [circle,draw] (\x) at (\x*\mult,0) {};
    }
    \foreach \x in {1,...,\M}{
        \pgfmathsetmacro\res{ int(\x + \N) }
        \node [circle,draw] (\res) at (\x*\mult+\mult,-\mult) {};
    }
    
    \path (5) edge (6);
    \path (4) edge (6);
    \path (1) edge (7);
    \path (1) edge (8);
    \path (2) edge (7);
    \path (3) edge (7);
    \path (5) edge (8);

    \draw [decorate, decoration={brace, amplitude=5pt}]
        (\mult-\mult/4, -\mult/4) -- (\mult-\mult/4, \mult/4)
        node[midway, left]{$V_1\ $};
    \draw [decorate, decoration={brace, amplitude=5pt}]
        (\mult-\mult/4, -\mult-\mult/4) -- (\mult-\mult/4, -\mult+\mult/4)
        node[midway, left]{$V_2\ $};
\end{tikzpicture}
    \legend{Fonte: produzido pelo autor.}
    \label{fig:grafo-bipartido}
\end{figure}

\begin{definition}
    Um grafo simples $\Gamma(V, E)$ é dito bipartido se $V$ for a união de dois conjuntos
    disjuntos -- \textit{i.e.} $V = V_1 \cup V_2$ e $V_1 \cap V_2 = \emptyset$ --
    de tal forma que os vértices de um conjunto \emph{não} são adjacentes a vértices do próprio conjunto
    -- \textit{i.e.} $v_j, u_j \in V_j \implies v_ju_j \notin E$ para $j \in {1, 2}$.
\end{definition}
Fig. \ref{fig:grafo-bipartido} ilustra um grafo bipartido com
$|V_1| = 5$ e $|V_2| = 3$.

\begin{figure}[htb]
    \centering
    \caption{Exemplo de grafo bipartido completo.}
    \begin{tikzpicture}
    \pgfmathsetmacro\N{5}
    \pgfmathsetmacro\M{3}
    \pgfmathsetmacro\mult{2}
    
    \foreach \x in {1,...,\N}{
        \node [circle,draw] (\x) at (\x*\mult,0) {};
    }
    \foreach \x in {1,...,\M}{
        \pgfmathsetmacro\res{ int(\x + \N) }
        \node [circle,draw] (\res) at (\x*\mult+\mult,-\mult) {};
    }
    
    \pgfmathsetmacro\Mstart{ int(\N + 1) }
    \pgfmathsetmacro\Mend{ int(\N + \M) }
    \foreach \v in {1,...,\N}
        \foreach \u in {\Mstart,...,\Mend}{
            \path (\v) edge (\u);
        }
    
\end{tikzpicture}
    \legend{Fonte: produzido pelo autor.}
    \label{fig:grafo-bipartido-completo}
\end{figure}
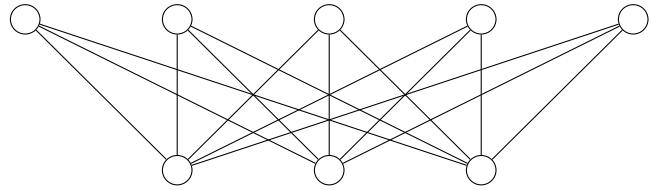

\begin{definition}
    Um grafo simples $\Gamma(V, E)$ é dito bipartido completo se ele for um grafo bipartido
    e todos os vértices de um conjunto disjunto forem adjacentes a
    todos os vértices do outro conjunto --
    \textit{i.e.} $v_1 \in V_1$ e $v_2 \in V_2 \implies \set{v_1, v_2} \in E$.
\end{definition}
Fig. \ref{fig:grafo-bipartido-completo} ilustra um grafo bipartido completo com
$|V_1| = 5$ e $|V_2| = 3$.

\begin{definition}
    Um grafo simples $\Gamma(V, E)$ é dito um grafo regular se todos os vértices
    possuem o mesmo grau.
    Se esse grau for $d$, chama-se o grafo de $d$-regular.
\end{definition}
Os grafos das Figs. \ref{fig:grafo-exemplo} e \ref{fig:grafo-completo}
são exemplos de grafos regulares.
Um último conceito que será necessário para o trabalho é o de coloração de arestas.
\begin{definition}
    Dado um grafo simples $\Gamma(V, E)$, uma coloração de arestas é
    uma rotulação das arestas onde cada aresta é associada com uma cor (ou número).
    Várias arestas podem ser associadas com uma cor, compondo uma classe de cor.
\end{definition}
\begin{definition}
    Um grafo simples $\Gamma(V, E)$ é $d$-colorível por arestas
    se existe uma coloração de arestas com $d$ cores
    tais que $\forall v \in V$ todas as arestas incidentes em $v$ possuem cores
    diferentes.
\end{definition}
\begin{definition}
    O índice cromático de um grafo simples $\Gamma(V, E)$ é o menor valor $d$ tal que
    $\Gamma$ seja $d$-colorível por arestas.
\end{definition}

\begin{figure}[htb]
    \centering
    \caption{Exemplo de coloração de arestas.}
    \begin{tikzpicture}
    \pgfmathsetmacro\N{3}
    \pgfmathsetmacro\M{\N}
    \pgfmathsetmacro\mult{2.5}
    
    \foreach \x in {1,...,\N}{
        \node [circle,draw, thick] (\x) at (\x*\mult,0) {};
    }
    \foreach \x in {1,...,\M}{
        \pgfmathsetmacro\res{ int(\x + \N) }
        \node [circle,draw, thick] (\res) at (\x*\mult,-\mult) {};
    }
    
    \def\colors{black, red, blue}
    \pgfmathsetmacro\Mstart{ int(\N + 1) }
    \foreach \v in {1,...,\N}
        \foreach \c [count=\i] in \colors {
            \pgfmathsetmacro\u{int(\N+1 + Mod(\v+\i-2, \N))}
            \path [color=\c, thick] (\v) edge (\u);
        }
        
\end{tikzpicture}
    \legend{Fonte: produzido pelo autor.}
    \label{fig:grafo-coloracao}
\end{figure}
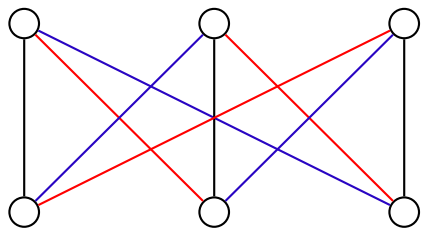

O grafo ilustrado na Fig. \ref{fig:grafo-coloracao} possui índice cromático 3,
logo, é 3-colorível, 4- colorível por arestas, etc.

Em particular, este trabalho foca em grafos simples regulares com número par de vértices.
Sabe-se que grafos $d$-regulares com número par de vértices
possuem índice cromático $d$.
Note que o grafo da Fig. \ref{fig:grafo-coloracao} é um exemplo desse tipo de grafo
com $d = 3$.
Grafos $d$-regulares com número ímpar de vértices possuem índice cromático $d+1$ e
fogem do escopo do trabalho.
\chapter{Algoritmo de Contagem} \label{cap:alg-cont}
Seja $S_N = \{0, \ldots, N - 1 \}$ para $N = 2^n$ e $n \in \mathbb{N}$ e
considere a função $f(x): S_N \to \{0, 1\}$.
O problema de contagem consiste em estimar o valor
$k = \left|\{x \in S_N| f(x) = 1\}\right|$;
\textit{i.e.} estimar a quantidade de elementos que possuem a propriedade descrita pela função $f$.
Esse problema foi estudado por Brassard, Høyer, Mosca e Tapp (BHMT)
-- na referência \cite{brassard2002quantum} -- e descrito nas próximas seções.
Seção \ref{sec:algoritmo-busca} descreve o problema de busca e o algoritmo de Grover
\cite{grover1996fast, grover1997quantum},
Seção \ref{sec:qft} descreve a Transformada Quântica de Fourier,
Seção \ref{sec:est-fase} descreve o algoritmo de estimativa de fase,
Seção \ref{sec:alg-cont} descreve o algoritmo de contagem, unindo todas as seções supracitadas.

\section{Algoritmo de Busca} \label{sec:algoritmo-busca}
Considerando a função $f(x): S_N \to \{0, 1\}$,
o problema de busca consiste em encontrar um elemento $x \in S_N$ tal que $f(x) = 1$.
Nesse tipo de problema, uma terceira pessoa é responsável por implementar a função $f$
(chamada de oráculo)
e o objetivo do problema é descobrir um elemento $x$ fazendo a menor quantidade de consultas
possíveis à $f$.
Evidentemente, num computador clássico são necessárias $O(N)$ invocações à $f$,
fazendo uma consulta para cada valor de $x$ possível.
Em contrapartida, num computador quântico podemos encontrar $x$ realizando $O(\sqrt N)$
invocações de $f$.
Esse processo será descrito nas Seções a seguir.

\subsection{Oráculo} \label{sec:oraculo}
Primeiramente, é necessário que uma terceira pessoa implemente uma porta
que simule o comportamento do oráculo $f$.
Assume-se que o oráculo age de acordo com
\begin{align}
    O_f: \ket{x}\ket{b} = \ket{x}\ket{b \oplus f(x)},
\end{align}
onde $\ket{x} \in \hilb^N$, $\ket{b} \in \hilb^2$ e $\oplus$ é a operação binária XOR, detalhada na Tabela \ref{tab:xor}.
\begin{table}[htb]
    \caption{Tabela verdade da operação XOR.}
    \IBGEtab{
        \label{tab:xor}
        }{
        \begin{tabular}{ccc}
        \toprule
        $a$ & $b$ & $a \oplus b$ \\
        \midrule \midrule
        0 & 0 & 0 \\
        \midrule
        0 & 1 & 1 \\
        \midrule
        1 & 0 & 1 \\
        \midrule
        1 & 1 & 0 \\
        \bottomrule
        \end{tabular}
    }{\fonte{autor.}}
\end{table}
O oráculo é unitário já que $O_f^\dagger = O_f$:
\begin{align}
    O_f^2 \ket{x}\ket{b} &= \ket{x}\ket{b \oplus f(x) \oplus f(x)} \\
    &= \ket{x}\ket{b \oplus 0} = \ket{x}\ket{b}.
\end{align}
Para utilizar o oráculo no algoritmo de busca, faz-se uso de \emph{phase kickback}.
Suponha que $x_0, x_1 \in S_N$ tais que $f(x_0) = 0$ e $f(x_1) = 1$;
então
\begin{align}
    O_f \ket{x_0} \otimes \frac{\ket 0 - \ket 1}{\sqrt 2}
        &= \ket{x_0} \otimes \frac{\ket{0 \oplus 0} - \ket{1 \oplus 0}}{\sqrt 2} \\
    &= \ket{x_0} \otimes \frac{\ket 0 - \ket 1}{\sqrt 2},
\end{align}
e
\begin{align}
    O_f \ket{x_1} \otimes \frac{\ket 0 - \ket 1}{\sqrt 2}
        &= \ket{x_1} \otimes \frac{\ket{1} - \ket{0}}{\sqrt 2} \\
    &= - \ket{x_1} \otimes \frac{\ket 0 - \ket 1}{\sqrt 2}.
\end{align}
Esse fenômeno é uma consequência direta das propriedades de produto tensorial.
Perceba que o segundo registrador permanece inalterado no estado
$\pr{\ket{0} - \ket{1}}/\sqrt 2$ e sua fase é ``transferida'' para o primeiro registrador.
Portanto é possível desconsiderar o segundo registrador no decorrer das contas.

\subsection{Operador de Evolução de Grover}
Para extrair informações de $O_f$ em menos de $O(N)$ passos,
é necessário realizar algum processamento entre invocações de $O_f$.
Tal processamento é descrito pela matriz de Grover, definida por
\begin{align}
    G &= H^{\otimes n} \pr{2 \ket{0}\bra{0} - I } H^{\otimes n} \\
    &= 2 H^{\otimes n} \ket 0 \pr{H^{\otimes n} \ket 0}^\dagger - \pr{H^2}^{\otimes n} \\
    &= \frac{2}{N} \sum_{i, j = 0}^{N - 1} \ket{i}\bra{j} - I .
\end{align}
A matriz de Grover também é chamada de operador de difusão devido ao seu comportamento
de espalhar a amplitude de um dos estados da base computacional para os demais:
\begin{align}
    G\ket{x} &= \frac{2}{N} \sum_{i, j = 0}^{N - 1} \ket{i} \braket{j|x} - I\ket{x} \\
    &= \frac{2}{N} \sum_{i = 0}^{N - 1}\ket{i} - \ket{x} \\
    &= \pr{\frac{2}{N} - 1} \ket{x} + \frac{2}{N} \sum_{i \neq x} \ket i.
\end{align}

Ao aplicar o oráculo seguido pela matriz de Grover,
obtém-se o operador de evolução de Grover
\begin{align}
    U_G = GO_f,
\end{align}
que é chamado múltiplas vezes durante o algoritmo de busca.
De fato, toda a análise do algoritmo de busca é focada em $U_G$.

\subsection{O Algoritmo e Sua Análise}
O algoritmo de busca está descrito em Algoritmo \ref{alg:busca}.

\begin{algorithm}
    \caption{Algoritmo de Busca}
    \label{alg:busca}
    \begin{algorithmic}[1]
        \REQUIRE $O_f$: Oráculo da função $f$;
            $n$: quantidade de qubits respeitando o domínio da função $f$
        \STATE Preparar o estado $\ket{\psi} = H^{\otimes n} \ket{0}$
        \STATE Aplicar $U_G^t \ket{\psi}$ onde $t = \floor{ \frac{\pi}{4}\sqrt{\frac{N}{k}} }$
        \STATE Realizar medição obtendo o resultado $\ket{\psi_f}$
        \RETURN $\ket{\psi_f}$
    \end{algorithmic}
\end{algorithm}

Note que o estado inicial $\ket{\psi_0}$ é uma sobreposição uniforme de todos
os valores de $x$ pertencentes ao domínio de $f$,
\begin{align}
    \ket{\psi} = \frac{1}{\sqrt{N}} \sum_{x = 0}^{N - 1} \ket{x}.
\end{align}
Portanto, é uma sobreposição de todos os valores
$x_0 \in \{x \in S_N | f(x) = 0\}$ e $x_1 \in \{x \in S_N | f(x) = 1\}$
(com cardinalidades $N - k$ e $k$, respectivamente).
Sejam
\begin{align}
    \ket{x_0} = \frac{1}{\sqrt{N - k}} \sum_{f(x) = 0} \ket{x}
\end{align}
e
\begin{align}
    \ket{x_1} = \frac{1}{\sqrt{k}} \sum_{f(x) = 1} \ket{x}.
\end{align}
Então o estado inicial pode ser representado como
\begin{align}
    \ket{\psi} = \sqrt{\frac{N - k}{N}} \ket{x_0} + \sqrt{\frac{k}{N}} \ket{x_1}.
\end{align}

Denotando $\cos\theta = \sqrt{\pr{N - k}/N}$ e $\sin\theta = \sqrt{k/N}$,
aplicando $U_G$ ao estado inicial, e usando as propriedades trigonométricas
para $\cosp{\alpha \pm \beta}$ e $\sinp{\alpha \pm \beta}$,
obtém-se
\begin{align}
    U_G \ket{\psi} &= GO_f \pr{ \cos\theta \ket{x_0} + \sin\theta \ket{x_1} } \\
    &= \pr{ 2\ket{\psi}\bra{\psi} - I} \pr{ \cos\theta \ket{x_0} - \sin\theta \ket{x_1} } \\
    &= 2 \pr{\cos^2\theta - \sin^2\theta} \ket{\psi} - \cos\theta\ket{x_0} +
        \sin\theta\ket{x_1} \\
    &= \cos\theta \pr{2\cosp{2\theta} - 1} \ket{x_0} +
        \sin\theta \pr{2\cosp{2\theta} + 1} \ket{x_1} \\
    &= \pr{ \cosp{3\theta} + \cos\theta - \cos\theta } \ket{x_0} +
        \pr{ \sinp{3\theta} - \sin\theta + \sin\theta } \ket{x_1} \\
    &= \cosp{3\theta} \ket{x_0} + \sinp{3\theta} \ket{x_1} .
\end{align}

\begin{figure}
    \centering
    \caption{Rotação de $2\theta$ no hiperplano definido por
        $\ket{x_0}$ e $\ket{x_1}$.}
    \begin{tikzpicture}

    \pgfmathsetmacro\mult{3}
    \pgfmathsetmacro\angle{15}
    
    \draw[->] (-\mult , 0) -- (1.1*\mult , 0) node[right] {$\ket{x_0}$};
    \draw[->] (0, -\mult ) -- (0, 1.1*\mult ) node[above] {$\ket{x_1}$};
    \draw (0, 0) circle (3);

    \draw[->] (0, 0) -- ({cos(\angle)*\mult}, {sin(\angle)*\mult})
        node[right] {$\ket{\psi}$};
    \draw[->] (0, 0) -- ({cos(3*\angle)*\mult}, {sin(3*\angle)*\mult})
        node[right] {$U_G\ket{\psi}$};
        
    \draw (\mult/2, 0) arc [start angle=0, end angle=\angle, radius={\mult/2}]
        node[midway, right]{$\theta$};
        
    \draw ({cos(\angle)*\mult/2}, {sin(\angle)*\mult/2})
        arc [start angle=\angle, end angle=3*\angle, radius={\mult/2}]
        node[midway, right]{$2\theta$};
        
\end{tikzpicture}
    \legend{Fonte: Produzido pelo autor.}
    \label{fig:rot-grover}
\end{figure}
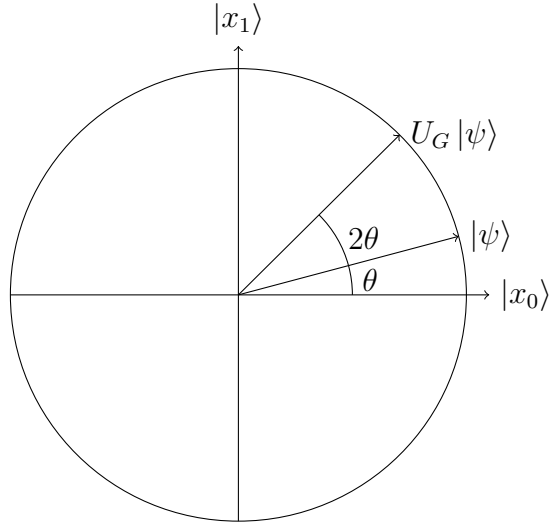

Esse resultado levanta suspeitas de que $U_G$ faz uma rotação de $2\theta$ graus
no hiperplano definido por $\ket{x_0}$ e $\ket{x_1}$
(conforme ilustrado na Figura \ref{fig:rot-grover}).
De fato, é possível fazer uma prova indutiva para isso.
Tome como caso base $U_G\ket{\psi} = \cosp{3\theta}\ket{x_0} + \sinp{3\theta}$; e
como hipótese indutiva
$U_G^{t'}\ket{\psi} = \cos\theta_{t'}\ket{x_0} + \sin\theta_{t'}\ket{x_1}$,
onde $\theta_{t'} = (2t'+1)\theta$ e $t' \in \mathbb{N}$.
Deseja-se demonstrar que
$U_G^{t' + 1}\ket{\psi} = \cos\theta_{t'+1}\ket{x_0} + \sin\theta_{t'+1}\ket{x_1}$;
\begin{align}
    U_G^{t' + 1}\ket{\psi} &= U_G \pr{ \cos\theta_{t'}\ket{x_0} + \sin\theta_{t'}\ket{x_1} }
    \\
    &= \pr{2\ket{\psi}\bra{\psi} - I} \pr{\cos\theta_{t'}\ket{x_0} - \sin\theta_{t'}\ket{x_1}}
    \\
    &= 2\cosp{\theta + \theta_{t'}}\ket{\psi} - \cos\theta_{t'}\ket{x_0} +
        \sin\theta_{t'}\ket{x_1}
    \\
    &= \cosp{2\theta + \theta_{t'}} \ket{x_0} + \sinp{2\theta + \theta_{t'}} \ket{x_1}
    \\
    &= \cos\theta_{t'+1}\ket{x_0} + \sin\theta_{t'+1}\ket{x_1},
\end{align}
concluindo a demonstração.

Dessa forma, também é possível interpretar a matriz $U_G$ como a matriz de rotação
$R(2\theta)$ definida na base $\ket{x_0}, \ket{x_1}$; onde
\begin{align}
    R(\theta') = \matrx{
        \cos\theta' & -\sin\theta' \\
        \sin\theta' & \cos\theta'
    } .
\end{align}
Matrizes de rotação têm autovetores e autovalores bem conhecidos. Especificamente,
\begin{align}
    \ket{\mp\ii} = \frac{\ket 0 \mp\ii \ket 1}{\sqrt{2}}
\end{align}
são os autovetores de $R(\theta')$ associados aos autovalores
$e^{\pm\ii\theta'}$, respectivamente.

Resta agora analisar a quantidade de iterações $t$ necessária no algoritmo.
Observe que deseja-se maximizar a amplitude dos elementos marcados $\ket{x_1}$ de
forma a maximizar a probabilidade de medir um desses elementos no passo subsequente.
Ou seja, deseja-se maximizar $\sinp{(2t + 1)\theta}$,
que ocorre quando $(2t + 1)\theta = \pi/2$.
Primeiro encontra-se uma expressão para $\theta$.
Supondo que $k \ll N$ e expandindo $\arcsin(\alpha)$ no ponto $\alpha = 0$,
obtém-se
\begin{align}
    \theta = \arcsin\pr{\sqrt{\frac{k}{N}}}
    = \sqrt{\frac{k}{N}} + \frac{1}{6} \pr{\frac{k}{N}}^{3/2} + O\pr{\pr{\frac{k}{N}}^{5/2}} .
    \label{eq:taylor-exp-arcsin}
\end{align}
Por último, isola-se $t$ na expressão $(2t+1)\theta = \pi/2$
e usando o termo mais dominante da eq. \ref{eq:taylor-exp-arcsin}:
\begin{align}
    t = \frac{\pi}{4\theta} - \frac{1}{2}
    = \frac{\pi}{4} \sqrt{\frac{N}{k}} - \frac{1}{2}.
\end{align}
Já que deseja-se $t \in \mathbb{N}$, tome
\begin{align}
    t = \floor{ \frac{\pi}{4} \sqrt{\frac{N}{k}} }.
\end{align}
Esse valor rotaciona a condição inicial em aproximadamente $\pi/2$ radianos
no hiperplano descrito por $\ket{x_0}$ e $\ket{x_1}$ (Fig. \ref{fig:rot-grover}).
Ao final, obtém-se $\ket{\psi_f} \approx \ket{x_1}$,
logo, há a probabilidade do algoritmo falhar.
No pior caso, as $t$ iterações rotacionam $\ket{\psi_0}$ em exatamente $\pi/2$ radianos --
na prática, serão menos de $\pi/2$ radianos
já que o segundo termo da expansão de $\theta$ é positivo (eq. \ref{eq:taylor-exp-arcsin}).
Nesse cenário, temos que a probabilidade de sucesso é dada por
\begin{align}
    \left| \braket{\psi_f | x_1} \right|^2 \geq \left| \braket{\psi_0 | x_0} \right|^2
    = \cos^2\theta = 1 - \frac{k}{N} .
\end{align}
\section{Transformada Quântica de Fourier} \label{sec:qft}
A Transformada de Fourier possui diversas aplicações na ciência,
sendo responsável por mudar o domínio de uma função,
\textit{e.g.} domínio do tempo para domínio da frequência temporal.
O domínio da frequência é útil na representação e processamento de sinais digitais,
no processamento de imagens, dentro outras aplicações.
Era de se esperar que uma contrapartida quântica da Transformada de Fourier
tivesse aplicações interessantes na Computação Quântica.
De fato, a Transformada Quântica de Fourier
(QFT \footnote{Quantum Fourier Transform})
tem um papel essencial em algoritmos quânticos \cite{coppersmith1994approximate, nielsen2002quantum},
por exemplo o Algoritmo de Shor \cite{shor1994algorithms}, o algoritmo de estimativa de fase e
o algoritmo de contagem;
sendo os dois últimos o foco deste trabalho.

É possível definir a QFT em torno do seguinte estado.
\begin{definition}
    Seja $P$ a dimensão do espaço de Hilbert e $\omega \in \mathbb{R}$;
    o estado $\ket{\F{P}{\omega}}$ é dado por
    \begin{align}
        \ket{\F{P}{\omega}} = \frac{1}{\sqrt{P}}
            \sum_{\ell = 0}^{P - 1} e^{2\pi\ii \omega \ell / P} \ket{\ell}.
    \end{align}
    \label{def:F_P(omega)}
\end{definition}
Note que é possível restringir os valores de $\omega$ no intervalo $0 \leq \omega < P$.
Entretanto, em alguns casos,
utiliza-se $0 \leq \omega \leq P$ pois a equivalência entre
$\ket{\F P 0}$ e $\ket{\F P P}$ facilita as demonstrações.

Essa definição genérica de $\ket{\F P \omega}$ será útil para
calcular a probabilidade de sucesso do algoritmo de estimativa de fase
(seção \ref{sec:est-fase}).
É possível utilizar os estados $\ket{\F P \omega}$ para descrever uma base de $\hilb^P$.
\begin{lemma}
    O conjunto de estados
    \begin{align}
        B_{\mathcal{F}} = \bigcup_{j=0}^{P - 1} \Set{\ket{\F P j}}
    \end{align}
    forma uma base ortonormal de $\hilb^P$, denominada base de Fourier.
\end{lemma}
\begin{proof}
    Primeiro, verifica-se a ortonormalidade dos estados de $B_{\mathcal F}$:
    \begin{align}
        \braket{\F P j | \F{P}{j'}} &=
            \pr{\frac{1}{\sqrt P} \sum_{\ell = 0}^{P - 1} e^{-2\pi\ii j \ell / P} \bra{\ell}}
            \pr{\frac{1}{\sqrt P} \sum_{\ell' = 0}^{P - 1} e^{2\pi\ii j' \ell' / P} \ket{\ell'}}
        \\
        &= \frac{1}{P} \sum_{\ell, \ell' = 0}^{P - 1}
            e^{2\pi\ii (-j\ell + j'\ell') / P} \delta_{\ell\ell'}
        \\
        &= \frac{1}{P} \sum_{\ell = 0}^{P-1} e^{2\pi\ii(j' - j) \ell / P} .
        \label{eq:proj-fourier-basis}
    \end{align}
    Caso $j = j'$,
    \begin{align}
        \braket{\F P j | \F{P}{j'}} &= \frac{1}{P} \sum_{k = 0}^{P-1} e^{0}
        \\
        &= \frac{1}{P} \cdot P = 1.
    \end{align}
    Caso $j \neq j'$, é possível expressar a Eq. \ref{eq:proj-fourier-basis}
    como uma série geométrica.
    Denotando $j_\Delta = j' - j$ e notando que $j_\Delta \in \mathbb{Z}$, obtém-se
    \begin{align}
        \braket{\F P j | \F{P}{j'}} &= \frac{1}{P} \cdot
            \frac{1 - e^{2\pi\ii j_\Delta}}{1 - e^{2\pi\ii j_\Delta/P}}
        \\
        &= \frac{1}{P} \cdot \frac{1 - 1}{1 - e^{2\pi\ii j_\Delta/P}}
        \\
        &= 0 .
    \end{align}
    Sendo assim os estados de $B_{\mathcal{F}}$ são ortonormais:
    \begin{align}
        \braket{\F P j | \F{P}{j'}} &= \delta_{jj'}.
    \end{align}
    
    Por último,
    \begin{align}
        \left| B_{\mathcal{F}} \right| = P \implies
        \textrm{span}\pr{B_{\mathcal{F}}} = \hilb^P .
    \end{align}
    Portanto, $B_{\mathcal{F}}$ é uma base ortonormal de $\hilb^P$.
\end{proof}
Agora, é possível definir a QFT e sua inversa.
\begin{definition}
    A Transformada Quântica de Fourier é descrita pelo operador que
    realiza a transformação
    \begin{align}
        \QFT \ket j = \ket{\F P j},
    \end{align}
    onde $\ket{j}$ são os estados da base canônica de $\hilb^P$.
    Analogamente, a Transformada Inversa de Fourier é descrita pelo operador
    que realiza a transformação
    \begin{align}
        \QFT^{-1} \ket{\F P j} = \ket{j}.
    \end{align}
\end{definition}
Note que $\QFT$ é unitário:
\begin{align}
    \QFT^{-1}\ \QFT &= \QFT^\dagger\ \QFT \\
    &= \sum_{j=0}^{P-1} \ket{j} \bra{\F P j} \sum_{j'=0}^{P-1} \ket{\F{P}{j'}} \bra{j'}
    \\
    &= \sum_{j,j'=0}^{P-1} \delta_{j,j'} \ket{j} \bra{j'}
    \\
    &= I .
\end{align}

\begin{figure}[htb]
 \centering
  \begin{minipage}{0.48\textwidth}
    \centering
    \caption{Valores de $\braket{k | \F 8 1}$ no plano complexo (ângulos base).}
    \includegraphics[width=\textwidth]{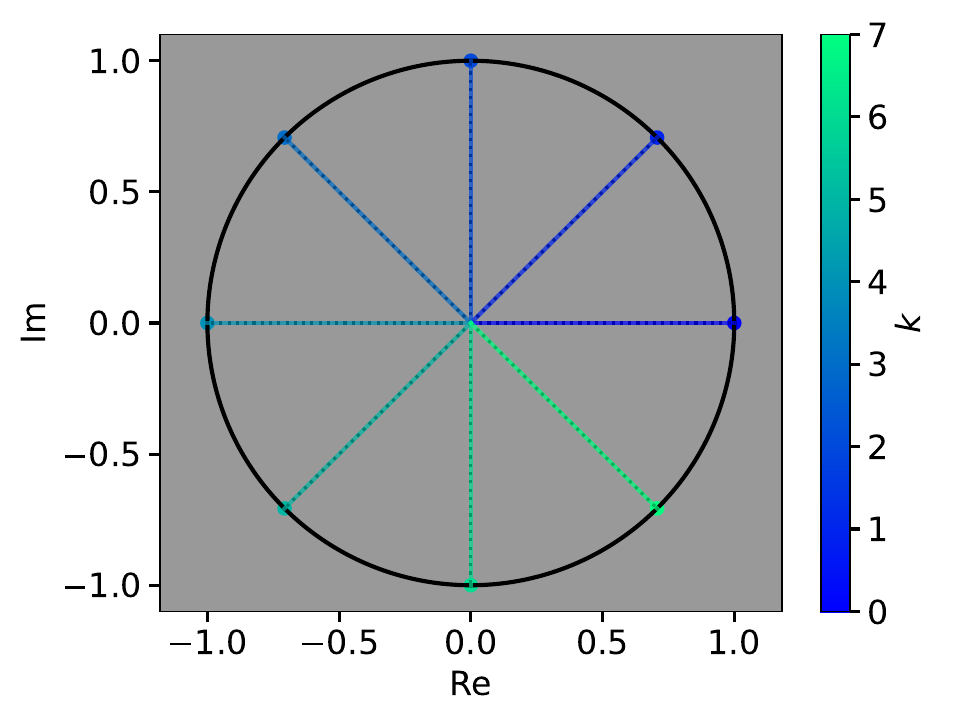}
    \legend{Fonte: Produzido pelo autor.}
    \label{fig:angulos-base}
  \end{minipage}
  \hfill
  \begin{minipage}{0.48\textwidth}
    \centering
    \caption{Valores de $\braket{k | \F 8 2}$ no plano complexo.}
    \includegraphics[width=\textwidth]{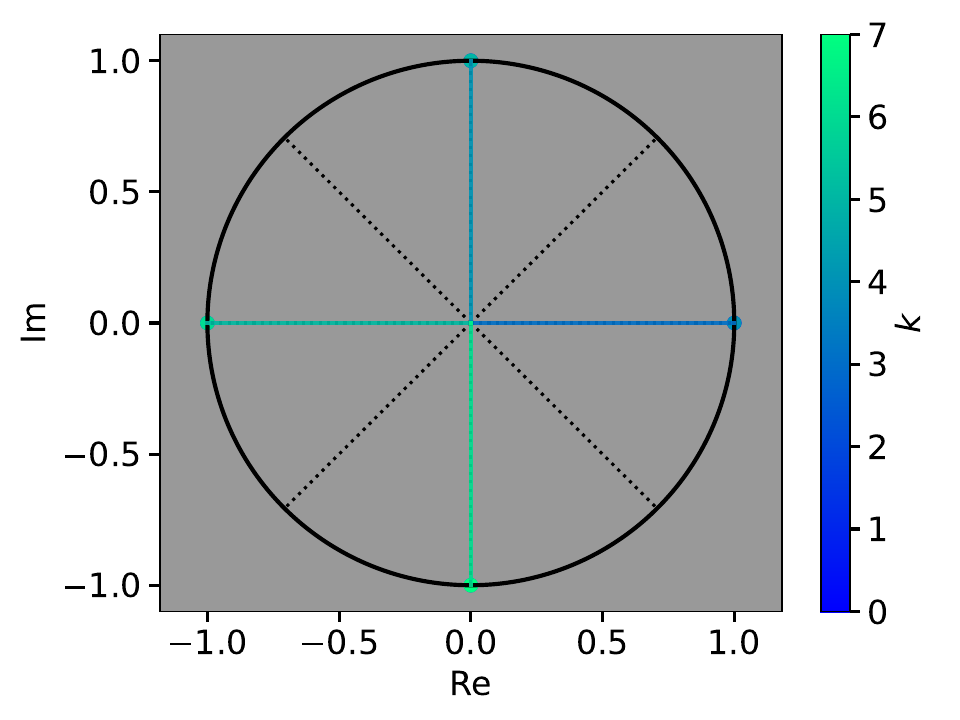}    
    \legend{Fonte: Produzido pelo autor.}
    \label{fig:angulos-F_8(2)}
  \end{minipage}
\end{figure}

Antes de prosseguir, é importante interpretar os estados $\ket{\F P \omega}$ geometricamente.
Ao dividir o círculo unitário uniformemente em $P$ partições,
todas as entradas do vetor $\ket{\F P \omega}$ podem ser representadas
em termos dos ângulos que delimitam essas partições.
Denomine os $P$ ângulos de $\ket{\F P 1}$ por \emph{ângulos base de Fourier}
(ou simplismente ângulos base),
\textit{i.e.} o $k$-ésimo ângulo base é dado por $2\pi k/P$.
Fig. \ref{fig:angulos-base} ilustra os 8 ângulos (base) de $\ket{\F 8 1}$.
De agora em diante, esses ângulos serão representados nas figuras por linhas pontilhadas.
Nessa interpretação, $\omega$ pode ser entendido como um multiplicador dos ângulos base.
Um exemplo disso são os pontos de $\ket{\F 8 2}$ na Fig. \ref{fig:angulos-F_8(2)},
onde o $k$-ésimo ângulo de $\ket{\F 8 2}$ coincide com o $(k+4)$-ésimo ângulo de $\ket{\F 8 1}$
(para $0 \leq k \leq 3$) e com alguns ângulos base.
De fato, sempre que $\omega \in \mathbb{N}$,
todos os ângulos $2\pi \omega k/P$ de $\ket{\F 8 \omega}$ coincidirão com ângulos base.

\begin{figure}[htb]
 \centering
  \begin{minipage}{0.48\textwidth}
    \centering
    \caption{Valores de $\braket{k | \F 8 {1.01}}$ \\no plano complexo.}
    \includegraphics[width=\textwidth]{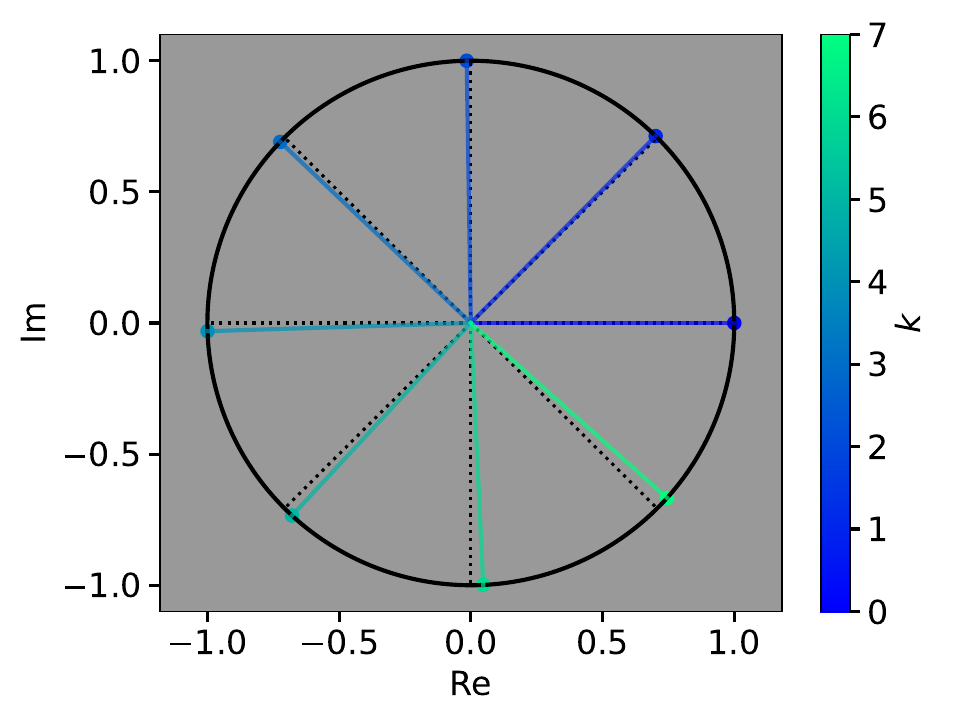}
    \legend{Fonte: Produzido pelo autor.}
    \label{fig:angulos-F_8(1.01)}
  \end{minipage}
  \hfill
  \begin{minipage}{0.48\textwidth}
    \centering
    \caption{Valores de $\braket{k | \F 8 {1.5}}$ no plano complexo.}
    \includegraphics[width=\textwidth]{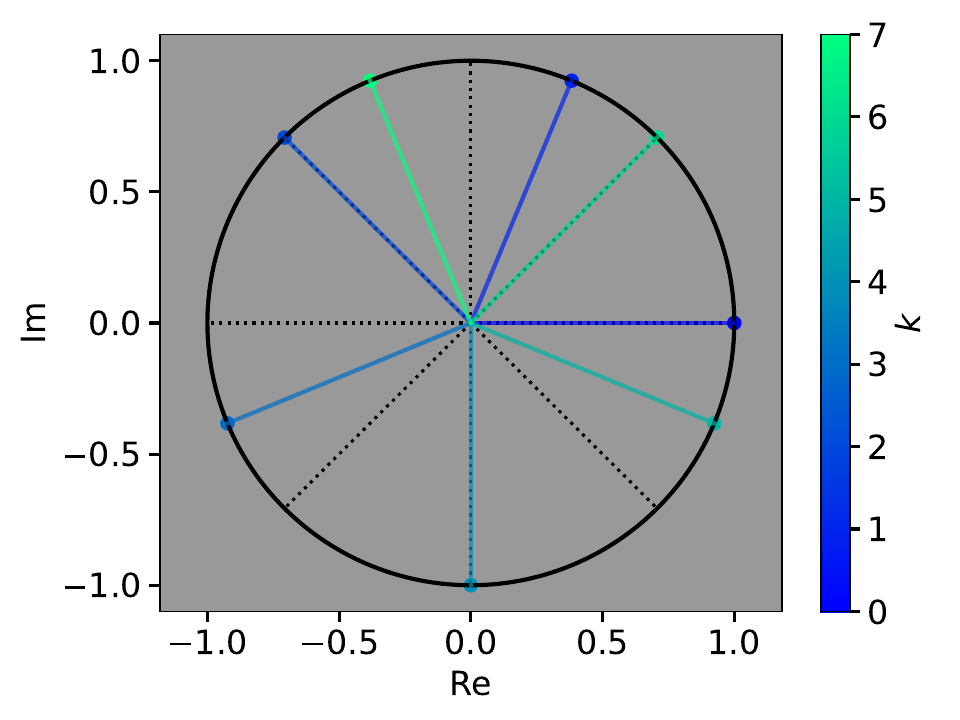}
    \legend{Fonte: Produzido pelo autor.}
    \label{fig:angulos-F_8(1.5)}
  \end{minipage}
\end{figure}

\begin{figure}[htb]
 \centering
  \begin{minipage}{0.48\textwidth}
    \centering
    \caption{Valores de $\braket{k | \F 8 {1.99}}$ \\no plano complexo.}
    \includegraphics[width=\textwidth]{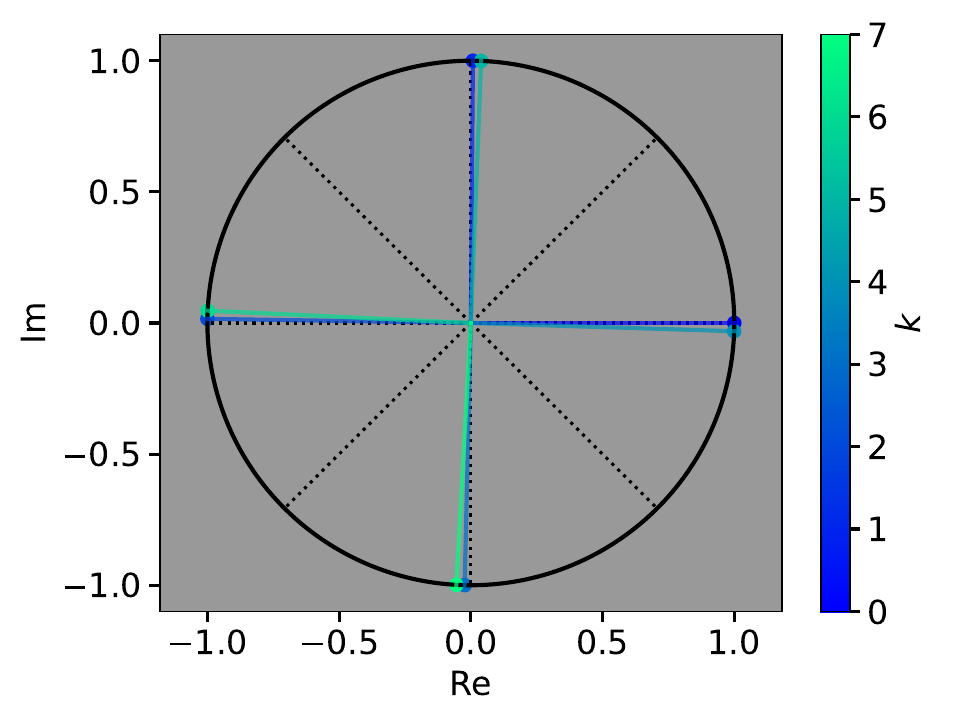}
    \legend{Fonte: Produzido pelo autor.}
    \label{fig:angulos-F_8(1.99)}
  \end{minipage}
  \hfill
  \begin{minipage}{0.48\textwidth}
    \centering
    \caption{Valor de $\braket{1 | \F 8 {1.5}}$ no plano complexo.}
    \includegraphics[width=\textwidth]{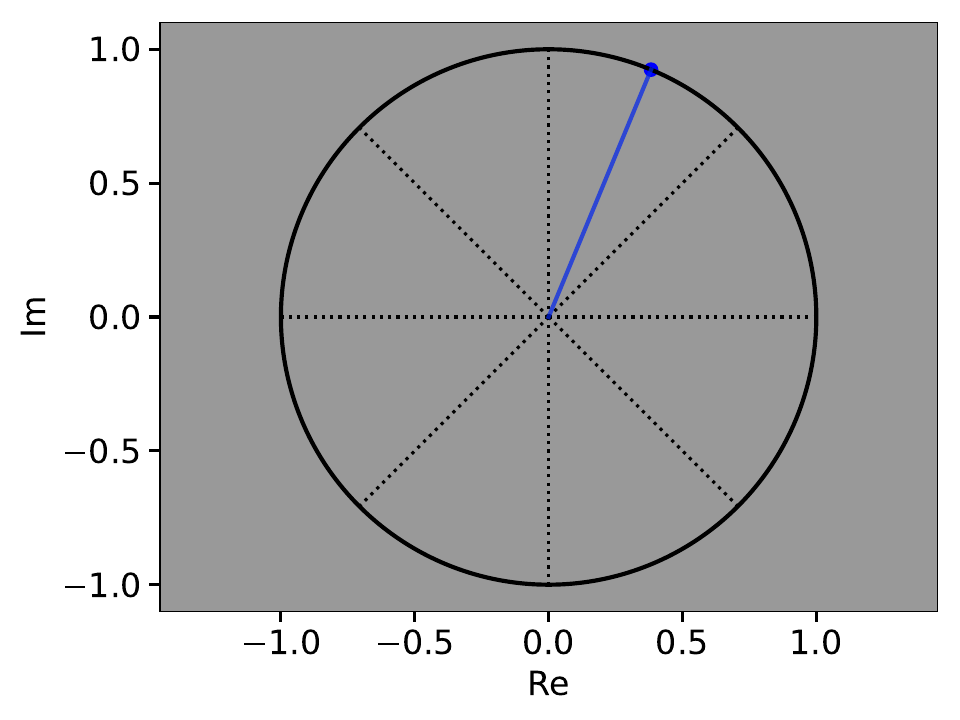}
    \legend{Fonte: Produzido pelo autor.}
    \label{fig:simpl-angulo-F_8(1.5)}
  \end{minipage}
\end{figure}

A definição genérica de $\ket{\F P \omega}$ permite analisar o cenário em que
$\omega \notin \mathbb{N}$, \textit{i.e.} $\ket{\F P \omega} \notin B_{\mathcal{F}}$.
Nesse caso, $\omega$ é um multiplicador dos ângulos base tal que $\floor\omega < \omega < \ceil\omega$,
e espera-se que os ângulos de $\ket{\F P \omega}$ estejam próximos dos ângulos
de $\ket{\F{P}{\floor\omega}}$ e $\ket{\F{P}{\ceil\omega}}$ --
vale frisar a equivalência entre $\ket{\F P 0}$ e $\ket{\F P P}$.
De fato, é possível ver que isso acontece ao comparar os ângulos de 
$\ket{\F{8}{1.01}}$ (Figura \ref{fig:angulos-F_8(1.01)}),
$\ket{\F{8}{1.5}}$ (Figura \ref{fig:angulos-F_8(1.5)}) e
$\ket{\F{8}{1.99}}$ (Figura \ref{fig:angulos-F_8(1.99)})
com os ângulos de $\ket{\F 8 1}$ (Figura \ref{fig:angulos-base}) e
$\ket{\F 8 2}$ (Figura \ref{fig:angulos-F_8(2)}).
Perceba que os ângulos de $\ket{\F{8}{1.5}}$ estão no meio do caminho entre
os ângulos de $\ket{\F 8 1}$ e $\ket{\F 8 2}$,
enquanto que os ângulos de $\ket{\F{8}{1.01}}$ estão bem mais próximos dos
ângulos de $\ket{\F 8 1}$ do que de $\ket{\F 8 2}$;
análogo para os ângulos de $\ket{\F 8 {1.99}}$.
Analisando as Figs. \ref{fig:angulos-F_8(1.01)} e \ref{fig:angulos-F_8(1.99)},
a interpretação de $\omega$ como multiplicador dos ângulos base fica mais clara.
Como todos os ângulos dependem de $\omega$,
é possível simplificar os gráficos ao representar apenas o valor de $\braket{1 | \F P \omega}$,
melhorando a visualização e facilitando interpretações futuras.
A Fig. \ref{fig:simpl-angulo-F_8(1.5)} ilustra a
representação simplificada de $\ket{\F 8 {1.5}}$.

É interessante notar também que alguns
ângulos de $\ket{\F{8}{1.5}}$ e $\ket{\F{8}{1.99}}$ não coincidem com os ângulos base --
\textit{i.e.} $\ket{\F{8}{1.5}}$, $\ket{\F{8}{1.99}} \notin B_{\mathcal{F}}$.
Essa observação levanta as seguintes perguntas (correlacionadas):
\begin{enumerate}
    \item Sendo $\ket{\F P \omega}$ representado na base $B_{\mathcal{F}}$,
    qual é o quadrado da norma das amplitudes de $\ket{\F{P}{\floor\omega}}$ e
    $\ket{\F{P}{\ceil\omega}}$? Ou, de forma geral, qual o valor de
    $\left|\braket{\F P \omega | \F{P}{\omega'}}\right|^2$ onde $\omega \neq \omega'$?
    \item Ao fazer uma medição de $QFT^{-1}\ket{\F P \omega}$ na base computacional,
    qual a probabilidade do resultado ser $\ket{\floor\omega}$ ou $\ket{\ceil\omega}$?
\end{enumerate}

A resposta da primeira pergunta é dada pelo
Lema 10 de BHMT \cite{brassard2002quantum}, adaptado a seguir.
\begin{lemma} \label{lema:proj-fourier}
Sejam $\hilb^P$ o espaço de Hilbert; $\omega$ e $\omega'$ valores reais tais que
$0 \leq \omega, \omega' \leq P$ e $\omega \neq \omega'$; e
$\omega_\Delta = \omega' - \omega$.
Então,
\begin{align}
    \left|\braket{\F P \omega | \F{P}{\omega'}}\right|^2 =
        \frac{\sin^2\pr{\omega_\Delta\pi}}{P^2\sin^2\pr{\omega_\Delta\pi/P}} .
\end{align}
\end{lemma}
\begin{proof}
    Análogo ao obtido na Eq. \ref{eq:proj-fourier-basis},
    obtém-se a série geométrica
    \begin{align}
        \left|\braket{\F P \omega | \F{P}{\omega'}}\right|^2 &= \left|
                \frac{1}{P^2} \sum_{\ell = 0}^{P-1} e^{2\pi\ii \omega_\Delta \ell / P}
            \right|^2
        \\
        &= \frac{1}{P^2}
            \pr{\frac{1 - e^{2\pi\ii \omega_\Delta}}{1 - e^{2\pi\ii\omega_\Delta/P}}}
            \pr{\frac{1 - e^{-2\pi\ii \omega_\Delta}}{1 - e^{-2\pi\ii\omega_\Delta/P}}}
        \\
        &= \frac{1}{P^2} \pr{
            \frac{2 - e^{2\pi\ii \omega_\Delta} - e^{-2\pi\ii \omega_\Delta}}{
            2 - e^{2\pi\ii \omega_\Delta/P} - e^{-2\pi\ii \omega_\Delta/P}}
        }
        \\
        &= \frac{1}{P^2} \pr{
            \frac{2 - 2\cosp{2\pi \omega_\Delta}}{2 - 2\cosp{2\pi \omega_\Delta/P}} }.
    \end{align}
    Usando a identidade trigonométrica $1 - \cosp{2\theta} = 2\sin^2\theta$,
    \begin{align}
        \left|\braket{\F P \omega | \F{P}{\omega'}}\right|^2 &=
        \frac{\sin^2\pr{\pi \omega_\Delta}}{P^2 \sin^2\pr{\pi \omega_\Delta/P}} ,
    \end{align}
    como esperado.
    
    Note que se $\omega, \omega' \in \mathbb{N}$, então
    $\ket{\F P \omega}$, $\ket{\F P {\omega'}} \in \BF$ e
    \begin{align}
        \sin^2(\omega_\Delta\pi) = 0,\ \sin^2(\omega_\Delta\pi/P) \neq 0
        \implies \left|\braket{\F P \omega | \F{P}{\omega'}}\right|^2 = 0 ,
    \end{align}
    conforme desejado.
\end{proof}

Já a resposta da segunda pergunta é dada pelo
Teorema 11 de BHMT \cite{brassard2002quantum} -- adaptado a seguir --,
que utiliza o Lema \ref{lema:proj-fourier}.
\begin{theorem} \label{teo:fourier-error}
    Seja $X$ a variável aleatória correspondente ao resultado da medição do estado
    $\QFT_P^{-1}\ket{\F P \omega}$ na base computacional onde
    $\omega \in \mathbb{R}$ e $0 \leq \omega \leq P$.
    Se $\omega$ for um inteiro, então $\prob{X = \omega} = 1$
    e o estado após a medição é $\ket\omega$ (denota-se $\ket P = \ket 0$).
    Caso contrário,
    \begin{align}
        \prob{X = \floor\omega} + \prob{X = \ceil\omega} =
        \prob{|X - \omega| \leq 1} \geq \frac{8}{\pi^2}
    \end{align}
    é a probabilidade associada a obter um dos estados
    $\ket{\floor\omega}$ ou $\ket{\ceil\omega}$ após a medição.
    
\end{theorem}
\begin{proof}
    Considere os operadores de medida da base computacional
    $\set{M_m} = \set{\ket{m}\bra{m}}$ para $m \in \mathbb{N}$ tal que $0 \leq m < P$.
    Então, de modo geral,
    \begin{align}
        \prob{X = m} &= \pr{\bra{\F P \omega} \QFT_P} M_m^\dagger
            M_m \pr{\QFT_P^\dagger\ket{\F P \omega}}
        \\
        &= \bra{\F P \omega} \QFT_P \ket m
            \bra m \QFT_P^\dagger\ket{\F P \omega}
        \\
        &= \braket{\F P \omega | \F P m} \braket{\F P m | \F P \omega}
        \\
        &= \card{ \braket{\F P m | \F P \omega} }^2
    \end{align}
    e após a medição o estado obtido ignorando-se a fase global é
    \begin{align}
        \frac{
            \pr{\ket{m}\bra{m}}\ \QFT^\dagger \ket{\F P \omega}
        }{
            \sqrt{\card{ \braket{\F P m | \F P \omega} }^2}
        }
        = \frac{
            \ket{m} \braket{\F P m | \F P \omega}
        }{
            \card{ \braket{\F P m | \F P \omega} }
        }
        = \ket{m} .
    \end{align}
    Como $\ket{\F P 0}$ e $\ket{\F P P}$ são equivalentes,
    ambos os estados geram os mesmos resultados;
    então denota-se $\ket P = \ket 0$ e $\prob{X = P} = \prob{X = 0}$.
    Usando-se esses fatos sistematicamente,
    vários cenários são considerados.
    
    Se $P = 1$, Lema \ref{lema:proj-fourier} dá que
    \begin{align}
        \card{ \braket{\F P \omega | \F P {\omega'}} }^2 =
        \frac{\sin^2\pr{\pi\omega_\Delta}}{\sin^2\pr{\pi\omega_\Delta}}
        = 1.
    \end{align}
    O caso $P = 2$ foi analisado no Apêndice \ref{apendice:minimo-funcao-teo-ang-int}.
    Caso $P > 2$ e $\omega$ for um inteiro, então $\ket{\F P \omega} \in \BF$ e
    \begin{align}
        \QFT_P^{-1}\ket{\F P \omega} = \ket{\omega}
    \end{align}
    é o estado obtido após a medida com probabilidade 1.
    
    Caso contrário ($P > 2$ e $\omega \notin \mathbb{N}$), deseja-se calcular
    \begin{align}
        \prob{|X - \omega| \leq 1} &= \prob{X = \floor\omega} + \prob{X = \ceil\omega}
        \\
        &= \left| \braket{\F P {\floor\omega} | \F P \omega} \right|^2 +
            \left| \braket{\F P {\ceil\omega} | \F P \omega} \right|^2.
    \end{align}
    Usando o Lema \ref{lema:proj-fourier} e donotando $w = |\omega - \floor\omega|$
    (note que $0 < w < 1$ e $\card{\omega - \ceil\omega} = 1 - w$),
    obtém-se
    \begin{align}
        \prob{|X - \omega| \leq 1} &= f(w) =
            \frac{\sin^2\pr{\pi w}}{P^2 \sin^2\pr{\pi w / P}} +
            \frac{\sin^2\pr{\pi (1-w)}}{
                P^2 \sin^2\pr{\pi (1-w) / P}} .
        \label{eq:teo-fourier-erro}
    \end{align}
    A função $f(w)$ tem mínimo quando $w = 1/2$,
    conforme ilustrado nas Figs. \ref{fig:prob-min-P=3} e \ref{fig:prob-min-P=30}
    para $P = 3$ e P = $30$, respectivamente.
    Uma demonstração formal pode ser encontrada no Apêndice \ref{apendice:minimo-funcao-teo-ang-int}.
    Usando também que $\sin^2\theta \leq \theta^2$, obtém-se
    \begin{align}
        \prob{|X - \omega| \leq 1} &\geq 2
            \frac{\sin^2\pr{\frac \pi 2}}{P^2 \sin^2\pr{ \frac{\pi}{2P} }}
        \\
        &\geq \frac{2}{P^2} \cdot \frac{4 P^2}{\pi^2}
        \\
        &= \frac{8}{\pi^2}.
    \end{align}
    Logo, uma medição de $\QFT_P^{-1}\ket{\F P \omega}$ na base computacional e
    ignorando a fase global resulta em um dos estados $\ket{\floor\omega}$
    ou $\ket{\ceil{\omega}}$ com probabilidade maior ou igual a $8/\pi^2$.
    
    \begin{figure}[htb]
     \centering
      \begin{minipage}{0.48\textwidth}
        \centering
        \caption{Gráfico de $f(w)$ com $P = 3$ e respectivo mínimo global.}
        \includegraphics[width=\textwidth]{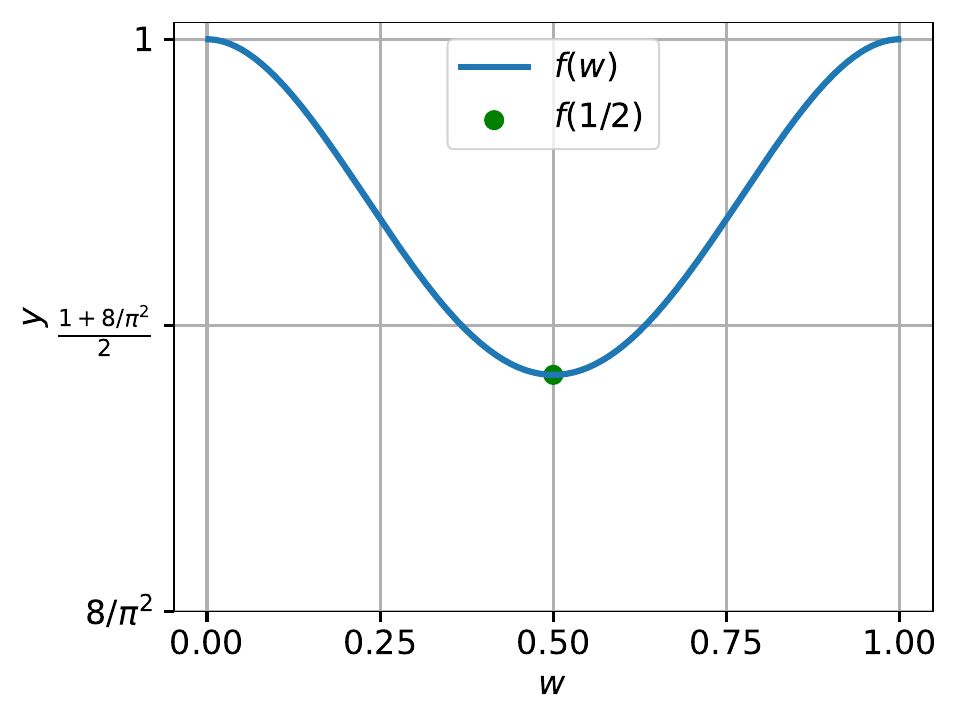}
        \legend{Fonte: Produzido pelo autor.}
        \label{fig:prob-min-P=3}
      \end{minipage}
      \hfill
      \begin{minipage}{0.48\textwidth}
        \centering
        \caption{Gráfico de $f(w)$ com $P = 30$ e respectivo mínimo global.}
        \includegraphics[width=\textwidth]{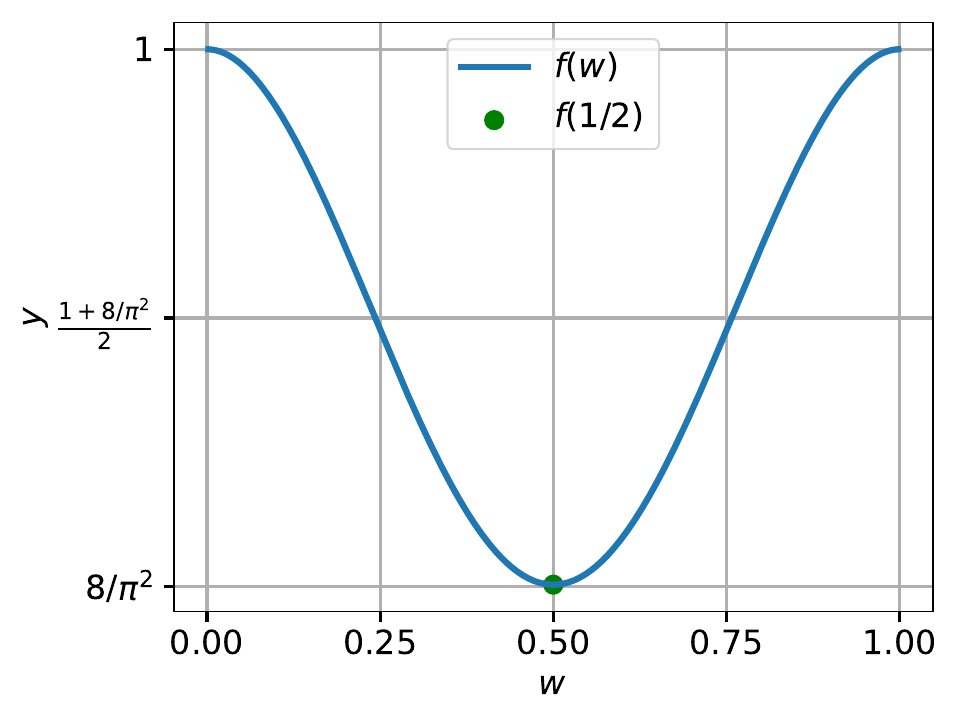}
        \legend{Fonte: Produzido pelo autor.}
        \label{fig:prob-min-P=30}
      \end{minipage}
    \end{figure}
\end{proof}

\subsection{Implementação da QFT e sua inversa}
O foco dessa seção é a implementação da QFT (e sua inversa) em portas quânticas.
Para tanto, considere ao longo dessa sessão que $P = 2^p$ onde $p \in \mathbb{N}$ é
a quantidade de qubits.
Quando $p = 1$, a porta $H$ implementa a QFT, já que
\begin{align}
    \ket{\F 2 0} &= \frac{1}{\sqrt 2} \sum_{\ell = 0}^1 e^{0} \ket\ell
    \\ &= \frac{\ket 0 + \ket 1}{\sqrt 2},
\end{align}
e
\begin{align}
    \ket{\F 2 1} &= \frac{1}{\sqrt 2} \sum_{\ell = 0}^1 e^{\pi \ii \ell} \ket\ell \\
    &= \frac{\ket 0 - \ket 1}{\sqrt 2};
\end{align}
ou, de forma geral,
\begin{align}
    \ket{\F 2 d} = 2^{-p/2} \pr{\ket0 + e^{2\pi\ii d/P} \ket 1},
\end{align}
onde $d = \Set{0, 1}$.

Para $p \geq 2$, será necessário utilizar a seguinte definição.
\begin{definition}
    A porta $\RF_k$ atua em vetores em $\hilb^2$ e é definida na base computacional como
    \begin{align}
        \RF_k &= \matrx{
            1 & 0 \\ 0 & e^{2\pi \ii / 2^k}
        }.
    \end{align}
    A ação dessa porta é multiplicar o segundo estado da base computacional pela fase
$e^{2\pi \ii / 2^k}$.
\end{definition}

\begin{figure}[htb]
    \centering
    \caption{Circuito básico para implementação da QFT.}
    \includegraphics[width=0.5\textwidth]{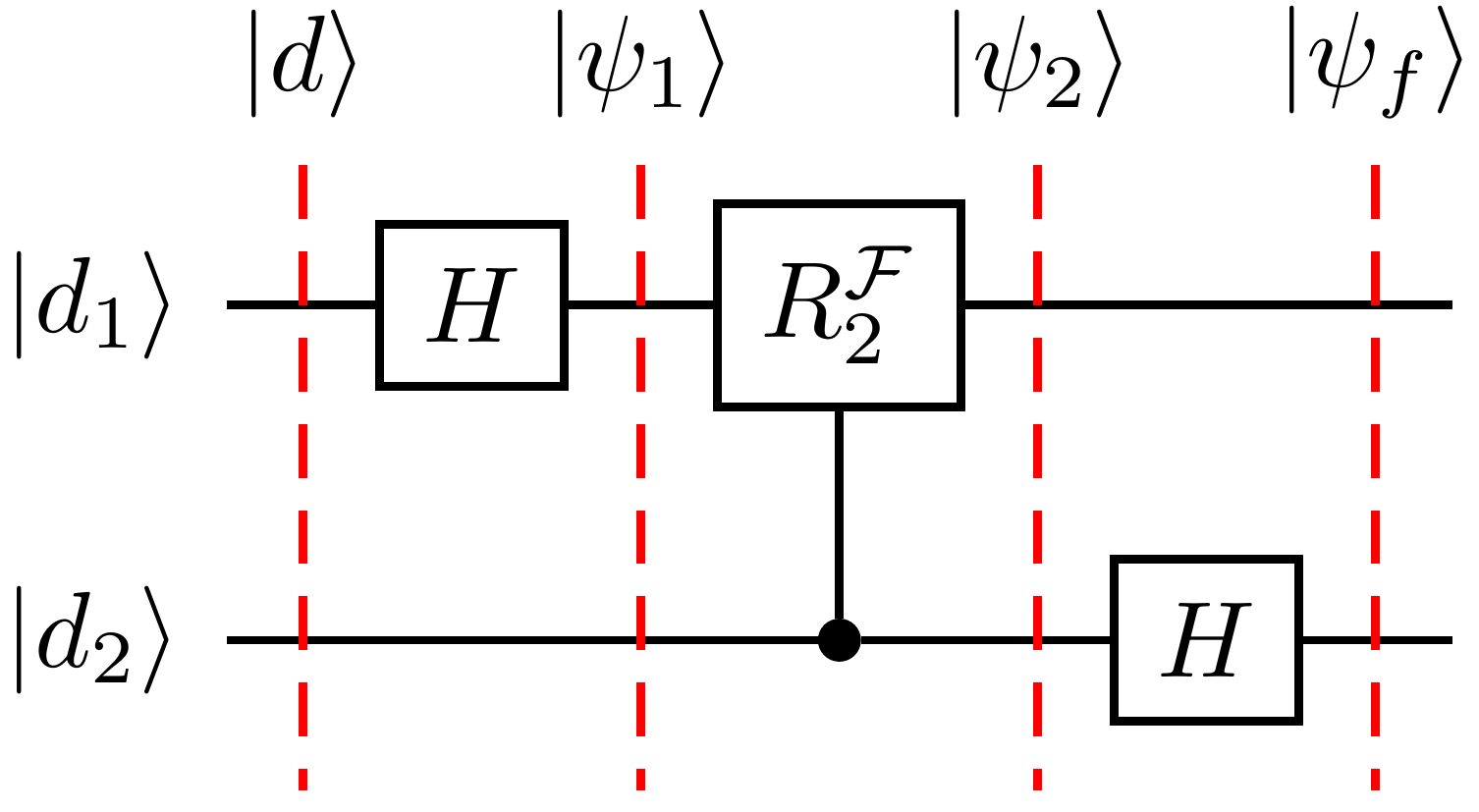}
    \legend{Fonte: Produzido pelo autor.}
    \label{fig:qft-rec-base}
\end{figure}

Para $p = 2$, considere o circuito da Fig. \ref{fig:qft-rec-base}
com entrada $\ket{d} = \ket{d_1}\ket{d_2} \in \hilb^4$ sendo
$\ket{d_1} e \ket{d_2}$ vetores da base computacional representando o valor $d$,
\textit{i.e.} $d_1d_2$ é a representação binária de $d$.
Primeiramente, aplica-se a porta $H$ no primeiro qubit, resultando no estado
\begin{align}
    \ket{\psi_1} &= (H \otimes I)\ket{d}
    \\
    &= 2^{-1/2} \pr{\ket0 + e^{2\pi\ii d_1 / 2} \ket{1}} \otimes \ket{d_2}.
    \label{eq:circ-qft-rec-base-ini}
\end{align}
A aplicação da porta $\RF_2$-controlada resulta em
\begin{align}
    \ket{\psi_2} &= \bctrl 2 1 {\RF_2} \ket{\psi_1}
    \\
     &= 2^{-1/2} \pr{\ket0 + e^{2\pi\ii d_1 / 2}e^{2\pi\ii d_2/4} \ket{1}} \otimes \ket{d_2}.
\end{align}
Note que a representação em binário da divisão $d/4$ é
$(d_1\cdot2^1 + d_2\cdot2^0)/2^2 = d_1\cdot2^{-1} + d_2\cdot2^{-2}$.
Logo, é possível reescrever o estado $\ket{\psi_2}$ como
\begin{align}
    \ket{\psi_2} = 2^{-1/2} \pr{\ket0 + e^{2\pi\ii d / 4} \ket{1}} \otimes \ket{d_2}.
\end{align}
Ao aplicar a porta $H$ no segundo qubit e notando que $d/2 = d_1\cdot2^0 + d_2\cdot2^{-1}$,
obtém-se o estado final
\begin{align}
    \ket{\psi_f} &= (I \otimes H) \ket{\psi_2}
    \\
    &= 2^{-1/2} \pr{\ket0 + e^{2\pi\ii d / 4} \ket{1}} \otimes
        2^{-1/2} \pr{\ket{0} + e^{2\pi\ii d_2 / 2} \ket{1}}
    \\
    &= 2^{-2/2} \pr{\ket0 + e^{2\pi\ii d / 4} \ket{1}} \otimes
        \pr{\ket{0} + e^{2\pi\ii d_1} e^{2\pi\ii d_2 / 2} \ket{1}}.
    \\
    &= 2^{-2/2} \pr{\ket0 + e^{2\pi\ii d / 4} \ket{1}} \otimes
        \pr{\ket{0} + e^{2\pi\ii d / 2} \ket{1}}.
    \label{eq:circ-qft-rec-base-fim}
\end{align}
A partir de $\ket{\psi_f}$, é possível obter um estado da base de Fourier
após aplicar a operação swap.
\begin{align}
    \textrm{SWAP} \ket{\psi_f} &= 2^{-2/2} \pr{\ket{0} + e^{2\pi\ii d / 2} \ket{1}} \otimes
        \pr{\ket0 + e^{2\pi\ii d / 4} \ket{1}}
    \\
    &= 2^{-2/2} \pr{\ket{00} + e^{2\pi\ii d / 4} \ket{01} +
        e^{2\pi\ii d\cdot2 / 4} \ket{10} + e^{2\pi\ii d\cdot3 / 4}\ket{11}}
    \\
    &= \frac{1}{2} \sum_{\ell = 0}^{3} e^{2\pi\ii d \ell / P} \ket{\ell}
    \\
    &= \ket{\F 4 d}.
\end{align}

\begin{figure}[hbt]
    \centering
    \caption{Circuito que inverte a ordem de sete qubits.}
    \includegraphics[width=0.2\textwidth]{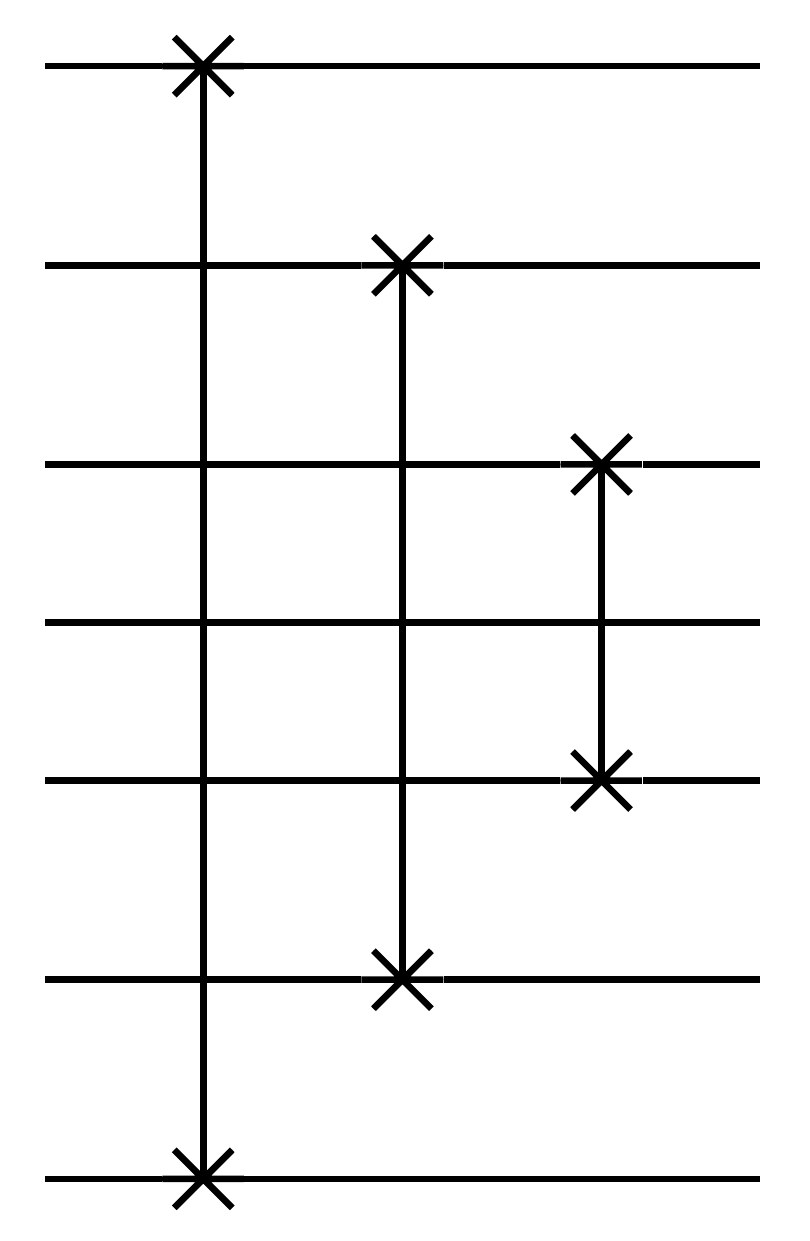}
    \legend{Produzido pelo autor.}
    \label{fig:inversao-qubits}
\end{figure}

Agora, deseja-se mostrar que é possível gerar os estados da base de Fourier para $p \geq 3$.
Para isso, as seguintes definições serão necessárias.
\begin{definition}
    \label{def:qft-pre-swap}
    Seja $p \geq 1$ e $d' \in \mathbb{N}$ cuja representação binária é $d_1 \ldots d_p$.
    Define-se a porta $\QFTrec_p$ como
    \begin{align}
        \QFTrec_{p}\ket{d'} &= 2^{-p/2} \bigotimes_{j=0}^{p-1} \pr{
            \ket 0 + \exp\pr{2\pi\ii d' / 2^{p-j}} \ket 1}.
    \end{align}
\end{definition}
Note que quando $p = 1$, obtém-se $\QFTrec_p \ket{d'} = H \ket{d'}$.
\begin{definition}
    Seja $p$ um inteiro maior ou igual a 2.
    Define-se $\SWAP_p'$ a sequência de operações SWAP que inverte a ordem de todos os qubits.
    Ou seja,
    \begin{align}
        \SWAP_p' = \prod_{t=1}^{\floor{p/2}} \SWAP_{t,p-t}.
    \end{align}
    \label{def:qft-swap}
\end{definition}
O circuito de $\SWAP_7'$ está ilustrado na Fig. \ref{fig:inversao-qubits}.

Em posse dessas definições, é possível descrever o circuito
que implementa $\QFT_P$ para $p \geq 2$.
\begin{theorem}
    Sejam $p, P, d \in \mathbb{N}$ tais que $p \geq 2$, $P = 2^p$ e $0 \leq d < P$
    cuja representação binária é $d_1 \ldots d_p$.
    Então,
    \begin{align}
        \QFTrec_{p} = \pr{I \otimes \QFTrec_{p-1}} \prod_{k=2}^{p}\bctrl k 1 {\RF_k}
            \pr{H \otimes I}
        \label{eq:circ-qft-rec}
    \end{align}
    e
    \begin{align}
        \SWAP_p'\ \QFTrec_p \ket{d} = \ket{\F P d}.
        \label{eq:circ-qft-final}
    \end{align}
    Ou seja, $\SWAP_p'\ \QFTrec_p$ descreve o circuito que implementa $\QFT_P$.
    \label{teo:circuito-qft}
\end{theorem}
\begin{proof}
    Prova-se a Eq. \ref{eq:circ-qft-rec} por indução.
    Note que o caso base $p=2$ já foi abordado da Eq.
    \ref{eq:circ-qft-rec-base-ini} à \ref{eq:circ-qft-rec-base-fim}.
    Hipótese indutiva: suponha que
    \begin{align}
        \QFTrec_{p'} = \pr{I \otimes \QFTrec_{p'-1}} \prod_{k=2}^{p'}\bctrl k 1 {\RF_k}
            \pr{H \otimes I}
    \end{align}
    é verdade para todos os valores $2 \leq p' \leq p-1$.
    Passo indutivo: demonstrar que a Eq. \ref{eq:circ-qft-rec} é verdadeira para $p$.
    O circuito correspondente está desenhado na Fig. \ref{fig:qft-rec}.
    
    \begin{figure}[hbt]
        \centering
        \caption{Circuito de $\QFTrec_p$.}
        \includegraphics[width=0.9\textwidth]{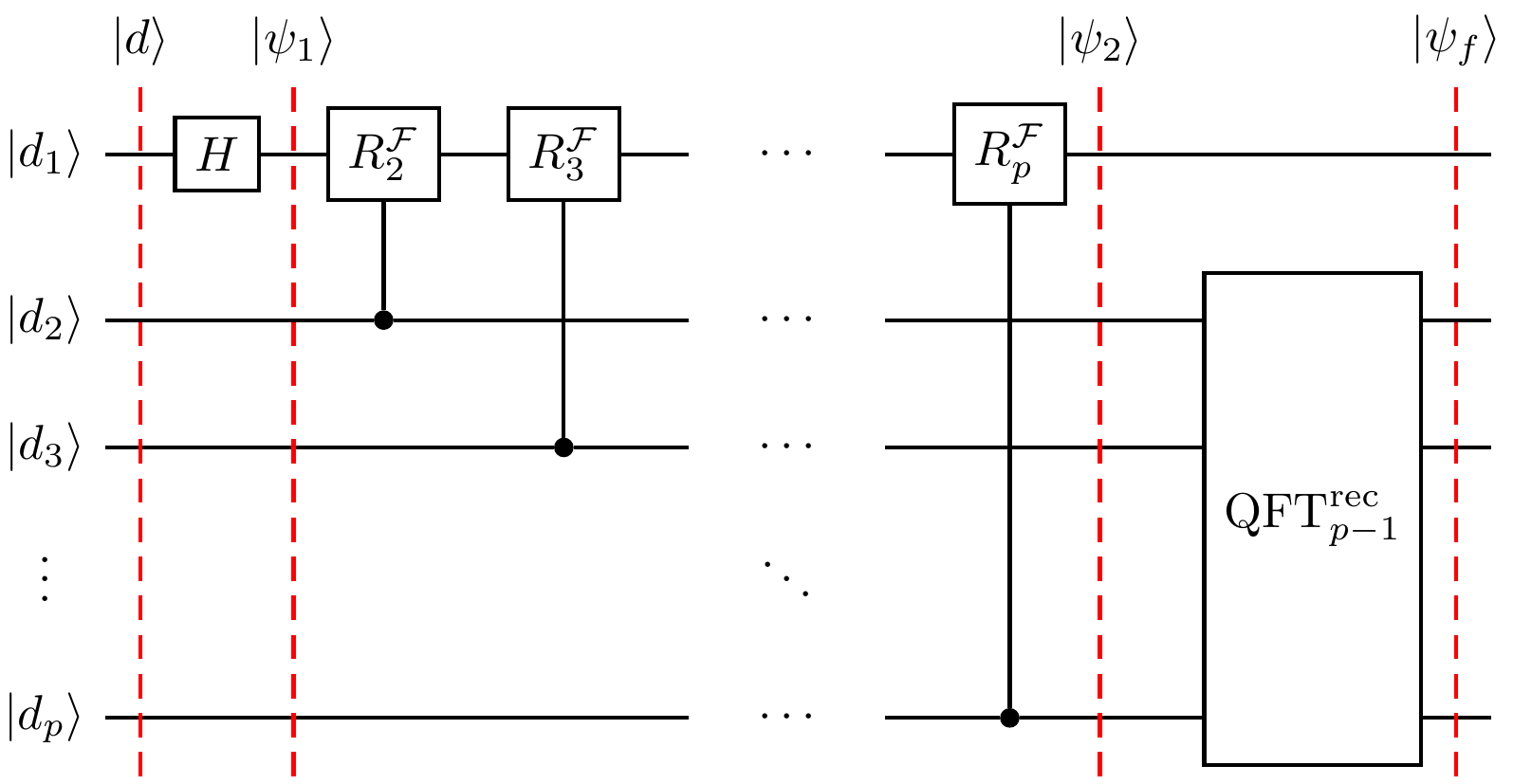}
        \legend{Fonte: Produzido pelo autor.}
        \label{fig:qft-rec}
    \end{figure}
    
    Primeiro aplica-se a porta $(H \otimes I)$ em $\ket d$, obtendo-se o estado
    \begin{align}
        \ket{\psi_1} &= (H \otimes I) \ket{d_1}\ket{d_2 \ldots d_{p}}
        \\
        &= 2^{-1/2} \pr{\ket 0 + \exp\pr{2\pi\ii d_1 / 2} \ket 1}
            \otimes \ket{d_2 \ldots d_{p}}.
    \end{align}
    Agora, aplica-se a sequência de portas controladas $\bctrl k 1 {\RF_k}$.
    Note que a ordem de aplicação dessas portas é irrelevante pois
    a ação de cada $\RF_k$ é de multiplicar a fase de $\ket 1$
    por $\exp\pr{2\pi\ii/2^k}$.
    A aplicação ou não da porta $\RF_k$ depende diretamente do valor $d_k$:
    a porta não atua caso $d_k = 0$ e atua caso $d_k = 1$,
    alterando o qubit alvo conforme
    \begin{align}
        \ket 1 \to \exp\pr{2\pi\ii d_k / 2^k} \ket 1.
    \end{align}
    Note também que a divisão $d / 2^p$ pode ser escrita como
    \begin{align}
        d / 2^p &= \pr{d_1 \cdot 2^{p-1} + \cdots + d_p \cdot 2^0} / 2^p
        \\
        &= \pr{d_1 / 2^{1} + \cdots + d_p / 2^p}
        \\
        &= \sum_{k=1}^{p} d_k / 2^k .
    \end{align}
    Juntado todos esses argumentos, a aplicação da sequência de $\bctrl k 1 {\RF_k}$ no
    estado $\ket{\psi_1}$ resulta em
    \begin{align}
        \ket{\psi_2} &=\prod_{k=2}^{p}\bctrl k 1 {\RF_k} \ket{\psi_1}
        \\
        &= 2^{-1/2} \pr{\ket 0 + \prod_{k=1}^p \exp\pr{2\pi\ii d_k / 2^k} \ket 1}
            \otimes \ket{d_2 \ldots d_p}
        \\
        &= 2^{-1/2} \pr{\ket 0 + \exp\pr{2\pi\ii \sum_{k=1}^p d_k / 2^k} \ket 1}
            \otimes \ket{d_2 \ldots d_p}
        \\
        &= 2^{-1/2} \pr{\ket 0 + \exp\pr{2\pi\ii d/2^p} \ket 1}
            \otimes \ket{d_2 \ldots d_p} .
    \end{align}
    Por último, seja o número binário $d' = d_2 \ldots d_p$.
    A aplicação de $I \otimes \QFTrec_{p-1}$ resulta em
    \begin{align}
        \ket{\psi_f} &= \pr{I \otimes \QFTrec_{p-1}} \ket{\psi_2}
        \\
        &= 2^{-1/2} \pr{\ket 0 + \exp\pr{2\pi\ii d/2^p} \ket 1} \otimes
            \QFTrec_{p-1} \ket{d_2 \ldots d_{p-1}}.
        \\
        &= 2^{-p/2} \pr{\ket 0 + \exp\pr{2\pi\ii d/2^p} \ket 1} \pr{
            \bigotimes_{j'=0}^{p-2} \pr{\ket 0 + \exp\pr{2\pi\ii d' / 2^{p-1-j'}} \ket 1}},
    \end{align}
    renomeando $j = j' + 1$,
    \begin{align}
        \ket{\psi_f} &= 2^{-p/2}\pr{\ket 0 + \exp\pr{2\pi\ii d/2^p} \ket 1}
            \bigotimes_{j=1}^{p-1} \pr{\ket 0 + \exp\pr{2\pi\ii d' / 2^{p-j}} \ket 1}
        \\
        &= 2^{-p/2} \bigotimes_{j=0}^{p-1} \pr{
            \ket 0 + \exp\pr{2\pi\ii d' / 2^{p-j}} \ket 1} .
    \end{align}
    Portanto, conclui-se que a Eq. \ref{eq:circ-qft-rec} é verdadeira.
    
    Resta demonstrar que a Eq. \ref{eq:circ-qft-final} realmente implementa
    $\QFT_P$.
    Sabendo que $\SWAP_p'$ apenas inverte a ordem dos qubits, tomando
    $\ell$ um número inteiro $0 \leq \ell < P$ com representação binária
    $\ell_1 \ldots \ell_p$ e notando que
    \begin{align}
        \ell/2^p = \sum_{k=1}^p \ell_k / 2^k = \sum_{\ell_k = 1} \ell_k / 2^k,
    \end{align}
    obtém-se
    \begin{align}
        \SWAP_p'\ \QFTrec_p \ket{d} &= 2^{-p/2}\ \SWAP_p'\ \bigotimes_{j=0}^{p-1} \pr{
            \ket 0 + \exp\pr{2\pi\ii d / 2^{p-j}} \ket 1}
        \\
        &= 2^{-p/2} \bigotimes_{k=1}^{p} \pr{\ket 0 + \exp\pr{2\pi\ii d / 2^k} \ket 1}
        \\
        &= 2^{-p/2} \sum_{\ell=0}^{P-1} \exp\pr{2\pi\ii d \ell/ 2^p} \ket{\ell}
        \\
        &= \ket{\F P d},
    \end{align}
    conforme desejado.
\end{proof}

\begin{figure}[hbt]
    \centering
    \caption{Circuito para $\QFT_P^{-1}$.}
    \includegraphics[width=\textwidth]{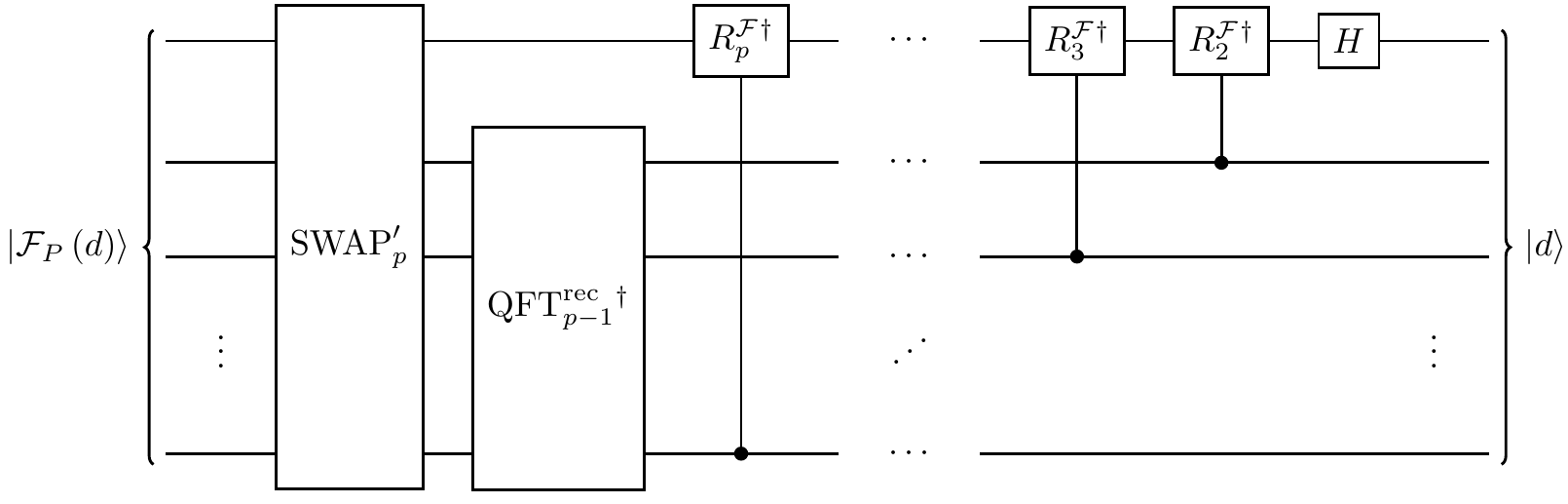}
    \legend{Fonte: Produzido pelo autor.}
    \label{fig:inv-qft}
\end{figure}

Convencendo-se que esse circuito implementa $\QFT_P$ num computador quântico,
o circuito para $\QFT_P^{-1}$ é obtido facilmente ao fazer
\begin{align}
    \QFT_P^{-1} &= \QFT_p^\dagger \\
    &= \QFTrec_p{^\dagger}\ \SWAP_p' \\
    &= \pr{H \otimes I} \prod_{k=2}^{p}\bctrl k 1 {\RF_{k}{^\dagger}}
        \pr{I \otimes \QFTrec_{p-1}{^\dagger}} \SWAP_p'.
\end{align}
originando o circuito da Fig. \ref{fig:inv-qft}.

\section{Estimativa de Fase} \label{sec:est-fase}
Seja $N \in \mathbb{N}$ tal que $N \geq 1$.
Nessa seção será abordado o problema de estimar a fase
$e^{2\pi\ii\lambda}$ de um autovetor $\ket\lambda \in \hilb^N$ de uma matriz unitária $U$.
A fase pode ser estimada a partir de uma estimativa do valor
$0 \leq \lambda < 1$ no expoente do autovalor.
Note que há uma similaridade na estrutura de $e^{2\pi\ii\lambda}$ com
os expoentes do estado $\ket{\F P {\lambda}}$ (Def. \ref{def:F_P(omega)}).
Essa similaridade levanta a possibilidade de utilizar a QFT e sua inversa para
tentar estimar $\lambda$.
Antes de utilizar a QFT inversa,
é necessário obter o estado $\ket{\F P {\lambda}}$ de alguma forma;
tópico abordado a seguir.

\subsection{Elevação de Matriz Controlada}

Seja $b \in \mathbb{N}$ cuja representação binária possui $p$ dígitos:
$b_{p-1}\ldots b_1b_0$.
Como usual na computação, bits menos significativos têm subíndices menores,
possibilitando a representação $b = \sum_{j=0}^{p-1} b_j \cdot 2^j$.
Sendo $\ket b = \ket{b_{p-1}} \otimes \cdots \otimes \ket{b_0}$
o estado correspondente ao número $b$.

\begin{figure}[htb]
    \centering
    \caption{Parte do circuito do algoritmo de estimativa de fase.}
    \includegraphics[width=0.8\textwidth]{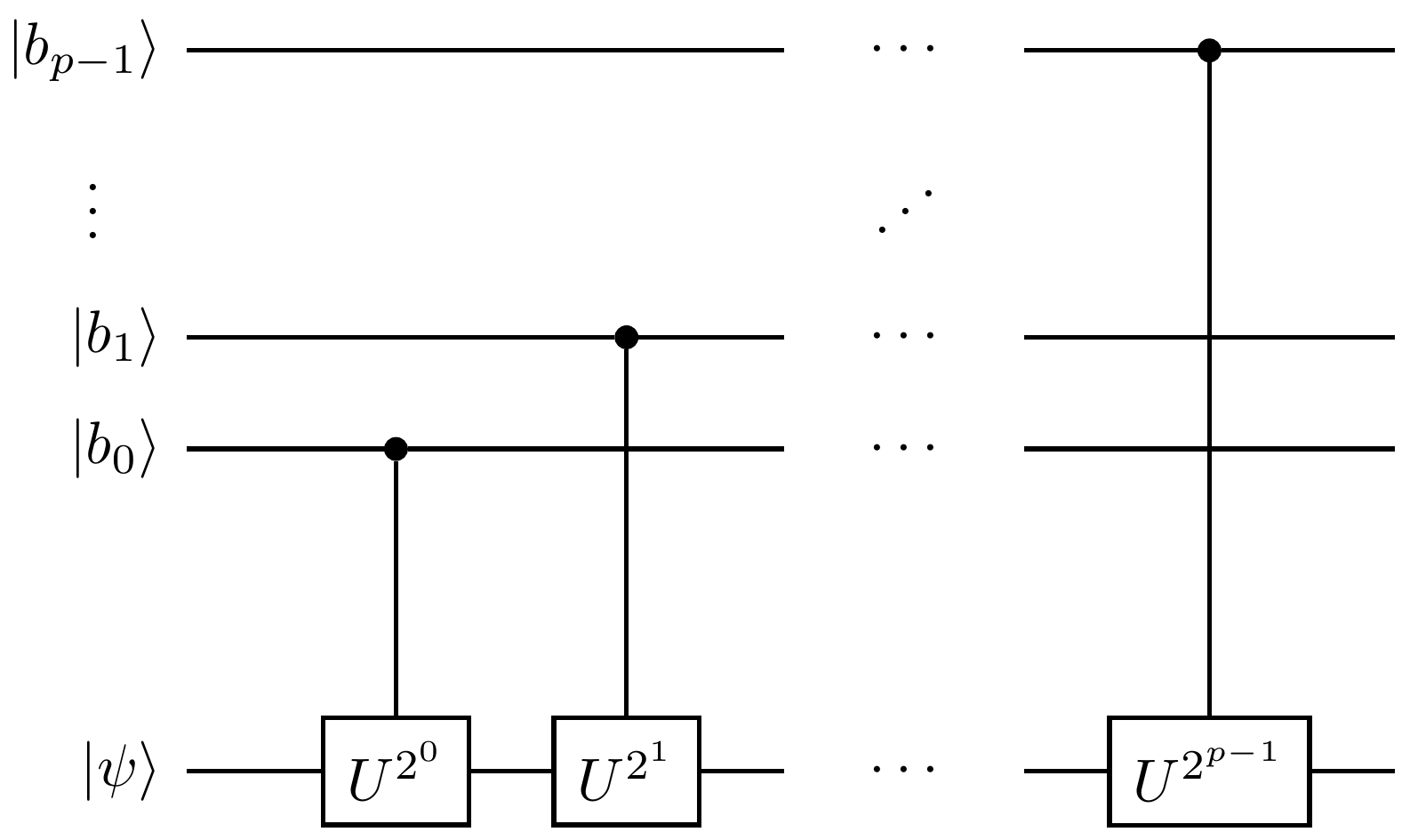}
    \legend{Fonte: Produzido pelo autor.}
    \label{fig:pot-ctrl}
\end{figure}

\begin{figure}[htb]
        \centering
        \caption{Representação do circuito $\pctrl{U}$ no espaço reduzido.}
        \includegraphics[width=0.3\textwidth]{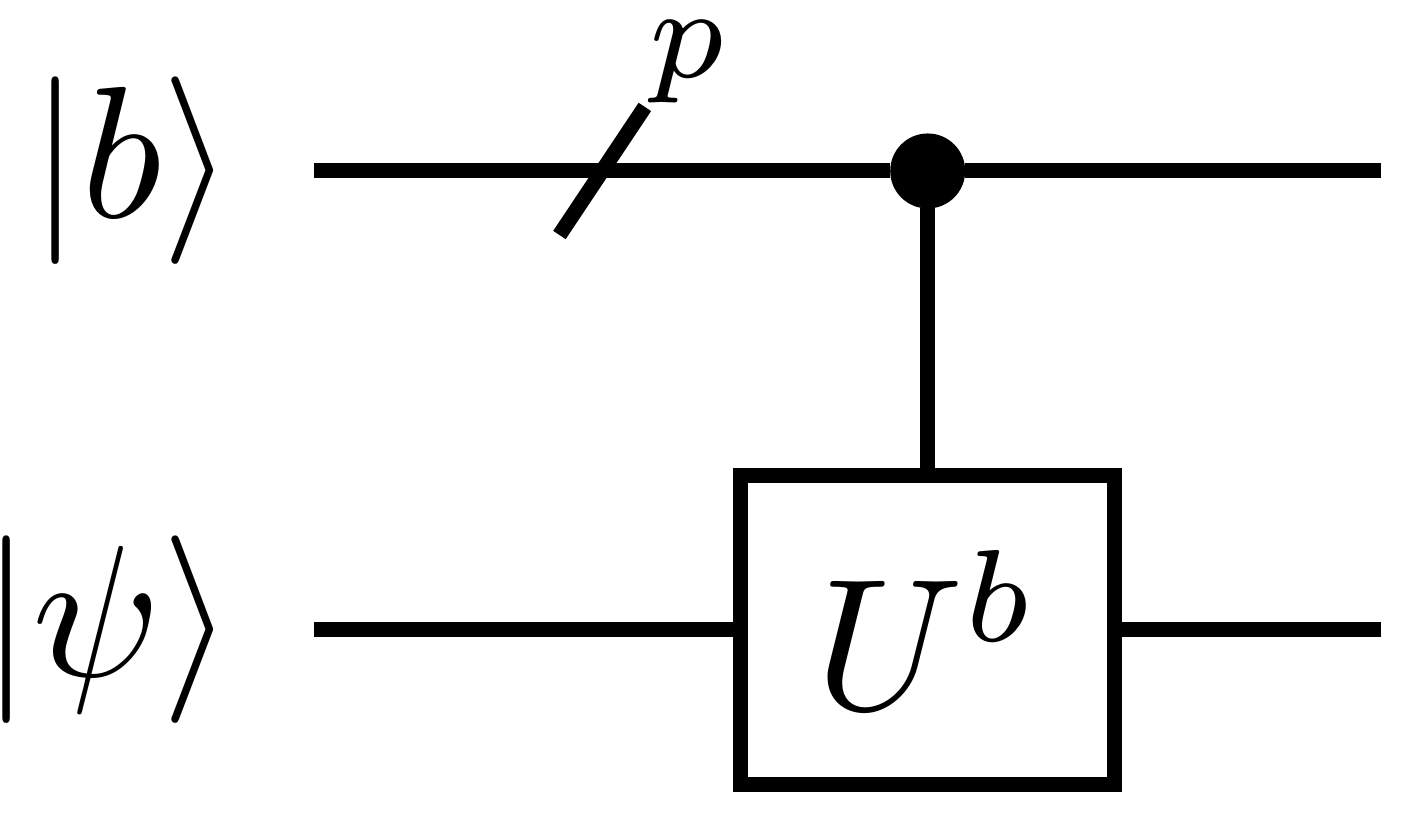}
        \legend{Fonte: Produzido pelo autor.}
        \label{fig:red-pot-ctrl}
    \end{figure}

Analisa-se o circuito da Fig. \ref{fig:pot-ctrl}, que possui dois registradores.
O primeiro registrador recebe $\ket{b}$ como entrada e o segundo registrador
um vetor $\ket\psi \in \hilb^N$.
Observe que o valor $b_j$ é responsável por
controlar a porta $U^{2^j}$ que atua no segundo registrador.
O segundo registrador não necessariamente é implementado com qubits,
já que $N$ não necessariamente é uma potência de 2 --
\textit{e.g.} pode ser um qudit \cite{thew2002qudit}.
Notando que $U^{0 \cdot 2^j} = I$ e $U^{1 \cdot 2^j} = U^{2^j}$,
a ação do circuito pode ser representada por
\begin{align}
    \ket{b} \otimes \prod_{j = 0}^{p-1} U^{b_j \cdot 2^j} \ket{\psi} &=
        \ket{b} \otimes U^{\sum_{j = 0}^{p-1} b_j \cdot 2^j} \ket{\psi}
    \\&= \ket b \otimes U^b \ket\psi .
\end{align}
Ou seja, esse circuito aplica $b$ vezes o operador $U$ no vetor $\ket\psi$.
Para simplificar a notação, faz-se a seguinte definição.
\begin{definition}
    O operador $\pctrl{U}$ atua em $\ket{b}\otimes\ket\psi \in \hilb^P \otimes \hilb^N$
    aplicando $b$ potências de $U$ em $\psi$.
    \begin{align}
        \pctrl{U} \ket{b} \otimes \ket{\psi} &= \ket{b}\otimes U^b \ket{\psi}.
    \end{align}
\end{definition}
O circuito de $\pctrl{U}$ no espaço reduzido é representado na Fig. \ref{fig:red-pot-ctrl}.

Sabendo do comportamento do circuito $\pctrl{U}$,
faz-se a seguinte pergunta:
``qual a saída do circuito se o vetor do segundo registrador for um autovetor de $U$?''
Considerando o $e^{2\pi\ii\lambda}$-autovetor $\ket\lambda$ obtém-se
\begin{align}
    \pctrl{U} \ket b \ket\lambda &= \ket{b} \otimes U^b\ket\lambda
    \\
    &= \ket{b} \otimes e^{b \cdot 2\pi\ii\lambda} \ket{\lambda}
    \\
    &= e^{b \cdot 2\pi\ii\lambda} \ket{b}\ket\lambda.
\end{align}
O fenômeno de \emph{phase kickback} ocorrido é o que possibilita o funcionamento
do algoritmo de estimativa de fase.

Uma pergunta análoga é:
``o que ocorre se a entrada do segundo registrador for uma sobreposição não trivial
da base ortonormal de autovetores,
\textit{i.e.} $\ket\psi = \sum_{j=1}^{N} \alpha_j \ket{\lambda_j}$?''
Nesse caso,
\begin{align}
    \pctrl{U} \ket b \ket\psi &= \ket{b} \otimes U^b \sum_{j=1}^N \alpha_j \ket{\lambda_j}
    \\
    &= \ket{b} \otimes \sum_{j = 1}^N e^{b \cdot 2\pi\ii\lambda_j} \alpha_j \ket{\lambda_j}
    \\
    &= \sum_{j = 1}^N \pr{ e^{b \cdot 2\pi\ii\lambda_j} \alpha_j \ket{b} \otimes \ket{\lambda_j}}.
\end{align}
Observe que o fenômeno de \emph{phase kickback} ainda ocorre,
mas as amplitudes $\alpha_j$ provenientes da sobreposição também influenciam
o primeiro registrador.
Tal qual na sobreposição $\ket\psi$, as amplitudes $\alpha_j$
influenciam nas probabilidades de cada estado.

\subsection{Circuito Completo}
\begin{figure}[hbt]
    \centering
    \caption{Circuito do algoritmo de estimativa de fase.}
    \includegraphics[width=0.9\textwidth]{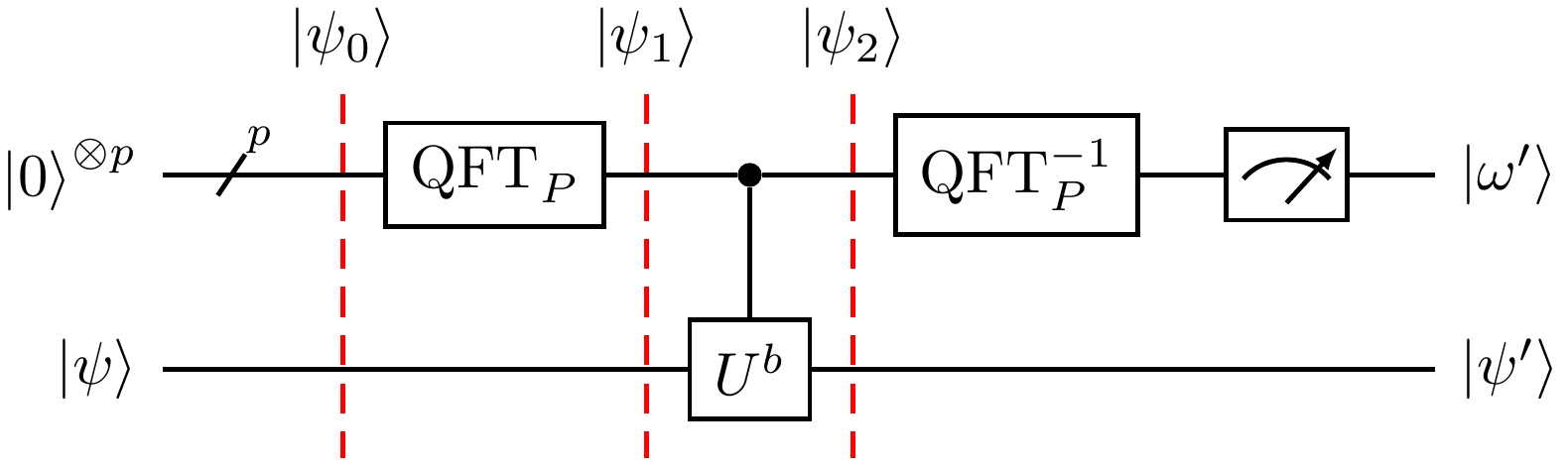}
    \legend{Fonte: Produzido pelo autor.}
    \label{fig:est-fase}
\end{figure}

Com o conhecimento mais aprofundado sobre a ação do operador $\pctrl{U}$,
é possível analisar o algoritmo de estimativa de fase (Alg. \ref{alg:est-fase})
cujo circuito está ilustrado na Fig. \ref{fig:est-fase}.

\begin{algorithm}
    \caption{Algoritmo de Estimativa de Fase -- est\_fase($p, U, \ket{\psi}$)}
    \label{alg:est-fase}
    \begin{algorithmic}[1]
        \REQUIRE $p$: tamanho do primeiro registrador (precisão);
            $U$: operador alvo;
            $\ket{\psi}$: (sobreposição de) autovetor(es) de $U$.
        \STATE $P \gets 2^p$, e seja
            $\ket{\psi_0} = \ket{0}^{\otimes p}\ket{\psi}$ o estado inicial
        \STATE Aplicar $\QFT_p$ no primeiro registrador
        \STATE Realizar operação de potências de $U$ controladas -- operador $\pctrl{U}$
        \STATE Aplicar $\QFT_p^{-1}$ no primeiro registrador
        \STATE Seja $\ket{\omega'}$ o resultado da medição do
            primeiro registrador na base computacional
        \RETURN $\omega'/P$
    \end{algorithmic}
\end{algorithm}

Analisando o algoritmo passo a passo, observa-se que
a aplicação de $\QFT_P$ no primeiro registrador resulta na sobreposição uniforme
de $\ket{0}, \ldots, \ket{P-1}$.
Isso ocorre porque a entrada do primeiro registrador é fixada em $\ket{0}^{\otimes p}$.
Sendo assim,
\begin{align}
    \ket{\psi_1} &= \pr{\QFT_P \otimes I} \ket{0}\ket{\psi}
    \\
    &= \frac{1}{\sqrt{P}} \sum_{j=0}^{P-1}\ket{j} \ket{\psi}.
\end{align}
Para simplificar a análise dos passos subsequentes,
suponha que $\ket{\psi}$ é o $e^{2\pi\ii \lambda}$-autovetor $\ket{\lambda}$ de $U$.
Após aplicar o operador de potências de $U$ controladas obtém-se o estado
\begin{align}
    \ket{\psi_2} &= \pctrl{U} \ket{\psi_1}
    \\
    &= \frac{1}{\sqrt P} \sum_{j=0}^{P-1}\ket{j} \otimes U^j \ket{\lambda}
    \\
    &= \frac{1}{\sqrt P} \sum_{j=0}^{P-1} e^{j \cdot 2\pi\ii\lambda} \ket j \ket\lambda
    \\
    &= \frac{1}{\sqrt P} \sum_{j=0}^{P-1}
        \exp\pr{2\pi\ii j (P\lambda)/P} \ket{j} \ket{\lambda}
    \\
    &= \ket{\F P {P\lambda}} \ket{\lambda}.
\end{align}

Agora, ignorando o segundo registrador,
deseja-se aplicar a $\QFT_P^{-1}$ e realizar uma medição na base computacional
para obter o resultado $\ket{P\lambda}$.
Entretanto, isso não necessariamente é verdade.
Recobre que $\lambda \in \mathbb{R}$ tal que $0 \leq \lambda < 1$.
Logo, o valor $\lambda$ pode ser representado como um número binário
$0.\lambda^b_1 \lambda^b_2 \lambda^b_3 \ldots$ não necessariamente finito.
Então, analisa-se dois cenários possíveis.

No primeiro caso, suponha que
$\lambda$ possa ser representado \emph{exatamente} com $p$ dígitos binários -- \textit{i.e.}
$\lambda = \lambda^b_1 \lambda^b_2 \ldots \lambda^b_{p-1} / 2^{p}
    = 0.\lambda^b_1 \lambda^b_2 \ldots \lambda^b_{p-1}$ --
conclui-se que $\ket{\F P {P\lambda}} \in B_{\mathcal{F}}$.
Logo, usando o Teorema \ref{teo:fourier-error},
\begin{align}
    \ket{\omega'} = \ket{P\lambda}
\end{align}
com probabilidade 1.
Caso $\lambda$ \emph{não} possa ser representado com $p$ dígitos binários,
$P\lambda \notin \mathbb{N}$;
mas deseja-se obter um resultado tão próximo quanto possível desse valor,
\textit{i.e.} ou $\floor{P\lambda}$ ou $\ceil{P\lambda}$.
Esse cenário já foi analisado no Teorema \ref{teo:fourier-error}.
Portanto,
\begin{align}
    \ket{\omega'}
    \in \set{ \ket{\floor{P\lambda}}, \ket{\ceil{P\lambda}} }
\end{align}
com probabilidade maior ou igual a $8/\pi^2$.

Resta ainda analisar o caso em que $\ket\psi$ é uma sobreposição
$\ket\psi = \sum_l \alpha_l \ket{\lambda_l}$ de autovetores de $U$.
Nesse caso,
\begin{align}
    \ket{\psi_2} &= \frac{1}{\sqrt P} \sum_l \alpha_l
        \sum_{j=0}^{P-1} \ket{j} \otimes U^j \ket{\lambda_l}
    \\
    &= \frac{1}{\sqrt P} \sum_l \alpha_l
        \sum_{j=0}^{P-1} \exp\pr{2\pi\ii j (P\lambda_l)/P} \ket{j} \otimes \ket{\lambda_l}
    \\
    &= \sum_l \alpha_l \ket{\F P {P\lambda_l}} \otimes \ket{\lambda_l}.
\end{align}

Deseja-se aplicar $(\QFT_P^{-1} \otimes I)$ e
fazer uma medição na base computacional no primeiro registrador.
Por enquanto, o segundo registrador será mantido.
Seja
\begin{align}
    \ket{\hat\omega} = \QFT_P^{-1} \sum_{l=1}^L \alpha_l \ket{\F P {P\lambda_l}}
        \otimes \ket{\lambda_l}.
\end{align}
Utilizando o conjunto de operadores de medida $\set{M_m}$ onde
\begin{align}
    M_m = \ket{m}\bra{m} \otimes I
\end{align}
com $m \in \mathbb{N}$ tal que $0 \leq m < P - 1$;
então, a probabilidade de medir $m$ é
\begin{align}
    \bra{\hat\omega} M_{m}^\dagger M_{m} \ket{\hat\omega} =& 
        \pr{\sum_l \alpha_l^* \bra{\F P {P\lambda_l}} \bra{\lambda_l}}
        \QFT_P \ket{m} \otimes I \notag \\
        &\bra{m} \QFT_P^\dagger \otimes I
        \pr{\sum_\ell \alpha_{\ell} \ket{\F P {P\lambda_\ell}} \ket{\lambda_\ell}}
    \\
    =& \pr{\sum_l \alpha_l^* \braket{\F P {P\lambda_l} | \F P m}}
        \delta_{l,\ell} \notag \\
        &\pr{\sum_\ell \alpha_\ell \braket{\F P m | \F P {P\lambda_\ell}}}
    \\
    =& \sum_l \card{\alpha_l}^2\ \card{\braket{\F P {P\lambda_l} | \F P m} }^2.
\end{align}
Note a semelhança entre essa equação e o que foi obtido no início da
demonstração do Teorema \ref{teo:fourier-error}.
Portanto, usando esse mesmo teorema, conclui-se que se
$P\lambda_l \in \mathbb{N}$, o estado após a medida será $\ket{P\lambda_l}$ com
probabilidade maior ou igual a $\card{\alpha_l}^2$ e
caso $P\lambda_l \notin \mathbb{N}$, o estado após a medida será
$\ket{\floor{P\lambda_l}}$ ou $\ket{\ceil{P\lambda_l}}$ com probabilidade
maior ou igual a $\card{\alpha_l}^2 \cdot 8 / \pi^2$.

\section{Algoritmo de Contagem} \label{sec:alg-cont}

O algoritmo de contagem elegantemente une todas as seções desse capítulo.
Relembre que o objetivo do algoritmo de busca (Alg. \ref{alg:busca})
é encontrar um elemento marcado pelo oráculo $O_f$.
Toda a análise do algoritmo foi feita em torno do ângulo $\theta$
definido por $\sin\theta = \sqrt{k/N}$,
onde $k$ é a quantidade de elementos marcados por $O_f$
e $N$ a dimensão do espaço de Hilbert.
Como a matriz $U_G$ faz uma rotação de $2\theta$
no hiperplano definido por $\ket{x_0}$ e $\ket{x_1}$,
é possível analisar o algoritmo em torno dos autovetores
\begin{align}
    \ket{\mp\ii_x} = \frac{\ket{x_0} \mp\ii \ket{x_1}}{\sqrt 2}
\end{align}
associados aos autovalores $e^{\pm 2\ii\theta}$, respectivamente.
A condição inicial do algoritmo de busca pode ser representada em termos desses
autovetores:
\begin{align}
    \ket\psi =& \cos\theta \ket{x_0} + \sin\theta \ket{x_1}
    \\
    =& \braket{-\ii_x | \psi} \ket{-\ii_x} + \braket{+\ii_x | \psi} \ket{+\ii_x}
    \\
    =& \frac{e^{\ii\theta}}{\sqrt 2}\ket{-\ii_x} + \frac{e^{-\ii\theta}}{\sqrt 2}\ket{+\ii_x}.
\end{align}

Agora, considere o problema de contar quantos elementos são marcados por $O_f$.
Observe que há uma relação direta entre $\theta$ e a quantidade de elementos marcados $k$.
Portanto, uma boa estimativa de $\theta$ deve resultar numa boa estimativa de $k$;
o que pode ser feito utilizando o algoritmo de estimativa de fase!
O algoritmo de contagem está detalhado em Alg. \ref{alg:contagem}.

\begin{algorithm}
    \caption{Algoritmo de Contagem}
    \label{alg:contagem}
    \begin{algorithmic}[1]
        \REQUIRE
            $O_f$: Oráculo da função $f$;
            $p$: número de qubits do primeiro registrador do algoritmo de estimativa de fase;
            $n$: número de qubits do segundo registrador respeitando o domínio de $f$.
        \STATE Construir operador $U_G = GO_f$
        \STATE Preparar o estado $\ket\psi = H^{\otimes n}\ket{0}$
        \STATE $\vartheta \gets$ est\_fase($p, U_G, \ket{\psi}$)
        \STATE $\theta' \gets \vartheta\pi$
        \STATE Se $\theta' > \pi/2$ então $\theta' \gets \theta' - \pi/2$
            \label{step:obtem-theta-posit}
        \RETURN $k = \sin^2\pr{\theta'} \cdot N$
    \end{algorithmic}
\end{algorithm}

Executar o algoritmo de estimativa de fase com $U_G$ como operador alvo resulta
numa estimativa equiprovável de um dos autovalores $e^{\pm 2\ii\theta}$ --
na realidade resulta num valor $\vartheta$ que é uma estimativa de $\pm \theta/\pi$.
Obtém-se o ângulo estimado $\theta'$ ao computar $\theta' = \vartheta\pi$.
Agora é necessário distinguir se $\theta'$ corresponde a
uma estimativa de $+\theta$ ou de $-\theta$.
Notando que
\begin{align}
    \sin\theta = \sqrt{\frac{k}{N}},\ \cos\theta = \sqrt{\frac{N - k}{N}}
    \implies 0 \leq \theta \leq \frac \pi 2,
\end{align}
obtém-se a estimativa de $+\theta$ facilmente.
Como $\sin\theta = \sqrt{k/N}$, calcular $\sin^2\pr{\theta'} \cdot N$ resulta
numa estimativa da quantidade de elementos marcados.
Resta analisar o quão precisa é essa estimativa.

\subsection{Precisão do Algoritmo de Contagem}
\label{sec:alg-cont-precisao}
O foco da análise da precisão será no valor $\sin^2(\theta')$.
A precisão da saída do algoritmo segue como um corolário simples.
A precisão é dada pelo seguinte Teorema
adaptado do Teorema 12 de BHMT \cite{brassard2002quantum}.

\begin{theorem}
    Sejam $p$ a entrada do do algoritmo; $P = 2^p$; e
    $\theta'$ a estimativa do ângulo $\theta$.
    Então $\sin^2\theta = 0 \implies \sin^2\theta' = 0$ e
    $\sin^2\theta=1 \implies \sin^2\theta' = 1$ com certeza;
    caso contrário,
    \begin{align}
        \left|\sin^2\theta' - \sin^2\theta\right| \leq
            2\pi \frac{\sin\theta\cos\theta}{P} + \frac{\pi^2}{P}
    \end{align}
    com probabilidade maior ou igual a $8/\pi^2$.
    Além disso, o algoritmo sempre realiza $P - 1$ consultas ao oráculo.
    \label{teo:est-ampl-error}
\end{theorem}
\begin{proof}
    Caso $\sin^2\theta = 0$ ou $\sin^2\theta = 1$,
    temos $\theta = 0$ ou $\theta = \pi/2$, respectivamente.
    Logo, $\vartheta = 0$ e $\vartheta = 1/2$ precisam ser saídas exatas do
    algoritmo de estimativa de fase.
    Isso acontece se for possível obter exatamente os estados
    $\ket{\F P 0}$ e $\ket{\F P {P/2}}$ após a aplicação da porta $\pctrl{U_G}$;
    e já que $P$ é par, isso de fato acontece.
    
    Caso contrário, suponha que $\eps \in \mathbb{R}$ tal que
    $\eps \geq 0$ e $|\theta - \theta'| \leq \eps$.
    Se $\theta' \geq \theta$,
    \begin{align}
        \sin^2(\theta + \eps) - \sin^2\theta =&
            \pr{\sin\theta\cos\eps + \cos\theta\sin\eps}^2 - \sin^2\theta
        \\
        =& \sin^2\theta\cos^2\eps - \sin^2\theta +
            \cos^2\theta\sin^2\eps +
            2 \sin\eps\cos\eps\sin\theta\cos\theta
        \\
        =& -\sin^2\theta\sin^2\eps + (1 - \sin^2\theta)\sin^2\eps +
            \sin\pr{2\eps}\sin\theta\cos\theta
        \\
        \leq&\ 2\eps\sin\theta\cos\theta + \eps^2.
    \end{align}
    Se $\theta' \leq \theta$,
    \begin{align}
        \sin^2\theta - \sin^2\pr{\theta - \eps} =&
            \sin^2\theta - \sin^2\theta\cos^2\eps - \sin^2\eps\cos^2\theta +
            2\sin\eps\cos\eps\sin\theta\cos\theta
        \\
        =& \sin^2\theta\sin^2\eps - \sin^2\eps\cos^2\theta + 
            2\eps\sin\theta\cos\theta
        \\
        \leq&\ 2\eps\sin\theta\cos\theta + \eps^2 .
    \end{align}
    Logo,
    \begin{align}
        |\theta - \theta'| \leq \eps \implies |\sin^2\theta - \sin^2\theta'| \leq
            \sin\pr{2\eps}\sin\theta\cos\theta + \eps^2.
    \end{align}
    O valor de $\eps$ depende do maior erro desejável no
    algoritmo de estimativa de fase, que é de $1/P$
    (ou seja, idealmente apenas o dígito menos significativo está errado).
    Usando o Teorema \ref{teo:fourier-error},
    esse erro é obtido com probabilidade maior ou igual a $8/\pi^2$.
    E como a saída do algoritmo é multiplicada por $\pi$, deseja-se que
    $\eps \leq \pi/P$.
    Logo,
    \begin{align}
        |\sin^2\theta - \sin^2\theta'| \leq
            2\pi \frac{\sin\theta\cos\theta}{P} + \frac{\pi^2}{P^2}.
    \end{align}
    
    Por último,
    o algoritmo sempre realiza $P - 1$ consultas ao oráculo por conta da ação do operador
    $\pctrl{U_G}$ no algoritmo de estimativa de fase.
\end{proof}

Usando esse Teorema, é possível obter uma estimativa do erro do algoritmo de contagem;
explicitada no seguinte corolário.

\begin{corollary}
    Sejam $p$ a entrada do algoritmo; $P = 2^p$;
    $\theta'$ a estimativa do ângulo $\theta$;
    $k$ a quantidade de elementos marcados pelo oráculo $O_f$ e
    $k'$ sua estimativa dada pelo algoritmo de contagem.
    Então $k = 0 \implies k' = 0$ e $k=N \implies k' = N$ com certeza;
    caso contrário,
    \begin{align}
        \left| k' - k \right| &\leq
            2\pi \frac{ \sqrt{k\pr{N-k}} }{P} + \frac{\pi^2 N}{P^2} .
        \label{eq:corolario}
    \end{align}
    com probabilidade maior ou igual a $8/\pi^2$.
    Além disso, o algoritmo sempre realiza $P - 1$ consultas ao oráculo.
    \label{cor:cont-erro}
\end{corollary}
\begin{proof}
    Maior parte das afirmações seguem trivialmente do Teorema \ref{teo:est-ampl-error}.
    Mostra-se apenas como obter a eq. \ref{eq:corolario}.
    Substituindo os valores de $\sin\theta$ e $\cos\theta$ e multiplicando ambos os lado por $N$:
    \begin{align}
        N \left|\sin^2\theta' - \sin^2\theta\right| &\leq
            2\pi N \frac{\sqrt{k/N}\sqrt{(N-k)/N}}{P} + N \frac{\pi^2}{P^2}
        \\
        \left| N\sin^2\theta' - N\sin^2\theta\right| &\leq
            2\pi \frac{ \sqrt{k\pr{N-k}} }{P} + \frac{\pi^2 N}{P^2} .
    \end{align}
\end{proof}

Analisa-se agora o significado do Corolário \ref{cor:cont-erro}.
Evidente que a precisão do algoritmo depende do tamanho do primeiro registrador
do algoritmo de estimativa de fase.
Sabe-se que o algoritmo de busca tem uma boa precisão com $O\pr{\sqrt N}$ iterações para $k = O(1)$,
e resultados similares são desejados para o algoritmo de contagem.
Suponha que $P = \sqrt{N}$ e $k = 1$, então, pelo Corolário \ref{cor:cont-erro}
o erro é menor ou igual a
\begin{align}
    2\pi \frac{\sqrt{1(N-1)}}{\sqrt{N}} + \frac{\pi^2N}{N}
    \approx 2\pi + \pi^2 = O(1).
\end{align}
Já se $k \approx N$ -- \textit{e.g.} $k = N - 1$, obtém um erro da mesma ordem:
\begin{align}
    2\pi \frac{\sqrt{(N-1)1}}{\sqrt{N}} + \frac{\pi^2N}{N}
    \approx 2\pi + \pi^2 = O(1).
\end{align}
Porém, ao tomar $k \approx N/2$, a ordem do erro aumenta significativamente:
\begin{align}
    2\pi \frac{\sqrt{(N/2)(N - N/2)}}{\sqrt{N}} + \frac{\pi^2N}{N}
    = \pi\sqrt{N} + \pi^2 = O(\sqrt{N}).
\end{align}

Esse comportamento é um pouco estranho para um algoritmo.
Poucos elementos marcados ($k \approx 1$) e muitos elementos marcados ($k \approx N$)
podem ser contados com precisão,
mas quando essa quantia se aproxima da metade $(k \approx N/2)$ a precisão diminui
significativamente.
Qual a razão desse comportamento?
Primeiro leve em conta a simetria do problema.
Se contar $k \approx 1$ pode ser feito com uma boa precisão,
contar $k \approx N - 1$ também pode ser feito com boa precisão
já que é equivalente a contar quantos elementos \emph{não} estão marcados --
\textit{i.e.} contar $N - k \approx 1$.
Mas ainda resta entender a razão para essa precisão diminuir quando $k \approx N/2$.
A causa está no comportamento da função $\arcsin$.

\begin{figure}[hbt]
    \centering
    \begin{minipage}{0.48\textwidth}
        \centering
        \caption{Gráfico da relação linear entre $\sin^2\theta$ e $k$.}
        \includegraphics[width=\textwidth]{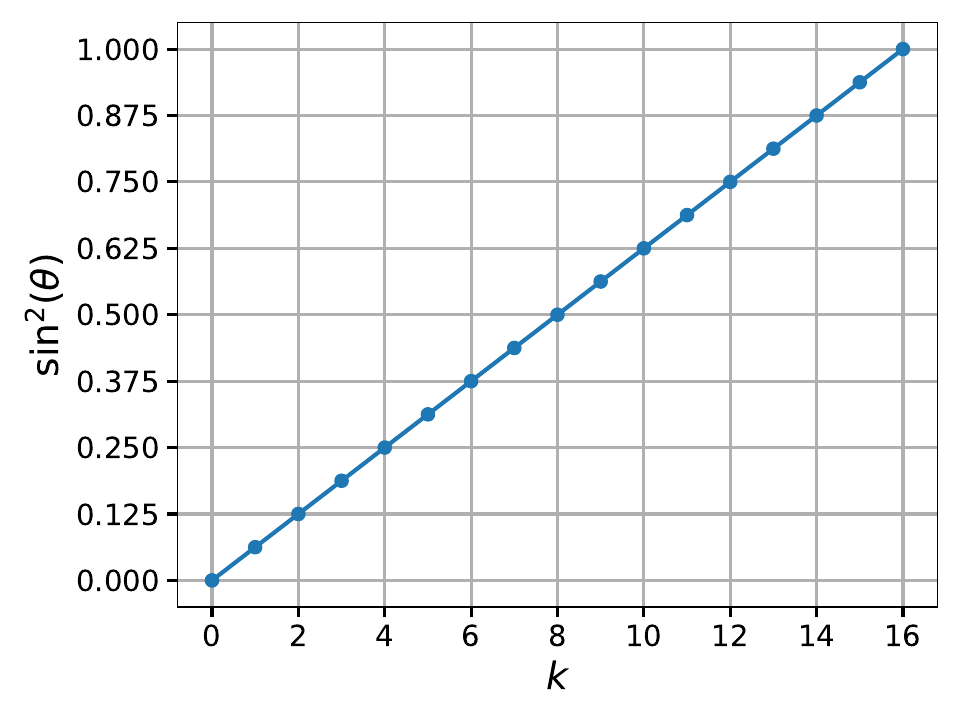}
        \legend{Fonte: Produzido pelo autor.}
        \label{fig:cont-sin-linear-k}
    \end{minipage}
    \hfill
    \begin{minipage}{0.48\textwidth}
        \centering
        \caption{Gráfico da relação não linear entre $\sin^2\theta$ e $\theta$.}
        \includegraphics[width=\textwidth]{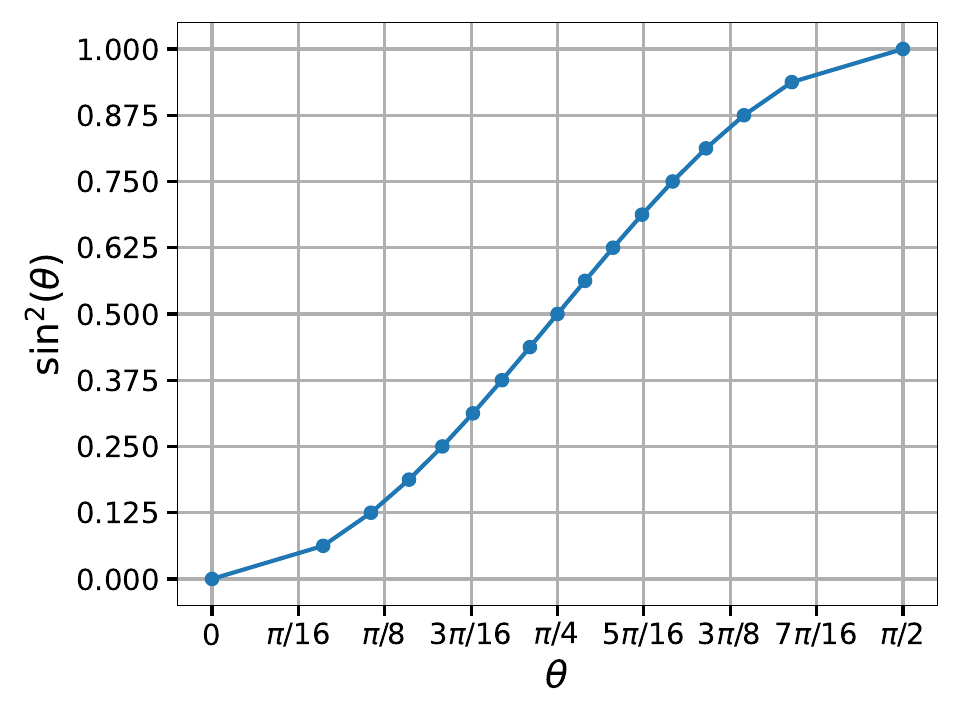}
        \legend{Fonte: Produzido pelo autor.}
        \label{fig:cont-sin-non-linear-theta}
    \end{minipage}
\end{figure}

\begin{figure}[hbt]
    \centering
    \begin{minipage}{0.48\textwidth}
        \centering
        \caption{Gráfico da relação não linear entre $k$ e $\theta = \arcsin\pr{\sqrt{k/N}}$.}
        \includegraphics[width=\textwidth]{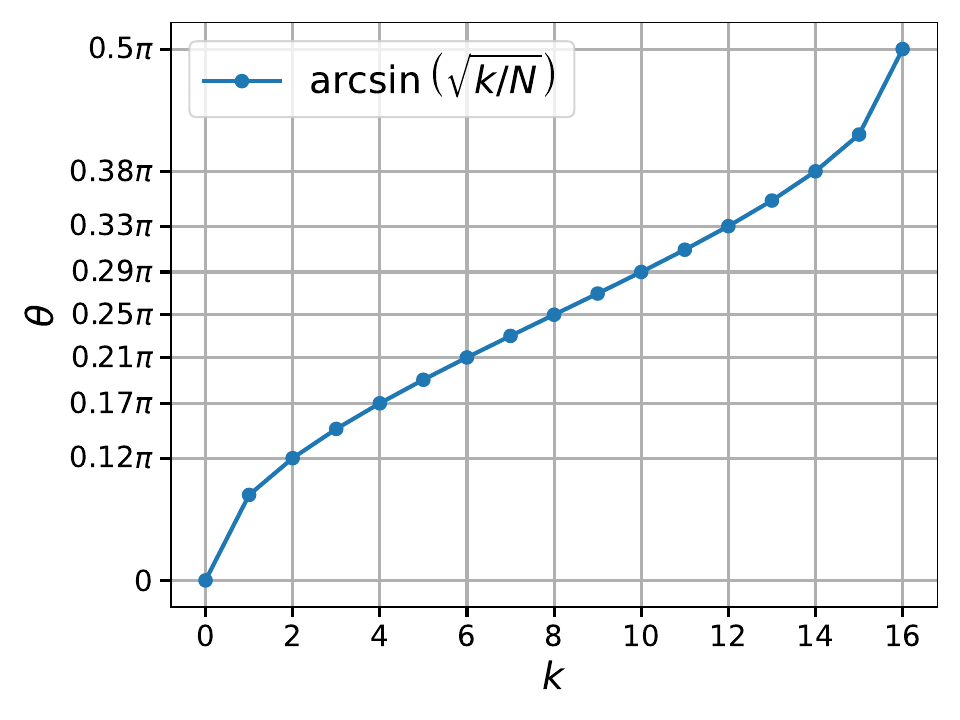}
        \legend{Fonte: Produzido pelo autor.}
        \label{fig:cont-arcsin}
    \end{minipage}
    \hfill
    \begin{minipage}{0.48\textwidth}
        \centering
        \caption{Gráfico das diferenças $\Delta\theta(k)$ com $N=16$.}
        \includegraphics[width=\textwidth]{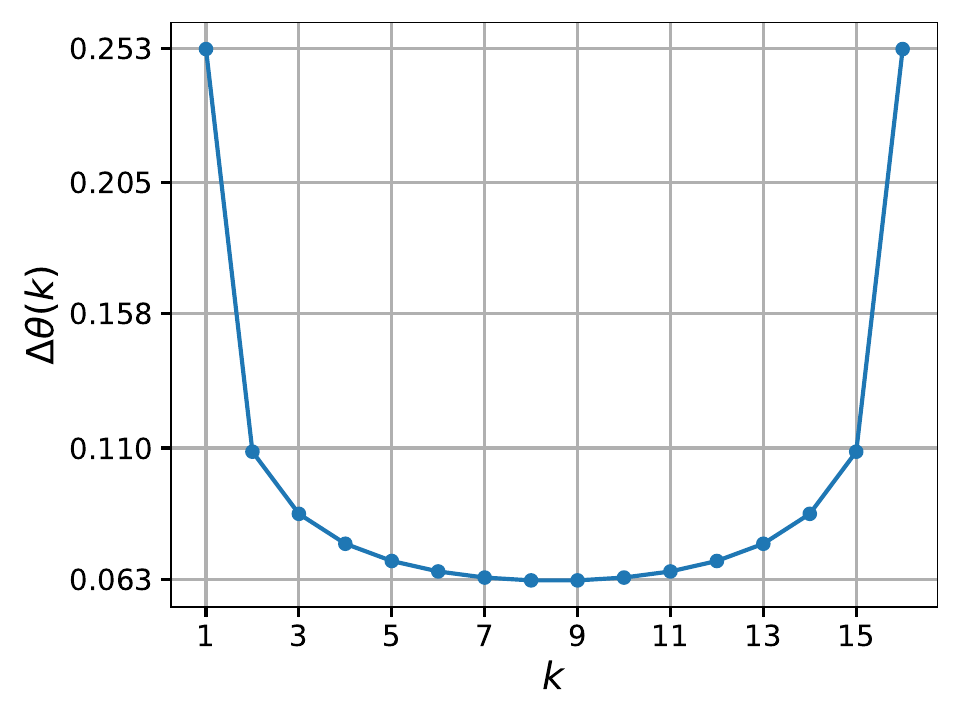}
        \legend{Fonte: Produzido pelo autor.}
        \label{fig:cont-diff-arcsin}
    \end{minipage}
\end{figure}

Note que $\sin^2\theta = k/N$ é linear em função de $k$
(Fig. \ref{fig:cont-sin-linear-k}),
mas não em função de $\theta$
(Fig. \ref{fig:cont-sin-non-linear-theta}).
Isso fica mais evidente ao compararmos $k$ com $\theta = \arcsin\sqrt{k/N}$
(Fig \ref{fig:cont-arcsin});
ou compararmos $k$ com a diferença de um ângulo com seu anterior:
$\Delta\theta(k) = \arcsin\sqrt{k / N} - \arcsin\sqrt{(k - 1)/N}$;
ilustrado na Fig. \ref{fig:cont-diff-arcsin}.
Para gerar os gráficos das
Figs. \ref{fig:cont-sin-linear-k} a  \ref{fig:cont-diff-arcsin},
utilizou-se $N = 16$.

\begin{figure}[htb]
    \centering
    \caption{Possíveis valores de $2\theta$ e $e^{\ii2\theta}$ com $N=16$.}
    \includegraphics[width=\textwidth]{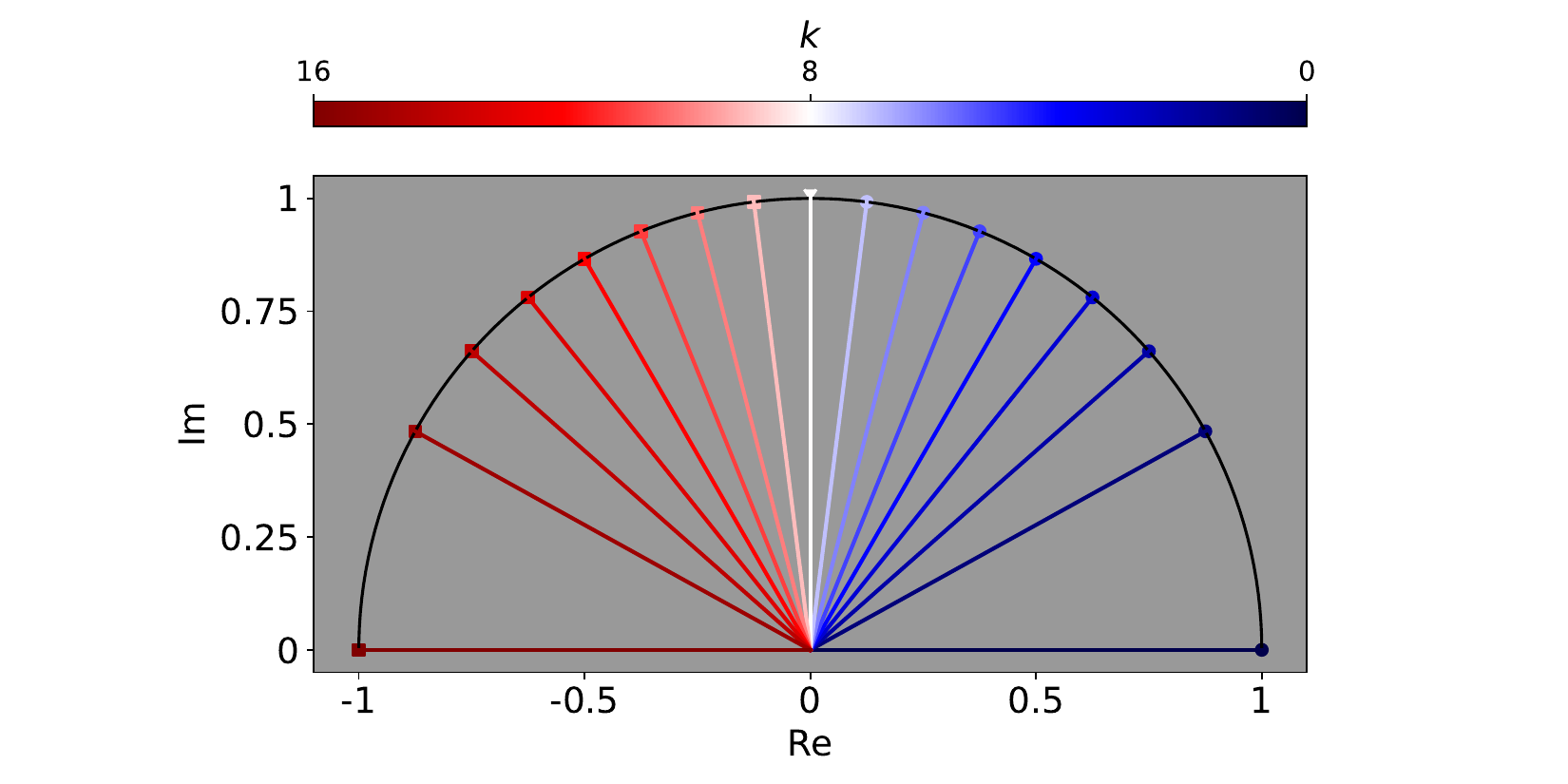}
    \legend{Fonte: Produzido pelo autor.}
    \label{fig:cont-eval}
\end{figure}

Analisa-se agora como essa relação entre $k$ e $\theta$ influencia o
comportamento do autovalor $e^{\ii 2\theta}$.
Fig. \ref{fig:cont-eval} ilustra como os ângulos $2\theta$ variam no círculo unitário
à medida que incrementa-se o valor de $k$.
Na Fig. \ref{fig:cont-eval}, usou-se $N = 16$.
Os ângulos correspondentes à $k < N/2$ são representados por cores azuladas
e o valor $e^{\ii2\theta}$ marcado com um círculo.
Os ângulos correspondentes à $k > N/2$ são representados por cores avermelhadas
e o valor $e^{\ii2\theta}$ marcado com um quadrado.
O ângulo correspondente à $k = N/2$ é representado pela cor branca
e o valor $e^{\ii2\theta}$ marcado por um triângulo.
Os ângulos $-2\theta$ preencheriam a parte inferior do círculo unitário.
Na Fig. \ref{fig:cont-eval},
é possível perceber como os ângulos correspondentes a $k \approx N/2$
são mais próximos entre si do que os ângulos $k \approx 1$ e $k \approx N$
(vide Fig. \ref{fig:cont-diff-arcsin});
o que ajuda a explicar o comportamento do Corolário \ref{cor:cont-erro}.

Um outro fator que influencia no erro do algoritmo de contagem
é a Transformada de Fourier.
Lembre-se que a Transformada de Fourier divide o círculo unitário em $P$ particões
igualmente espaçadas (os ângulos base).
E como os ângulos representados na Fig. \ref{fig:cont-eval}
não são igualmente espaçados,
um valor de $P > N$ é necessário para distinguir exatamente entre todos os ângulos.

\begin{figure}[htb]
    \centering
    \caption{Possíveis valores de $2\theta$ e ângulos base de Fourier com $N=P=16$.}
    \includegraphics[width=\textwidth]{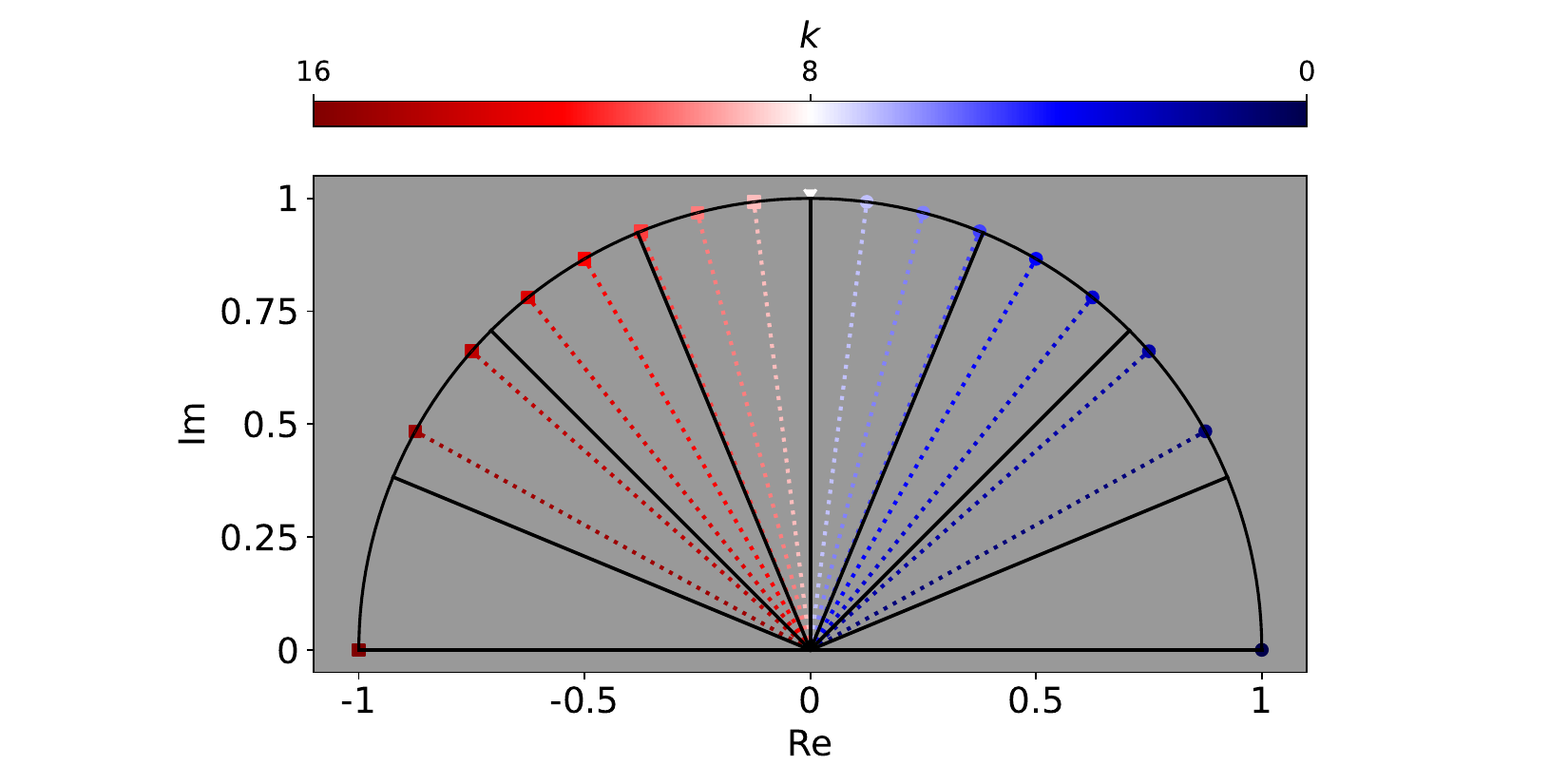}
    \legend{Fonte: Produzido pelo autor.}
    \label{fig:cont-fourier-eval}
\end{figure}

A Fig \ref{fig:cont-fourier-eval} ilustra os ângulos da
Fig. \ref{fig:cont-eval} em linhas pontilhadas.
Os ângulos base de Fourier com $P = 16$ estão representados pela cor preta
em linhas sólidas, sobrepondo os ângulos $2\theta$.

É possível observar que mesmo um valor $P = O(N)$
engloba múltiplos valores possíveis para $\theta$ numa única partição.
Ainda assim, utilizar $P = O\pr{\sqrt{N}}$ permite
distinguir valores $k \approx 1$ ou $k \approx N$
já que $\Delta\theta(k)$ é maior nesses casos (Fig. \ref{fig:cont-diff-arcsin}).
\chapter{Algoritmo de Contagem no Grafo Bipartido Completo} \label{cap:alg-cont-grafo-bip-compl}
Nesse capítulo, entra-se em detalhes da aplicação do algoritmo de contagem
no passeio quântico de um grafo bipartido completo.
A Seção \ref{sec:passeio-grafo-regular} descreve passeios quânticos em grafos regulares;
Seção \ref{sec:passeio-busca} entra em detalhes sobre o operador de evolução do passeio de busca
no grafo bipartido completo; e
a Seção \ref{sec:passeio-contagem} analisa os autovalores e autovetores de interesse, e
seu impacto no algoritmo de contagem.

\section{Passeios Quânticos em Grafos Regulares} \label{sec:passeio-grafo-regular}
Considere primeiramente o cenário clássico: um passeio aleatório.
Um passeio aleatório ocorre num grafo simples $\Gamma\pr{V, E}$.
O grafo descreve possíveis posições onde um caminhante pode estar --
\textit{i.e.} vértices $V$ de $\Gamma$ --,
e descreve caminhos que o caminhante pode tomar dependendo da sua posição --
\textit{i.e.} arestas $E$ de $\Gamma$.
O termo aleatório provém de que a escolha do caminho a ser tomado é feita de maneira aleatória e,
normalmente, equiprovável.

Considere o grafo apresentado na Fig. \ref{fig:passeio-aleatorio}.
Suponha que numa determinada etapa do passeio aleatório,
o caminhante está no vértice de cor preta.
Nessa situação, o caminhante pode ir para quatro direções possíveis:
$d_0$, $d_1$, $d_2$ ou $d_3$.
Então, joga-se uma \emph{moeda} (4-dimensional, no caso) e
o caminhante segue aquele caminho,
\textit{i.e.} sua próxima posição será o vértice adjacente ao atual.
Por exemplo, se, na situação da Fig. \ref{fig:passeio-aleatorio},
o resultado após jogar a moeda for $d_0$, o caminhante move-se ``para cima'' --
sua próxima posição está ilustrada pelo vértice preto na
Fig. \ref{fig:passeio-aleatorio-pos-moeda}.

\begin{figure}[hbt]
    \centering
    \begin{minipage}{0.48\textwidth}
        \centering
        \caption{Etapa de um passeio aleatório.}
        \begin{tikzpicture}
    \pgfmathsetmacro\mult{1.5}
    \pgfmathsetmacro\N{4}
    \pgfmathsetmacro{\pos}{9}
    \pgfmathsetmacro{\cima}{\pos+\N}
    \pgfmathsetmacro{\baixo}{\pos-\N}
    \pgfmathsetmacro{\dir}{\pos+1}
    \pgfmathsetmacro{\esq}{\pos-1}
    
    \foreach \xx in {1,...,\N}{
        \foreach \yy in {1,...,\N} {
            \pgfmathsetmacro{\x}{int(\xx - 1)}
            \pgfmathsetmacro{\y}{int(\yy - 1)}
            \pgfmathsetmacro{\res}{int(\y*\N + \x)}
            \ifnum \res=\pos
                \node[circle,draw,fill] (\res) at (\x*\mult, \y*\mult) {};
            \else
                \node[circle,draw] (\res) at (\x*\mult, \y*\mult) {};
            \fi
        }
    }
    \foreach \xx in {1,...,\N}{
        \foreach \yy in {1,...,\N} {
            \pgfmathsetmacro{\x}{int(\xx - 1)}
            \pgfmathsetmacro{\y}{int(\yy - 1)}
            \pgfmathsetmacro{\curr}{int(\y*\N + \x)}
            \pgfmathsetmacro{\r}{ int(\y*\N + mod(\x+1, \N)) }
            \pgfmathtruncatemacro{\compr}{\curr<\r?1:0}
            \pgfmathsetmacro{\u}{ int(mod(\y+1,\N)*\N + \x) }
            \pgfmathtruncatemacro{\compu}{\curr<\u?1:0}
            \ifnum \compr=1
                \ifnum \curr=\pos
                    \path (\curr) edge node[below] {$d_1$} (\r);
                \else \ifnum \r=\pos
                    \path (\curr) edge node[above] {$d_3$} (\r);
                \else
                    \path (\curr) edge (\r);
                \fi \fi
            \fi
            \ifnum \compu=1
                \ifnum \curr=\pos
                    \path (\curr) edge node[right] {$d_0$} (\u);
                \else \ifnum \u=\pos
                    \path (\curr) edge node[left] {$d_2$} (\u);
                \else
                    \path (\curr) edge (\u);
                \fi \fi
            \fi
        }
    }
\end{tikzpicture}
        \legend{Fonte: produzido pelo autor.}
        \label{fig:passeio-aleatorio}
    \end{minipage}
    \hfill
    \begin{minipage}{0.48\textwidth}
        \centering
        \caption{Possível etapa seguinte.}
        \begin{tikzpicture}
    \pgfmathsetmacro\mult{1.5}
    \pgfmathsetmacro\N{4}
    \pgfmathsetmacro{\pos}{13}
    \pgfmathsetmacro{\cima}{\pos+\N}
    \pgfmathsetmacro{\baixo}{\pos-\N}
    \pgfmathsetmacro{\dir}{\pos+1}
    \pgfmathsetmacro{\esq}{\pos-1}
    
    \foreach \xx in {1,...,\N}{
        \foreach \yy in {1,...,\N} {
            \pgfmathsetmacro{\x}{int(\xx - 1)}
            \pgfmathsetmacro{\y}{int(\yy - 1)}
            \pgfmathsetmacro{\res}{int(\y*\N + \x)}
            \ifnum \res=\pos
                \node[circle,draw,fill] (\res) at (\x*\mult, \y*\mult) {};
            \else
                \node[circle,draw] (\res) at (\x*\mult, \y*\mult) {};
            \fi
        }
    }
    \foreach \xx in {1,...,\N}{
        \foreach \yy in {1,...,\N} {
            \pgfmathsetmacro{\x}{int(\xx - 1)}
            \pgfmathsetmacro{\y}{int(\yy - 1)}
            \pgfmathsetmacro{\curr}{int(\y*\N + \x)}
            \pgfmathsetmacro{\r}{ int(\y*\N + mod(\x+1, \N)) }
            \pgfmathtruncatemacro{\compr}{\curr<\r?1:0}
            \pgfmathsetmacro{\u}{ int(mod(\y+1,\N)*\N + \x) }
            \pgfmathtruncatemacro{\compu}{\curr<\u?1:0}
            \ifnum \compr=1
                \path (\curr) edge (\r);
            \fi
            \ifnum \compu=1
                \path (\curr) edge (\u);
            \fi
        }
    }
\end{tikzpicture}
        \legend{Fonte: produzido pelo autor.}
        \label{fig:passeio-aleatorio-pos-moeda}
    \end{minipage}
\end{figure}

\begin{figure}[hbt]
    \centering
    \caption{Sobreposição de posições.}
    \begin{tikzpicture}
    \pgfmathsetmacro\mult{1.5}
    \pgfmathsetmacro\N{4}
    \pgfmathsetmacro{\pos}{9}
    \pgfmathsetmacro{\cima}{\pos+\N}
    \pgfmathsetmacro{\baixo}{\pos-\N}
    \pgfmathsetmacro{\dir}{\pos+1}
    \pgfmathsetmacro{\esq}{\pos-1}
    
    \foreach \xx in {1,...,\N}{
        \foreach \yy in {1,...,\N} {
            \pgfmathsetmacro{\x}{int(\xx - 1)}
            \pgfmathsetmacro{\y}{int(\yy - 1)}
            \pgfmathsetmacro{\res}{int(\y*\N + \x)}
            \ifthenelse{\res=\cima \OR \res=\baixo \OR
                \res=\dir \OR \res=\esq}{
                \node[circle,draw,fill=lightgray] (\res) at (\x*\mult, \y*\mult) {};
            }{
                \node[circle,draw] (\res) at (\x*\mult, \y*\mult) {};
            }
        }
    }
    \foreach \xx in {1,...,\N}{
        \foreach \yy in {1,...,\N} {
            \pgfmathsetmacro{\x}{int(\xx - 1)}
            \pgfmathsetmacro{\y}{int(\yy - 1)}
            \pgfmathsetmacro{\curr}{int(\y*\N + \x)}
            \pgfmathsetmacro{\r}{ int(\y*\N + mod(\x+1, \N)) }
            \pgfmathtruncatemacro{\compr}{\curr<\r?1:0}
            \pgfmathsetmacro{\u}{ int(mod(\y+1,\N)*\N + \x) }
            \pgfmathtruncatemacro{\compu}{\curr<\u?1:0}
            \ifnum \compr=1
                \path (\curr) edge (\r);
            \fi
            \ifnum \compu=1
                \path (\curr) edge (\u);
            \fi
        }
    }
\end{tikzpicture}
    \legend{Fonte: produzido pelo autor}
    \label{fig:superposicao-posicoes}
\end{figure}
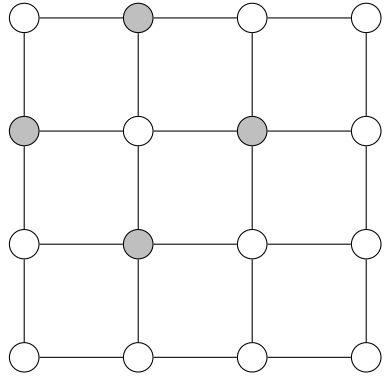

Como esperado de um algoritmo clássico, em cada etapa,
o caminhante ocupa somente uma posição.
Em contrapartida, num passeio quântico, deseja-se tirar proveito
das propriedades da Mecânica Quântica;
fazendo com que o caminhante esteja numa sobreposição de posições.
Esse cenário é ilustrado na Fig. \ref{fig:superposicao-posicoes}.
Os vértices em cinza indicam uma possível sobreposição das posições do caminhante.
Com essa intuição de como o passeio quântico funciona,
formaliza-se o passeio quântico em grafos regulares.

Seja $\Gamma(V, E)$ um grafo $d$-regular com número par de vértices
e com índice cromático $d$ e $|V| = N$;
então, é possível associar cada uma das $d$ cores com uma direção
(se o grafo tivesse um número ímpar de vértices, o índice cromático seria $d + 1$ e
não seria possível associar cada cor a uma direção de modo geral).
Logo, descreve-se passeio quântico no espaço de Hilbert $\hilb^N \otimes \hilb^d$,
onde $\hilb^N$ é denomidado ``espaço das posições'' e
$\hilb^d$ é denominado ``espaço da moeda''.
Associa-se, bijetivamente, os vértices de $\Gamma$ com os $N$ vetores da base computacional.
O \emph{operador de evolução do passeio} $U_w$ descreve a progressão do passeio é dado por
\begin{align}
    U_w = S \pr{I \otimes C}, \label{eq:operador-evol-passio}
\end{align}
onde $S$ é o operador de \emph{flip-flop shift} $Nd$-dimensional e
$C$ o operador da moeda $d$-dimensional.
O operador $S$ é responsável por fazer o caminhante andar,
mudando sua posição de acordo com o caminho apontado pela moeda.
Note que cada aplicação de $U$ simula o comportamento descrito anteriormente:
primeiro joga-se a moeda ($C$) para escolher a direção e
posteriormente o caminhante se move naquela direção ($S$).
Para definir o operador $S$ formalmente,
considere $v, u \in V$ e $e_c = vu \in E$ uma aresta cujo rótulo da coloração é $0 \leq c < d$,
\textit{i.e.} $\cor{vu} = c$;
a ação de $e_c$ em $v$ resulta num vértice adjacente a $v$ e incidente a $e_c$,
\textit{i.e.} $e_c(v) = u$ se e somente se $e_c = vu$.
Sendo assim, o operador $S$ é dado por
\begin{align}
    S \ket{v, c} = \ket{e_c(v), c} = \ket{u, c},
\end{align}
onde $\ket{v,c} = \ket{v}\otimes\ket{c}$ com
$\ket{v} \in \hilb^N$ e $\ket{c} \in \hilb^d$.
Esse operador é unitário já que $S = S^\dagger$,
\begin{align}
    S^2\ket{v, c} = S\ket{u, c} = \ket{e_c(u), c} = \ket{v, c}.
\end{align}

\subsection{Passeio com Moeda de Grover}
Para esse passeio, será utilizado a moeda de Grover,
\begin{align}
    C = \frac{2}{d} \sum_{c,c' = 0}^{d-1} \ket{c}\bra{c'} - I.
\end{align}

Nesse caso, a ação do operador de evolução é
\begin{align}
    U_w \ket{v, c} &= S \pr{\ket{v} \otimes
       \pr{\frac{2}{d}  \sum_{a, b=0}^{d-1} \ket{a} \bra{b} - I} \ket{c}
    }
    \\
    &= S \pr{ \ket{v} \otimes \pr{
        \frac{2}{d} \sum_{a,b = 0}^{d-1} \ket{a} \delta_{b,c} - \ket{c}
    } }
    \\
    &= \pr{\frac{2}{d}-1} S \ket{v, c} +
        \frac{2}{d}\ S \sum_{a \neq c} \ket{v, a}
    \\
    &= \pr{\frac{2}{d}-1} \ket{e_c(v), c} +
        \frac{2}{d} \sum_{a \neq c} \ket{e_a(v), a}
    \\
    &= \pr{\frac{2}{d}-1} \ket{u, \cor{vu}} +
        \frac{2}{d} \sum_{u' \neq u} \ket{u', \cor{vu'}}
    \label{eq:passeio-grover}
\end{align}
Essa Equação será útil ao analisar o passeio quântico de busca.
\section{Passeio de Busca no Grafo Bipartido Completo}
\label{sec:passeio-busca}
Considere um grafo bipartido completo $\Gamma\pr{V, E}$
de tal modo que $V = V_1 \cup V_2$, $V_1 \cap V_2 = \emptyset$,
$\card V = N$, $|V_1| = N_1$ e $|V_2| = N_2$.
Sejam $K_1$ e $K_2$ os conjuntos de vértices marcados tais que
$K_1 \subseteq V_1$, $K_2 \subseteq V_2$, $K = K_1 \cup K_2$, $|K_1| = k_1$,
$|K_2| = k_2$, e $|K| = k$.

\begin{figure}[htb]
    \caption{Exemplo de grafo bipartido completo.}
    \begin{center}
        \begin{tikzpicture}
    \def\N1{5}
    \pgfmathparse{\N1-1}
    \foreach \y in {0, 1}
        \foreach \x in {0,...,\pgfmathresult}{
            \pgfmathsetmacro\z{ int(\N1*\y+\x) }
            \ifthenelse{\z=0 \OR \z=1 \OR \z=\N1 \OR \z=\intcalcNum{\N1+1}}{
                \node [circle,draw,fill=black] (\z) at (\x,-2*\y) {};
            }{
                \node [circle,draw] (\z) at (\x,-2*\y) {};
            }
        }
        
    \pgfmathsetmacro\vend{ int(\N1-1) }
    \pgfmathsetmacro\uend{ int(2*\N1-1) }
    \foreach \v in {0,...,\vend}
        \foreach \u in {\N1,...,\uend}{
            \path (\v) edge (\u);
        }

    \draw [decorate, decoration={brace, amplitude=5pt}] (4.25, 0.3) -- (4.25, -0.3)
        node[midway, xshift=1.em]{$V_1$};
    \draw [decorate, decoration={brace, amplitude=5pt}] (4.25, -1.7) -- (4.25, -2.3)
        node[midway, xshift=1.em]{$V_2$};
        
    \draw [decorate, decoration={brace, amplitude=5pt}] (-0.25, 0.25) -- (1.25, 0.25)
        node[midway, yshift=1.2em]{$K_1$};
    \draw [decorate, decoration={brace, amplitude=5pt, mirror}] (-0.25, -2.25) -- (1.25, -2.25)
        node[midway, yshift=-1.2em]{$K_2$};
\end{tikzpicture}
    \end{center}
    \legend{Fonte: produzido pelo autor.}
    \label{fig:grafo-bipartido-marcado}
\end{figure}
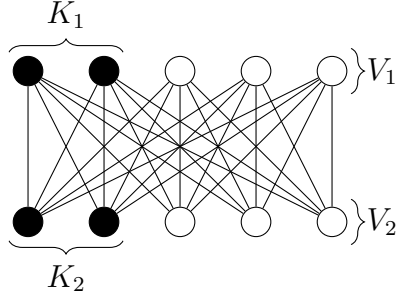

Este trabalho, entretanto, restringe-se ao subcaso em que $k_1 = k_2$ e $N_1 = N_2$.
Esse grafo é $N_1$-regular e possui um número par de vértices.
Ou seja, a quantidade de vértices em cada partição é a mesma.
A Fig. \ref{fig:grafo-bipartido-marcado} ilustra um grafo com essas restrições onde
$N_1 = 5$ e $k_1 = 2$; os vértices marcados têm cor preta.

O operador de evolução de busca nesses grafos é dado por
\begin{align}
    U = U_w O_f,
    \label{eq:bipartido-op-evol-busca}
\end{align}
onde $U_w$ é dado pela Eq. \ref{eq:operador-evol-passio}; e
$O_f$ é o oráculo que inverte a amplitude dos elementos procurados,
\begin{align}
    O = \paren{I - 2 \sum_{j \in K} \ket{j}\bra{j}}
        \otimes I .
\end{align}

Pode-se analisar a evolução do passeio num espaço reduzido.
Note que algumas arestas ilustradas na Fig. \ref{fig:grafo-bipartido-marcado}
desempenham um papel em comum durante o passeio --
\textit{e.g.} arestas que vão de vértices não marcados em um conjunto para
não marcados no outro conjunto.
Antes de analisar a ação de $U$, considera-se apenas a ação de $U_w$ --
já que $O_f$ simplesmente inverte o sinal em alguns casos.

Denote por $K_j^C = V_j \setminus K_j$ para $j \in \set{1, 2}$ e considere o estado
\begin{align}
    \ket{K_1^C, K_2^C} = \frac{1}{\sqrt{\left|K_1^C\right|}} \sum_{v \in K_1^C} \ket{v}
        \otimes \frac{1}{\sqrt{\left|K_2^C\right|}} \sum_{u \in K_2^C} \ket{\cor{vu}}
    \label{eq:K1C-K2C}
\end{align}
a sobreposição uniforme dos vértices no conjunto $K_1^C$ cuja moeda aponta para
vértices no conjunto $K_2^C$.
Esse estado é unitário já que
\begin{align}
    \braket{K_1^C, K_2^C | K_1^C, K_2^C} &= \frac{1}{\left|K_1^C\right| \left|K_2^C\right|}
        \sum_{v, v' \in K_1^C} \braket{v | v'}
        \sum_{u, u' \in K_2^C} \braket{\cor{vu} | \cor{v'u'}}
    \\
    &= \frac{1}{\left|K_1^C\right| \left|K_2^C\right|}
        \sum_{v\in K_1^C}
        \sum_{u, u' \in K_2^C} \braket{\cor{vu} | \cor{vu'}}
    \\
    &= \frac{1}{\left|K_1^C\right| \left|K_2^C\right|} \cdot
        \left|K_1^C\right| \cdot \left|K_2^C\right|
    \\
    &= 1.
\end{align}
Aplicando $U_w$ nesse estado e usando a Eq. \ref{eq:passeio-grover},
\begin{align}
    U_w \ket{K_1^C, K_2^C} &= \frac{1}{\sqrt{\left|K_1^C\right| \left|K_2^C\right|}}
        \sum_{\substack{v \in K_1^C \\ u \in K_2^C}} U_w \ket{v, \cor{vu}}
    \\
    &= \frac{1}{\sqrt{\left|K_1^C\right| \left|K_2^C\right|}}
        \sum_{\substack{v \in K_1^C \\ u \in K_2^C}} \pr{
            \pr{\frac{2}{d}-1} \ket{u, \cor{v,u}} +
            \frac{2}{d} \sum_{u' \neq u} \ket{u', \cor{vu'}}
        }
    \\
    &= \frac{1}{\sqrt{\left|K_1^C\right| \left|K_2^C\right|}} \pr{
            \ket{\varphi_1} + \ket{\varphi_2}
        },
\end{align}
note que $u \in K_2^C$ e que $u' \in V_2$;
logo, separou-se o somatório em dois casos:
$\ket{\varphi_1}$ quando $u, u' \in K_2^C$, e
$\ket{\varphi_2}$ quando $u' \in K_2$.

\begin{figure}[htb]
    \centering
    \caption{Alguns termos do somatório quando $u, u' \in K_2^C$.}
    \begin{tikzpicture}
    \def\N{3}
    \pgfmathsetmacro\mult{1.5}
    
    \pgfmathsetmacro{\shft}{5}
    \pgfmathsetmacro{\mult}{2}
    \pgfmathparse{\N-1}
    \foreach \s in {0, 1, 2} {
        \foreach \y in {0, 1} {
            \foreach \x in {0,...,\pgfmathresult}{
                \pgfmathsetmacro\curr{ int(\N*\y+\x+\s*2*\N) }
                \pgfmathsetmacro\v{ int(1+\s*2*\N) }
                \pgfmathsetmacro\u{ int(3+\s*2*\N +\s)}
                \pgfmathsetmacro{\ustart}{int(3+\s*2*\N}
                \pgfmathsetmacro{\uend}{int(3+\s*2*\N +\N-1}
                
                \ifthenelse{\y=0} {
                    \ifthenelse{\curr=\v}{
                        \node [circle,draw,minimum size=8mm] (\curr) at (\s*\shft+\x,-\mult*\y) {$v$};
                    }{
                        \node (\curr) at (\s*\shft+\x,-\mult*\y) {};
                    }
                }{
                    \ifthenelse{\curr=\u}{
                        \node [circle,draw,minimum size=8mm] (\curr) at (\s*\shft+\x,-\mult*\y) {$u$};
                    }{
                        \node [circle,draw,dashed,minimum size=8mm] (\curr) at (\s*\shft+\x,-\mult*\y) {$u'$};
                    }
                }
            }
        }
    }
    
    \foreach \s in {0, 1, 2} {
        \pgfmathsetmacro\v{ int(1+\s*2*\N) }
        \pgfmathsetmacro\u{ int(3+\s*2*\N +\s)}
        \pgfmathsetmacro{\ustart}{int(3+\s*2*\N}
        \pgfmathsetmacro{\uend}{int(3+\s*2*\N +\N-1}
        \foreach \ktwo in {\ustart,...,\uend} {
            \ifthenelse{\ktwo=\u}{
                \path (\v) edge (\ktwo);
            }{
                \path [dashed] (\v) edge (\ktwo);
            }
        }
    }
\end{tikzpicture}
    \legend{Fonte: produzido pelo autor.}
    \label{fig:grafo-bipartido-subcaso-1}
\end{figure}
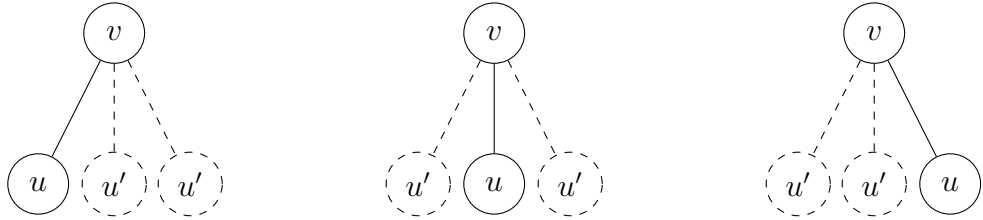

No primeiro caso, tem-se
\begin{align}
    \ket{\varphi_1} =
    \sum_{\substack{v \in K_1^C \\ u \in K_2^C}} \pr{
            \pr{\frac{2}{d}-1} \ket{u, \cor{vu}} +
            \frac{2}{d} \sum_{\substack{u' \neq u \\ u' \in K_2^C}} \ket{u', \cor{vu'}}
        } .
\end{align}
Considere o grafo de exemplo apresentado na Fig. \ref{fig:grafo-bipartido-marcado}.
Ao fixar $v \in K_1^C$, temos três possibilidades de $u$ que contribuem no somatório mais externo;
rotuladas e representadas por vértices com contornos sólidos na Fig. \ref{fig:grafo-bipartido-subcaso-1}.
Os vértices com cortornos tracejados indicam os valores de $u'$ que contribuem no somatório mais interno
dados $v$ e $u$ fixos.
Observe que cada aresta $vu$ é contabilizada outras
$|K_2^C| - 1$ vezes pelo somatório mais interno.

Logo,
\begin{align}
    \ket{\varphi_1} &=
    \sum_{\substack{v \in K_1^C \\ u \in K_2^C}} \pr{
            \pr{\frac{2}{d}-1} \ket{u, \cor{vu}} +
            \frac{2}{d} \pr{|K_2^C| - 1} \ket{u, \cor{vu}}
        }
    \\
    &= \sum_{\substack{v \in K_1^C \\ u \in K_2^C}}
            \pr{\frac{2}{d} |K_2^C| - 1} \ket{u, \cor{vu}} .
    \label{eq:varphi-1-final}
\end{align}
No segundo caso, é possível seguir um raciocínio análogo, obtendo
\begin{align}
    \ket{\varphi_2} &= \frac{2}{d} \sum_{\substack{v \in K_1^C \\ u \in K_2^C}} 
             \pr{ \sum_{u' \in K_2} \ket{u', \cor{v,u'}} }
    \\
    &= \frac{2}{d} \card{K_2^C} \sum_{v \in K_1^C} 
             \sum_{u' \in K_2} \ket{u', \cor{v,u'}} .
    \label{eq:varphi-2-final}
\end{align}
Note que há uma similaridade entre os somatórios das Eqs.
\ref{eq:varphi-1-final} e \ref{eq:varphi-2-final} com o
somatório da Eq. \ref{eq:K1C-K2C}.
De fato, ao definir a sobreposição uniforme de vértices no $K_2^C$ cuja moeda
aponta para vértices em $K_1^C$
\begin{align}
    \ket{K_2^C, K_1^C} = \frac{1}{\sqrt{\left|K_2^C\right|}} \sum_{v \in K_2^C} \ket{v}
            \otimes \frac{1}{\sqrt{\left|K_1^C\right|}} \sum_{u \in K_1^C} \ket{\cor{vu}}    
\end{align}
e a sobreposição uniformee de vértices em $K_2$ cuja moeda
aponta para vértices em $K_1^C$
\begin{align}
    \ket{K_2, K_1^C} = \frac{1}{\sqrt{\left|K_2\right|}} \sum_{v \in K_2} \ket{v}
        \otimes \frac{1}{\sqrt{\left|K_1^C\right|}} \sum_{u \in K_1^C} \ket{\cor{vu}} ,
\end{align}
obtém-se
\begin{align}
    U_w \ket{K_1^C, K_2^C} &=
        \frac{1}{\sqrt{\left|K_1^C\right| \left|K_2^C\right|}} \pr{
            \sum_{\substack{v \in K_2^C \\ u \in K_1^C}}
            \pr{\frac{2}{d} |K_2^C| - 1} \ket{v, \cor{vu}}
            +
            \frac{2}{d} \left|K_2^C\right|
            \sum_{\substack{v \in K_2 \\ u \in K_1^C}} \ket{v, \cor{vu}}
        }
    \\
    &= \pr{\frac{2}{d} |K_2^C| - 1} \ket{K_2^C, K_1^C} +
        \frac{2}{d} \sqrt{\left|K_2\right| \left|K_2^C\right|} \ket{K_2, K_1^C} .
\end{align}

Esse comportamento se repete para outros estados similares a $\ket{K_1^C, K_2^C}$.
Considere a definição a seguir.
\begin{definition}
    Denote por
    \begin{align}
        \ket{S_i, S_j} = \frac{1}{\sqrt{\left|S_i\right|}} \sum_{v \in S_i} \ket{v} \otimes
            \frac{1}{\sqrt{\left|S_j\right|}} \sum_{u \in S_j} \ket{\cor{vu}}
    \end{align}
    a sobreposição uniforme dos vértices em $S_i$
    cuja moeda aponta para vértices em $S_j$,
    onde $S_i \in \set{K_i, K_i^C}$ com
    $i, j \in \set{1, 2}$ de tal forma que $i \neq j$.
\end{definition}

Note que esses vetores formam uma base ortonormal $B$ já que
\begin{align}
    \braket{S_i, S_j | S_{i'}, S_{j'}} &=
        \frac{1}{\sqrt{\card{S_i} \card{S_{i'}}}}
            \sum_{\substack{v \in S_i \\ v' \in S_{i'}}} \braket{v | v'}
        \cdot \frac{1}{\sqrt{\card{S_j} \card{S_{j'}}}}
            \sum_{\substack{u \in S_j \\ u' \in S_{j'}}} \braket{\cor{vu} | \cor{v'u'}}
    \\
    &= \frac{1}{\sqrt{\card{S_i} \card{S_{i'}}}} \delta_{i, i'} \card{S_i} \cdot
        \frac{1}{\sqrt{\card{S_j} \card{S_{j'}}}}
            \sum_{\substack{u \in S_j \\ u' \in S_{j'}}} \braket{\cor{vu} | \cor{vu'}}
    \\
    &= \delta_{i, i'} \cdot  \frac{1}{\sqrt{\card{S_j} \card{S_{j'}}}}
        \delta_{j, j'} \card{S_j}
    \\
    &= \delta_{i, i'} \delta_{j, j'} .
\end{align}
Além disso,
\begin{align}
    U_w \ket{S_i, S_j} = \pr{\frac{2}{d} |S_j| - 1} \ket{S_j, S_i} +
        \frac{2}{d} \sqrt{\left|S_j^C\right| \left|S_j\right|} \ket{S_j^C, S_i} ,
    \label{eq:passeio-bipartido-subespaco}
\end{align}
desde que $\pr{K_j^C}^C = K_j$.
Ou seja, se um estado $\ket{\psi} \in \textnormal{span}(B)$,
então a ação de $U_w$ fica restrita ao subespaço definido por $B$.

A partir da Eq. \ref{eq:passeio-bipartido-subespaco},
a ação de $U$ segue trivialmente.
Se $S_i \in \set{K_1, K_2}$, então $U \ket{S_i, S_j} = - U_w \ket{S_i, S_j}$ e,
caso contrário, $U \ket{S_i, S_j} = U_w \ket{S_i, S_j}$.
Com isso em mãos, ordene $B$ por
$\ket{K_1, K_2}$, $\ket{K_1, K_2^C}$, $\ket{K_1^C, K_2}$, $\ket{K_1^C, K_2^C}$,
$\ket{K_2, K_1}$, $\ket{K_2, K_1^C}$, $\ket{K_2^C, K_1}$ e $\ket{K_2^C, K_1^C}$.
Dadas as restrições do problema, sabe-se também que a dimensão da moeda é $d = N_1$.
Então, constrói-se o operador de evolução no subespaço $\textnormal{span}(B)$ como
a matriz por blocos
\begin{align}
    U' = \matrx{
        0 & \U'(\theta_1) \\ \U'(\theta_2) & 0
    },
    \label{eq:op-evol-busca-grafo-bip-compl}
\end{align}
onde
\begin{align}
    \U' (\theta) = \matrx{
        \cos\theta  & -\sin\theta   &       0       &       0   \\
            0       &       0       & -\cos\theta    & \sin\theta \\ 
        -\sin\theta & -\cos\theta   &       0       &       0   \\
            0       &       0       & \sin\theta    & \cos\theta
    },
\end{align}
e $\theta_j$ são ângulos tais que
$\cos\theta_j = 1 - 2k_j/N_j$ e $\sin\theta_j = \frac{2}{N_j} \sqrt{k_j(N_j - k_j)}$
para $j \in \set{1, 2}$.

A motivação para utilizar $N_2$ e $k_2$ (ao invés de somente $N_1$ e $k_1$) é
que a mesma matriz é obtida para qualquer grafo bipartido completo,
mesmo sendo necessário alterar o espaço da moeda \cite{rhodes2019quantum}.
Portanto, decidiu-se manter a utilização de $N_2$ e $k_2$ para
o cálculo de autovalores e autovetores.

\subsection{Autovalores e Autovetores do Operador no Subespaço}
\label{sec:autov-subespaco}
A obtenção dos autovetores e autovalores é essencial para o algoritmo de contagem.
Como a análise ficará restrita ao subespaço $\textnormal{span}(B)$,
analisa-se os autovetores de $U'$.
Os autovalores $\lambda$ são as soluções para a equação
\begin{align}
    \det(U' - \lambda I) = 0.
\end{align}
Como
\begin{align}
    U' - \lambda I = \matrx{
        \lambda I & \U'(\theta_1) \\
        \U'(\theta_2) & \lambda I},
\end{align}
e as duas matrizes por bloco debaixo comutam -- \textit{i.e.}
$\lambda I\ \U'(\theta_2) = \U'(\theta_2)\ \lambda I$ --,
o determinante é dado por \cite{silvester2000determinants}
\begin{align}
    \det(U' - \lambda I) = \det\pr{\lambda^2 I - \U'(\theta_1)\U'(\theta_2)} = 0.
    \label{eq:U-eval}
\end{align}
Note que
\begin{align}
    \U'(\theta_1)\U'(\theta_2) &= \matrx{
        \cos\theta_1 \cos\theta_2 & -\cos\theta_1 \sin\theta_2 &
        \sin\theta_1 \cos\theta_2 & -\sin\theta_1 \sin\theta_2
        \\
        \cos\theta_1 \sin\theta_2 & \cos\theta_1 \cos\theta_2 &
        \sin\theta_1 \sin\theta_2 & \sin\theta_1 \cos\theta_2
        \\
        -\sin\theta_1 \cos\theta_2 & \sin\theta_1 \sin\theta_2 &
        \cos\theta_1 \cos\theta_2 & -\cos\theta_1 \sin\theta_2
        \\
        -\sin\theta_1 \sin\theta_2 & \sin\theta_1 \cos\theta_2 &
        \cos\theta_1 \sin\theta_2 & \cos\theta_1 \cos\theta_2
    } \\
    &= \matrx {
        \cos\theta_1 R(\theta_2) & \sin\theta_1 R(\theta_2) \\
        -\sin\theta_1 R(\theta_2) & \cos\theta_1 R(\theta_2)
    } \\
    &= R(\theta_1)^T \otimes R(\theta_2), \label{eq:matriz-bloco-eval}
\end{align}
onde
\begin{align}
    R(\theta) = \matrx{
        \cos\theta & -\sin\theta \\
        \sin\theta & \cos\theta
    }.
\end{align}
Como visto anteriormente,
$R(\theta)$ é uma matriz de rotação conhecida que possui autovalores $e^{\pm\ii\theta}$
associados aos autovetores  $\ket{\mp\ii} = (\ket{0} \mp \ii\ket{1})/\sqrt{2}$, respectivamente.
Analogamente, $R(\theta)^T$ tem autovalores $e^{\pm\ii\theta}$ associados
aos autovetores $\ket{\pm\ii}$.
Juntando esses fatos com a Eq. \ref{eq:matriz-bloco-eval} resulta que
$\U'(\theta_1)\U'(\theta_2)$ tem autovalores
$e^{\ii(\theta_1 \pm \theta_2)}$ e seus complexos conjugados, e
autovetores $\ket{\pm\ii}\otimes\ket{\pm\ii}$.
Entretanto, esse resultado é a solução para a equação
\begin{align}
    \det\pr{ \lambda I - \U'(\theta_1) \U'(\theta_2) } = 0,
\end{align}
não para a Eq. \ref{eq:U-eval}.
As soluções para a Eq. \ref{eq:U-eval} são as raízes quadradas
(positivas e negativas) dos valores obtidos --
\textit{i.e.} $e^{\ii(\theta_1 \pm \theta_2)/2}$, $-e^{\ii(\theta_1 \pm \theta_2)/2}$ e
seus complexos conjugados.

Para calcular os autovetores, suponha que
$\ket\lambda = \ket{\lambda_1} \oplus \ket{\lambda_2}$ é um $\lambda$-autovetor de $U'$,
onde $\oplus$ denota a soma direta. Então,
\begin{align}
    U' \ket\lambda &= \matrx{0 & \U'(\theta_1) \\ \U'(\theta_2) & 0}
        \ket{\lambda_1} \oplus \ket{\lambda_2}
    \\
    &= \U'(\theta_1)\ket{\lambda_2} \oplus \U'(\theta_2)\ket{\lambda_1}
\end{align}
implica que os autovetores obedecem o sistema de equações
\begin{align}
    \begin{cases}
            \U'(\theta_1)\ket{\lambda_2} &= \lambda \ket{\lambda_1} \\
            \U'(\theta_2)\ket{\lambda_1} &= \lambda \ket{\lambda_2}
    \end{cases}.
\end{align}
Da primeira equação, tem-se que
$\ket{\lambda_1} = \U'(\theta_1) \frac{1}{\lambda} \ket{\lambda_2}$.
Substituindo isso na segunda equação obtém-se
\begin{align}
    \U'(\theta_2) \U'(\theta_1) \frac{1}{\lambda} \ket{\lambda_2} &= \lambda \ket{\lambda_2} \\
    R(\theta_2)^T \otimes R(\theta_1) \ket{\lambda_2} &= \lambda^2 \ket{\lambda_2},
\end{align}
\textit{i.e.} $\ket{\lambda_2}$ é um $\lambda^2$-autovetor de $R(\theta_2)^T \otimes R(\theta_1)$.
Cálculos anteriores mostram que esses autovetores são
$\ket{\lambda_2} = \ket{s_2 \ii} \otimes \ket{s_1 \ii}$
associados com os autovalores $\pm e^{ \ii \pr{s_2\theta_2 - s_1\theta_1} }$
para $s_1, s_2 \in \{+1, -1\}$.
Com $\ket{\lambda_2}$ e seu autovalor em mãos,
é possível calcular
\begin{align}
    \ket{\lambda_1} &= \frac{1}{\lambda} \U'(\theta_1) \ket{\lambda_2}
    \\
    &= \frac{1}{\lambda} \matrx{
        \cos\theta_1  & -\sin\theta_1   &       0       &       0   \\
            0       &       0       & -\cos\theta_1    & \sin\theta_1 \\ 
        -\sin\theta_1 & -\cos\theta_1   &       0       &       0   \\
            0       &       0       & \sin\theta_1    & \cos\theta_1
    } \frac{1}{2}\matrx{
        1 \\ s_1\ii \\ s_2\ii \\ -s_1 s_2
    }
    \\
    &= \frac{1}{\lambda} \cdot \frac{1}{2} \matrx{
        \cos\theta_1 - s_1\ii \sin\theta_1 \\
        -s_2\ii \paren{\cos\theta_1 - s_1\ii \sin\theta_1} \\
        -s_1\ii \paren{\cos\theta_1 - s_1\ii \sin\theta_1} \\
        -s_1s_2 \paren{\cos\theta_1 - s_1\ii \sin\theta_1}
    }
    \\
    &= \frac{e^{-s_1\ii\ \theta_1}}{\lambda} \ket{-s_1\ii}\otimes\ket{-s_2\ii} .
\end{align}
Como $\ket{\lambda_2}$ é um $\lambda^2$-autovetor de $R(\theta_2)^T \otimes R(\theta_1)$,
cada $\ket{\lambda_2}$ gera dois resultados possíveis para $\ket{\lambda_1}$.
Por exemplo, tome $\ket{\lambda_2} = \ket{+\ii}\otimes\ket{-\ii}$,
então $\lambda = \pm e^{\ii(\theta_1 + \theta_2)/2}$ e
$\ket{\lambda_1} = \pm e^{\ii(\theta_1 - \theta_2)/2} \ket{+\ii}\otimes\ket{-\ii}$;
consequentemente,
\begin{align}
    \ket{\Sigma_\pm} = \frac{1}{\sqrt 8} \matrx{
        \pm e^{\ii\Delta} \\ \mp\ii e^{\ii\Delta} \\ \pm\ii e^{\ii\Delta} \\ \pm e^{\ii\Delta} \\
        1 \\ -\ii \\ \ii \\ 1
    }
    \label{eq:evecs-Sigma}
\end{align}
é um $\pm e^{\ii\Sigma}$-autovetor de $U'$,
onde $\Delta = (\theta_1 - \theta_2)/2$ e $\Sigma = (\theta_1 + \theta_2)/2$.
Analogamente,
\begin{align}
    \ket{\Delta_\pm} = \frac{1}{\sqrt 8} \matrx{
        \pm e^{\ii\Sigma} \\ \pm\ii e^{\ii\Sigma} \\ \pm\ii e^{\ii\Sigma} \\ \mp e^{\ii\Sigma}\\
        1 \\ -\ii \\ -\ii \\ -1
    },
    \label{eq:evecs-Delta}
\end{align}
é um $\pm e^{\ii\Delta}$-autovetor de $U'$.

Seguindo esse mesmo raciocínio, obtém-se todos os autovalores e autovetores de $U'$:
$\ket{\Sigma_\pm}$, $\ket{\Delta_\pm}$ e seus complexos conjugados --
denotados respectivamente por $\ket{\Sigma_\pm^*}$ e $\ket{\Delta_\pm^*}$.
Esses autovetores estão respectivamente associados aos autovalores
$\pm e^{\ii\Sigma}$, $\pm e^{\ii\Delta}$  e seus complexos conjugados.

\subsubsection{Comportamento dos Autovalores}
Esta Seção dedica-se a analisar o comportamento dos autovalores.
Uma intuição desse comportamento é útil para facilitar o entendimento
dos resultados provenientes de usar $U$ (Eq. \ref{eq:bipartido-op-evol-busca})
como entrada no algoritmo de estimativa de fase ou contagem.

Denomine os autovalores $\pm e^{\ii \Sigma}$ e $\pm e^{-\ii \Sigma}$ por $\Sigma$-autovalores.
Analogamente, denote $\pm e^{\ii \Delta}$ e $\pm e^{-\ii \Delta}$ por $\Delta$-autovalores.
Note que $-e^{\pm\ii\Sigma} = e^{\pm\ii\pr{\Sigma + \pi}}$
(análogo para os $\Delta$-autovalores).
Primeiro, nota-se que os valores de $\Sigma$ e $\Delta$ dependem de $\theta_1$ e $\theta_2$,
sendo que $\Sigma$ se relaciona com a soma desses ângulos,
enquanto $\Delta$ se relaciona com a diferença.
Pelo modo como $\theta_j$ para $j \in \set{1,2}$ foram definidos, tem-se que
$-1 \leq \cos\theta_j \leq 1$ e $0\leq \sin\theta_j \leq 1$;
logo, $0 \leq \theta_j \leq \pi$.
Consequentemente, $0 \leq \Sigma \leq \pi$ e $-\pi/2 \leq \Delta \leq \pi/2$.

\begin{figure}[hbt]
    \centering
    \begin{minipage}{0.48\textwidth}
        \centering
        \caption{Autovalores para $N_1=N_2=40$, $k_1 = 2$ e $k_2 = 1$.}
        \includegraphics[width=\textwidth]{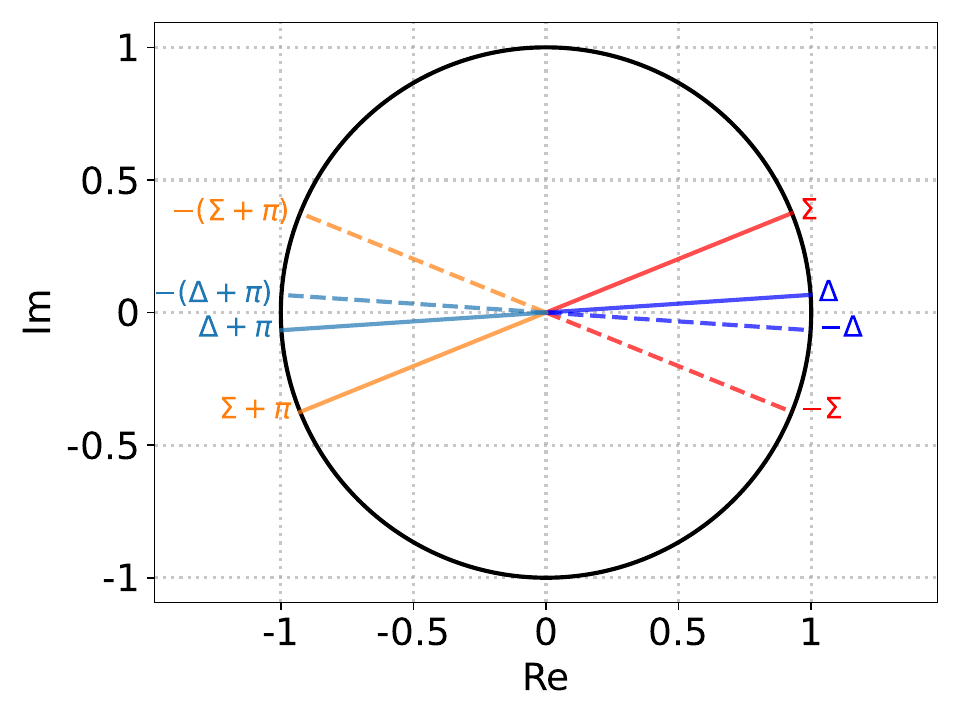}
        \legend{Fonte: Produzido pelo autor.}
        \label{fig:N1=40-N2=40-k1=2-k2=1}
    \end{minipage}
    \hfill
    \begin{minipage}{0.48\textwidth}
        \centering
        \caption{Autovalores para $N_1=N_2=40$, $k_1 = 8$ e $k_2 = 4$.}
        \includegraphics[width=\textwidth]{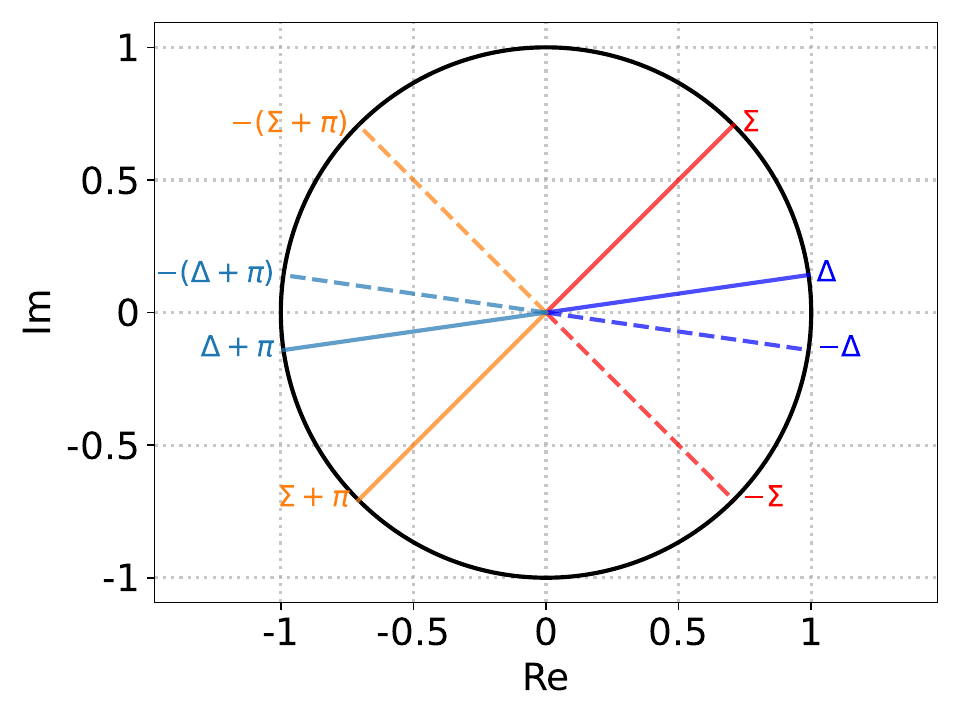}
        \legend{Fonte: Produzido pelo autor.}
        \label{fig:N1=40-N2=40-k1=8-k2=4}
    \end{minipage}
\end{figure}

Figs. \ref{fig:N1=40-N2=40-k1=2-k2=1} e \ref{fig:N1=40-N2=40-k1=8-k2=4} ilustram
os $\Sigma$ e $\Delta$-autovalores no círculo unitário com $N_1 = N_2 = 40$.
Os $\Sigma$-autovalores são representados por cores vermelho-alaranjadas
enquanto que os $\Delta$-autovalores são representados por cores azuladas.
Ângulos positivos são representados por linhas contínuas (\textit{e.g.} $\Sigma$ e $\Sigma + \pi$),
enquanto que ângulos negativos são representados por linhas tracejadas
(\textit{e.g.} $-\Sigma$ e $-(\Sigma + \pi)$).
Na Fig. \ref{fig:N1=40-N2=40-k1=2-k2=1}, utilizou-se $k_1 = 2$ e $k_2 = 1$,
enquanto que na Fig. \ref{fig:N1=40-N2=40-k1=8-k2=4}, $k_1 = 8$ e $k_2 = 4$.
Conforme esperado, $\Sigma > \card\Delta$.
Também é possível notar que o ângulo $\Sigma$ aumenta à medida que $k_1$ ou $k_2$ aumentam e
que se $k_1$ for próximo a $k_2$, $\Delta$ é próximo de $0$.

\section{Contagem no Grafo Bipartido Completo}
\label{sec:passeio-contagem}

Relembre que o algoritmo de contagem original utiliza
o operador de evolução de Grover como uma das entradas para o algoritmo de estimativa de fase;
além de um autovetor dessa matriz (na realidade uma sobreposição de autovetores).

Analogamente ao Algoritmo de Grover,
os autovalores de $U$ (Eq. \ref{eq:bipartido-op-evol-busca}) no espaço reduzido
possuem informações sobre a quantidade de elementos marcados.
Essa informação pode ser obtida facilmente a partir dos ângulos $\theta_1$ e $\theta_2$.
Sendo assim,
deseja-se utilizar o algoritmo de estimativa de fase para obter estimativas de
$\Sigma$ e de $\Delta$ para então extrair a informação desejada.
Para isso, entretanto, seria necessário preparar os autovetores
$\ket{\Sigma_+}$ e $\ket{\Delta_+}$,
que exigem conhecimento prévio dos valores de $\Sigma$ e $\Delta$.
A solução é utilizar uma sobreposição de autovetores como entrada do segundo registrador.

Relembre que $N_1$ é a dimensão do espaço da moeda.
Então, a sobreposição uniforme é dada por
\begin{align}
    \ket{D} &= \frac{1}{\sqrt{N \cdot N_1}}
        \sum_{j = 0}^{N-1} \sum_{c = 0}^{N_1 - 1} \ket{j, c}
    \\
    &= \frac{1}{\sqrt{2 N_1 N_2}} \sum_{\substack{S_i, S_j \\ i \neq j}}
        \sqrt{\card{S_i} \card{S_j}}\ \ket{S_i, S_j}.
    \label{eq:sup-uni-arestas-geral}
\end{align}
Apesar do escopo desse trabalho estar reduzido a $N_1 = N_2$ e $k_1 = k_2$,
trabalha-se com a expressão \ref{eq:sup-uni-arestas-geral} para
facilitar uma futura extensão dos resultados para o caso geral.
Já que $\ket D$ pode ser escrito como uma combinação linear de $B$,
tanto $\ket D$ quanto $U\ket{D}$ ficam restritos nesse subespaço.
O algoritmo de contagem no grafo bipartido está descrito em \ref{alg:contagem-bipartido}.
Note que é bastante similar ao algoritmo de contagem original,
especialmente o pós-processamento clássico.

\begin{algorithm}[hbt]
    \caption{Algoritmo de Contagem no Grafo Bipartido.}
    \label{alg:contagem-bipartido}
    \begin{algorithmic}[1]
        \REQUIRE
            $O_f$: Oráculo da função $f$;
            $p$: número de qubits do primeiro registrador do algoritmo de estimativa de fase;
            $N$: tamanho do segundo registrador respeitando o domínio de $f$;
            $t$: quantidade máxima de iterações.
        \STATE Construir operador $U = U_w O_f$
        \STATE Preparar o estado $\ket D$; $t' \gets 0$
        \WHILE{$t' < t\ \AND\ \theta_1' \in \{0, \pi\}$} \label{alg:cont-bip-loop}
            \STATE $\vartheta \gets$ est\_fase($p, U, \ket D$)
            \STATE $\theta_1' \gets 2\pi\vartheta$; $t' \gets t' + 1$
        \ENDWHILE
        \IF{$\theta_1' \in \{0, \pi\}$} \label{alg:qc-k1=k2-check-full-marked-begin}
            \STATE Se um elemento aleatório estiver marcado então $k' \gets N$;
                senão $k' \gets 0$
            \RETURN $k'$
        \ENDIF
        \RETURN $k' = \sin^2\pr{\theta_1' / 2} \cdot N$ \label{alg:cont-bip-est-final}
    \end{algorithmic}
\end{algorithm}

Para prosseguir com a análise,
é necessário representar $\ket{D}$ como uma sobreposição dos autovetores do subespaço de $U$.
Essa análise é essencial pois influencia na probabilidade de obter uma determinada estimativa,
conforme visto na Seção \ref{sec:est-fase}.
Ou seja, deseja-se obter as amplitudes de
\begin{align}
    I \ket{D} = \sum_{\lambda} \ket{\lambda} \bra{\lambda} D
    = \sum_{\lambda} \braket{\lambda | D} \ket{\lambda},
\end{align}
onde $\lambda$ e $\ket{\lambda}$ representam respectivamente os autovalores e autovetores
apresentados na Seção \ref{sec:autov-subespaco}.

Calculando-se a projeção de $\ket D$ em $\ket{\Sigma_\pm}$ obtém-se
\begin{align}
    \braket{\Sigma_\pm|D} &= \paren{ \braket{D|\Sigma_\pm} }^* \\
    &= \frac{1}{4 \sqrt{N_1 N_2}} \paren{
        \pm e^{\ii\Delta} \chi + \chi^*
    }^*,
\end{align}
onde
\begin{align}
    \chi &= \sqrt{k_1 k_2} - \ii\sqrt{k_1 (N_2 - k_2)} +
        \ii\sqrt{(N_1 - k_1)k_2} + \sqrt{(N_1 - k_1)(N_2 - k_2)} \\
    &= \paren{\sqrt{k_1} + \ii\sqrt{N_1 - k_1}} \paren{\sqrt{k_2} - \ii \sqrt{N_2 - k_2}}.
\end{align}
Conforme visto na Seção \ref{sec:est-fase},
calcular $\card{ \braket{\Sigma_\pm | D} }^2$ é necessário para
obter a probabilidade de estimar os autovalores $\pm e^{\ii\Sigma}$,
\begin{align}
    \left|\braket{\Sigma_\pm | D}\right|^2 &= \frac{1}{16 N_1 N_2}
        \paren{\pm e^{\ii\Delta} \chi + \chi^*} \paren{\pm e^{\ii\Delta} \chi + \chi^*}^* \\
    &= \frac{1}{16 N_1 N_2} \paren{2\chi\chi^* \pm 2 \Re\paren{e^{\ii\Delta}\chi^2}} \\
    &= \frac{1}{8}\paren{ 1 \pm \Re\paren{\frac{e^{\ii\Delta}\chi^2}{N_1 N_2}} },
\end{align}
e usando
\begin{align}
    \frac{e^{\ii\Delta}\chi^2}{N_1 N_2} &= e^{\ii\Delta}
        \paren{ \frac{2k_1 - N_1 + 2\ii\sqrt{N_1 - k_1}}{N_1} }
        \paren{ \frac{2k_2 - N_2 - 2\ii\sqrt{N_2 - k_2}}{N_2} } \\
    &= e^{\ii\Delta} \paren{-\cos\theta_1 +\ii\sin\theta_1} \paren{-\cos\theta_2 -\ii\sin\theta_2} \\
    &= e^{\ii\Delta} e^{-\ii\theta_1} e^{\ii\theta_2} \\
    &= e^{-\ii\Delta}
\end{align}
resulta em
\begin{align}
    \left|\braket{\Sigma_\pm | D}\right|^2 &= \frac{1}{8} \paren{1 \pm \cos\Delta}.
\end{align}

Fazendo cálculos similares para os autovetores restantes e
usando as identidades trigonométricas
$1 + \cos\theta = 2\cos^2\paren{\theta/2}$, $1 - \cos\theta = 2\sin^2\paren{\theta/2}$;
obtém-se a probabilidade de de estimar os autovalores restantes
(Apêndice \ref{apendice:proj-arestas-autovecs}).
Esse resultado relaciona-se diretamente com a estimativa do ângulo de um autovalor de $U'$ e
estão resumidos na Tabela \ref{tab:prob-autovals}.
Então, a probabilidade de obter uma estimativa de algum ângulo do tipo $\Sigma$ é
\begin{align}
    2 \cdot \frac{1}{4}\paren{
        \cos^2\frac{\Delta}{2} + \sin^2\frac{\Delta}{2}
    }
    = \frac{1}{2}.
\end{align}
Análogo para os ângulos do tipo $\Delta$.

\begin{table}[htb]
    \caption{Probabilidade de estimativa de cada ângulo dos autovalores de $U'$.}
    \IBGEtab{
    \label{tab:prob-autovals}
    }{
    \begin{tabular}{cc}
    \toprule
    Ângulo estimado & Probabilidade de medida \\
    \midrule \midrule
    $\pm\Sigma$ & $\frac{1}{4}\cos^2\paren{\Delta/2}$ \\
    \midrule
    $\pm(\Sigma + \pi)$ & $\frac{1}{4}\sin^2\paren{\Delta/2}$ \\
    \midrule
    $\pm\Delta$ & $\frac{1}{4}\cos^2\paren{\Sigma/2}$ \\
    \midrule
    $\pm(\Delta + \pi)$ & $\frac{1}{4}\sin^2\paren{\Sigma/2}$  \\
    \bottomrule
    \end{tabular}
    }{\fonte{autor.}}
\end{table}

Analisando os dados da Tabela \ref{tab:prob-autovals}, percebe-se que
a probabilidade de obter uma estimativa para um dado ângulo
varia de acordo com a quantidade de elementos marcados.
Por exemplo, se $k_1 \neq k_2$ então $\Delta \neq 0$ e a probabilidade de estimar
tanto $\pm\Sigma$ quanto $\pm(\Sigma + \pi)$ é não nula.
Além disso, considerando os ângulos apresentados nas Figs.
\ref{fig:N1=40-N2=40-k1=2-k2=1} e \ref{fig:N1=40-N2=40-k1=8-k2=4},
observa-se que no caso geral, pode ser necessário fazer a distinção entre os ângulos
do tipo $\Sigma$ e do tipo $\Delta$;
e fazer a distinção dos ângulos $\Sigma$ e $-(\Sigma + \pi)$,
já que eles sempre estão no intervalo $[0, \pi]$.
Esses empecilhos motivaram a redução do escopo do problema para $k_1 = k_2$ e $N_1 = N_2$.

Uma consequência da redução do escopo é que $\Delta = 0$ e o valor de $\Sigma$ varia.
Logo, os ângulos $\pm{\Sigma + \pi}$ nunca serão estimados,
e a probabilidade de estimar cada um dos ângulos tipo $\Delta$ varia.
Porém, no caso de estimar um ângulo $\Delta$,
obtém-se os valores $\theta_1' = \{0, \pi\}$ exatamente,
facilitando a identificação de um ângulos tipo $\Sigma$.
Essa é a razão para o laço no passo \ref{alg:cont-bip-loop}.

Dois cenários podem fazer o algoritmo sair do laço:
estimar um ângulo $\theta_1' \notin \set{0, \pi}$ ou após $t$ iterações.
No primeiro caso, garante-se que uma estimativa de $\pm\Sigma$ foi obtida.
Por conta da redução de escopo, $\Sigma = (\theta_1 + \theta_2)/2 = \theta_1$ e $k = 2k_1$.
Então, a estimativa da quantidade de elementos marcados pode
ser obtida a partir de $\cos(\pm\theta_1)$ --
logo, independentemente do sinal de $\theta_1$.
Note que
\begin{align}
    \cos\pr{\theta_1} &= 1 - \frac{2k_1}{N_1}
    \\
    k_1 &= \pr{1 - \cos\pr{\theta_1}} \frac{N_1}{2}
    \\
    k_1 &= \sin^2\pr{\theta_1 / 2} \cdot N_1
    \\
    \implies k &= \sin^2\pr{\theta_1 / 2} \cdot N.
\end{align}
Isso justifica o passo \ref{alg:cont-bip-est-final} do algoritmo.
Ainda no contexto do primeiro caso, usar o Teorema \ref{teo:est-ampl-error} resulta
numa estimativa do erro:
\begin{align}
    \card{\sin^2\frac{\theta_1'}{2} - \sin^2\frac{\theta_1}{2}} &\geq
        2\pi\frac{\sin\frac{\theta_1}{2} \cos\frac{\theta_1}{2}}{P} +
        \frac{\pi^2}{P}
    \\
    &= 2\pi\frac{\sinp{\theta_1}/2}{P} +
        \frac{\pi^2}{P}
    \\
    &= 2\pi\frac{\frac{1}{N_1}\sqrt{k_1 \pr{N_1 - k_1}}}{P} +
        \frac{\pi^2}{P}
    \\
    \implies \card{k' - k} &\leq 2\pi\frac{2\sqrt{k_1 \pr{N_1 - k_1}}}{P} +
        \frac{\pi^2 N}{P}
    \\
    &= 2\pi\frac{\sqrt{k \pr{N - k}}}{P} +
        \frac{\pi^2 N}{P}
\end{align}
com probabilidade maior ou igual a $8/\pi^2$.

O segundo caso é subdividido em outros dois cenários:
(1) caso $k \in \set{0, N}$;
(2) caso $k \notin \set{0, N}$ e $t$ iterações foram executadas.
No primeiro cenário, o ângulo $\theta_1$ foi estimado corretamente,
mas coincide com os valores possíveis de $\Delta$.
Uma consulta adicional ao oráculo é suficiente para definir se nenhum
ou todos os elementos estão marcados.
No segundo cenário, não estimou-se um ângulo $\Sigma$ após $t$ iterações,
ocasionando um erro.
A probabilidade disso acontecer é de $(1/2)^t$.

Juntando todos esses argumentos, conclui-se que
o alg. \ref{alg:contagem-bipartido} realiza $t(P - 1)$ consultas ao oráculo;
estima os casos $k = 0$ e $k = N$ exatamente com probabilidade de sucesso $1$;
e caso $0 < k < N$, o algoritmo retorna uma estimativa $k'$ de $k$ com erro
\begin{align}
    \left| k' - k \right| &\leq
                2\pi \frac{ \sqrt{k\pr{N-k}} }{P} + \frac{\pi^2 N}{P^2}
\end{align}
e com probabilidade de sucesso maior ou igual a $\pr{1 - \frac{1}{2^t}}\frac{8}{\pi^2}$.
Portanto, os mesmos comentários feitos na Seção \ref{sec:alg-cont-precisao}
sobre a ordem do erro dependendo do valor $k$
estendem-se para o alg. \ref{alg:contagem-bipartido}.



\phantompart

\chapter{Conclusão} \label{cap:conclusao}

O algoritmo de contagem original é utilizado para estimar a quantidade de
entradas $k$ que satisfazem uma busca num banco de dados não ordenado.
Essa estimativa é obtida ao usar o operador de busca de Grover como
entrada do algoritmo de estimativa de fase.
Nesse trabalho, mostrou-se que é possível utilizar a mesma estratégia
para estimar a quantidade $k$ de elementos marcados num grafo;
ou seja, utilizando o operador do passeio de busca como entrada do algoritmo de estimativa de fase.
É necessário que os autovalores do operador dependam de $k$,
cuja estimativa é obtida depois de um pós-processamento clássico.

Essa técnica foi aplicada utilizando o operador de evolução de busca do grafo bipartido completo
($U'$, Eq. \ref{eq:op-evol-busca-grafo-bip-compl}) no subcaso $k_1 = k_2$ e $N_1 = N_2$;
onde $k_1$ e $k_2$ são as quantidades de vértices marcados dos conjuntos disjuntos $V_1$ e $V_2$
(respectivamente), e
$N_1$ e $N_2$ são as cardinalidades de $V_1$ e $V_2$ (respectivamente).
Obteve-se uma complexidade de $O\pr{t\sqrt{N}}$ chamadas ao oráculo
onde $t$ é a quantidade máxima de vezes que o algoritmo de estimativa de fase é executado e
$N = N_1 + N_2$;
em comparação com $O\pr{\sqrt{N}}$ do algoritmo de contagem original
\cite{brassard2002quantum}.
Obteve-se também um erro da ordem de $O\pr{\sqrt k}$ --
mais precisamente de $O\pr{\sqrt{k \pr{1 - \frac k N }}}$ --,
a mesma ordem de erro do algoritmo de contagem original;
porém com uma probabilidade de sucesso maior ou igual a $\pr{1 - 2^{-t}}\frac{8}{\pi^2}$,
enquanto a probabilidade de sucesso do algoritmo original é maior ou igual a $\frac{8}{\pi^2}$.

Uma consequência desse resultado foi a obtenção dos autovetores e autovalores exatos do
operador $U'$.
Previamente tinha-se apenas uma aproximação quando
$k_1 = o(N_1)$ e $k_2 = o(N_2)$ \cite{rhodes2019quantum}.
Os autovalores e autovetores de $U'$ também são os mesmos do operador de busca no caso geral
(\textit{e.g.} $N_1 \neq N_2$),
desde que o espaço da moeda seja alterado adequadamente.

Para trabalhos futuros,
ainda é necessário analisar outras situações no grafo bipartido completo.
Particularmente, o caso em que $k_1 \neq k_2$ e $N_1 = N_2$
parece bem mais simples que o restante dos casos.
Outras possibilidades abrangem a análise de como
o algoritmo de contagem se comporta em outros grafos;
por exemplo, o hipercubo e a malha.
Também é interessante analisar como (e se) o algoritmo de contagem é influenciado por
configurações excepcionais, \textit{e.g.} a diagonal na malha. 

\postextual

\bibliography{bibliografia}

%
%


\begin{apendicesenv}

\partapendices



\chapter{Mínimo da Equação \ref{eq:teo-fourier-erro}}
\label{apendice:minimo-funcao-teo-ang-int}

Deseja-se demonstrar que o ponto $w = 1/2$ é o único mínimo da função
\begin{align}
    f(w) = \frac{\sin^2\pr{\pi w}}{P^2 \sin^2\pr{\pi w / P}} +
        \frac{\sin^2\pr{\pi (1-w)}}{P^2 \sin^2\pr{\pi (1-w) / P}} ,
\end{align}
onde $0 < w < 1$ e $P \in \mathbb{N}^+$.
Serão considerados dois casos particulares ($P = 1$ e $P = 2$) e
então o caso geral será analisado.

\section{P = 1}
Nesse caso, $f(w)$ é constante:
\begin{align}
    f(w) &= \frac{\sin^2\pr{\pi w}}{\sin^2\pr{\pi w}} +
        \frac{\sin^2\pr{\pi (1-w)}}{\sin^2\pr{\pi (1-w)}}
    \\
    &= 2.
\end{align}
Esse cenário é desconsiderado no Teorema \ref{teo:fourier-error} porque
$P = 1$ implica um espaço de Hilbert unidimensional.
Além disso, o círculo unitário é dividido em 1 uma única partição,
ou seja, $f(w)$ contabiliza a probabilidade de obter $w$ duas vezes.

\section{P = 2}
Usando a identidade trigonométrica
$\sin^2\pr{\theta/2} = \frac{1}{2}(1 - \cos\theta)$, obtém-se
\begin{align}
    \frac{\sin^2\theta}{\sin^2\pr{\theta/2}} &=
        \frac{(1 + \cos\theta)(1-\cos\theta)}{\frac{1}{2}(1 - \cos\theta)}
    \\
    &= 2 \cdot 2 \cos^2\frac{\theta}{2} .
\end{align}
Usando esse resultado e a identidade trigonométrica para $\cos(\theta + \theta')$
calcula-se $f(w)$:
\begin{align}
    f(w) &= \frac{\sin^2\pr{\pi w}}{4 \sin^2\pr{\pi w / 2}} +
        \frac{\sin^2\pr{\pi (1-w)}}{4 \sin^2\pr{\pi (1-w) / 2}}
    \\
    &= \cos^2{\frac{\pi w}{2}} + \cos^2\frac{\pi(1 - w)}{2}
    \\
    &= \cos^2{\frac{\pi w}{2}} + \pr{
        \sin\frac{\pi}{2}\sin\frac{\pi w}{2} + \cos\frac{\pi}{2}\cos\frac{\pi w}{2}}^2
    \\
    &= \cos^2\frac{\pi w}{2} + \sin^2\frac{\pi w}{2}
    \\
    &= 1.
\end{align}
Esse resultado é esperado porque divide-se o círculo unitário em duas partições,
demarcadas pelos ângulos 0 e $\pi$.
Evidentemente que ao fazer a medição, apenas um desses resultados será obtido,
abrangendo todas as opções.
Como $f(w)$ é constante nesse caso,
qualquer ponto do domínio corresponde ao mínimo global.

\section{Caso Geral}
Analisa-se agora o caso de $P \geq 3$.
Sendo $\eps \in \mathbb{R}$ tal que $0 < \eps < 1/2$,
mostra-se que a função $f(w)$ é simétrica em torno do ponto $w = 1/2$.
\begin{align}
    f\pr{\frac{1}{2} + \eps} &= \frac{\sin^2\pr{\pi \pr{\frac{1}{2} + \eps}}}{
        P^2 \sin^2\pr{\frac{\pi}{P} \pr{\frac{1}{2} + \eps}}}
        +
        \frac{\sin^2\pr{\pi \pr{\frac{1}{2} - \eps}}}{
        P^2 \sin^2\pr{\frac{\pi}{P} \pr{\frac{1}{2} - \eps}}}
    \\
    &= \frac{\sin^2\pr{\pi \pr{ 1 - \pr{\frac{1}{2} - \eps}}}}{
        P^2 \sin^2\pr{\frac{\pi}{P} \pr{ 1 - \pr{\frac{1}{2} - \eps}}}}
        +
        \frac{\sin^2\pr{\pi \pr{\frac{1}{2} - \eps}}}{
        P^2 \sin^2\pr{\frac{\pi}{P} \pr{\frac{1}{2} - \eps}}}
    \\
    &= f\pr{\frac{1}{2} - \eps} .
\end{align}
Resta demonstrar que $f(1/2)$ é um mínimo global.
A análise será feita em torno da função par $f_\pi(\theta)$ obtida a partir de $f(w)$
da seguinte maneira.
Seja $w' = w - 1/2$, desloca-se a função $f(w)$ para a esquerda tomando
$f_0(w') = f(w' + 1/2)$.
Note que $f_0(w')$ é uma função par e que $-1/2 < w' < 1/2$.
Para obter $f_\pi(\theta)$, faz-se uma outra conversão simples de domínio:
$\theta = w'\pi$ e $f_\pi(\theta) = f_0(\theta/\pi)$.
Assim, toda análise será feita em torno da função
\begin{align}
    f_\pi(\theta) &= f\pr{\frac{\theta}{\pi} + \frac{1}{2}}
    \\
    &= \frac{\sin^2\pr{\pi \pr{\frac{\theta}{\pi} + \frac{1}{2}}}}{
        \sin^2\pr{\frac{\pi}{P} \pr{\frac{\theta}{\pi} + \frac{1}{2}}}}
        +
        \frac{\sin^2\pr{\pi (1-\pr{\frac{\theta}{\pi} + \frac{1}{2}})}}{
        \sin^2\pr{\frac{\pi}{P} \pr{1-\pr{\frac{\theta}{\pi} + \frac{1}{2}}}}}
    \\
    &= \frac{\sin^2\pr{\theta + \frac{\pi}{2}}}{
        \sin^2\pr{\frac{\theta}{P} + \frac{\pi}{2P}}}
        +
        \frac{\sin^2\pr{\theta - \frac{\pi}{2}}}{
        \sin^2\pr{- \frac{\theta}{P} + \frac{\pi}{2P}}}
    \\
    &= \cos^2(\theta) \pr{
        \csc^2\pr{\frac{\pi}{2P} + \frac{\theta}{P}} +
        \csc^2\pr{\frac{\pi}{2P} - \frac{\theta}{P}}
    },
\end{align}
com domínio $\theta \in (-\pi/2, \pi/2)$.
Como a função é par, restringe-se a análise para $0 \leq \theta < \pi/2$
e deseja-se provar que o ponto $f_\pi(0)$ é o mínimo global.
Ou seja,
\begin{align}
    f_\pi(\theta) &\geq f_\pi(0)
    \\
    &= \cos^2(0) \pr{\csc^2\pr{\frac \pi {2P}} + \csc^2\pr{\frac \pi {2P}}}
    \\
    &= 2 \csc^2\pr{\frac \pi {2P}} .
\end{align}
Divide-se a análise em dois casos:
$P = 3$ e $P \geq 4$.

\subsection{Caso 1 -- P igual a 3}

Para $P = 3$, usa-se as identidades
\begin{enumerate}
    \item $\sinp{\frac \pi 6 - x} = \cosp{\frac \pi 3 + x}$;
    \item $\sinp{a \pm b} = \sin a \cos b \pm \cos a \sin b$;
    \item $\cosp{a \pm b} = \cos a \cos b \mp \sin a \sin b$;
    \item $2\cos^2 x = \cosp{2x} + 1$
\end{enumerate}
Então,
\begin{align}
    f_\pi(\theta) =&\ \cos^2\pr\theta \pr{
        \frac{1}{\sin^2\pr{\frac \pi 6 + \frac x 3}}
        +
        \frac{1}{\sin^2\pr{\frac \pi 6 - \frac x 3}}
    }
    \\
    =&\ \frac{ \pr{\cos\theta \sin\pr{\frac \pi 6 + \frac \theta 3}}^2 +
            \pr{\cos\theta \cos\pr{\frac \pi 3 + \frac \theta 3}}^2
        }{
            \sin^2\pr{\frac \pi 6 + \frac \theta 3}
            \sin^2\pr{\frac \pi 6 - \frac \theta 3}
        }
    \\
    =&\ \frac{
        \frac{1}{4}\pr{
            \sinp{\frac \pi 6 + \frac{4\theta}{3}}
            + \sinp{\frac \pi 6 - \frac{2\theta}{3}}}^2
        +
        \frac{1}{4}\pr{
            \cosp{\frac{4\theta}{3} + \frac{\pi}{3}} +
            \cosp{\frac{2\theta}{3} - \frac{\pi}{3}}
        }^2
    }{\frac{1}{4}\pr{
        \cos\frac{2\theta}{3} - \cos\frac{\pi}{3}
    }^2}
    \\
    =&\ \frac{\gamma_+^2 + \gamma_-^2
        }{
            \pr{\cos\frac{2\theta}{3} - \frac{1}{2}}^2
        }
    \\
    =&\ \frac{
            -\cos^2\frac{2\theta}{3} - \cos^2\frac{4\theta}{3} + 3 -
            \cos\frac{2\theta}{3} + 2\cos 2\theta
        }{
            \frac{1}{4}\pr{2\cos\frac{2\theta}{3} - 1}^2
        }
    \\
    =&\ \frac{
            - \pr{\cos\frac{2\theta}{3} + \cos\frac{4\theta}{3}}^2 +
            \pr{\cosp{\frac{2\theta}{3}} + \cos{2\theta} } +
            3 - \cos\frac{2\theta}{3} + 2\cos 2\theta
        }{
            \frac{1}{4}\pr{2\cos\frac{2\theta}{3} - 1}^2
        }
    \\
    =&\ 4 \cdot \frac{
            -\pr{
                \pr{2\cos^2\frac{\theta}{3} - 1} +
                \pr{8\cos^4\frac{\theta}{3} - 8\cos^2\frac{\theta}{3} + 1}
            }^2 +
            3 \pr{1 + \cos2\theta}
        }{
            \pr{2\cos\frac{2\theta}{3} - 1}^2
        }
    \\
    =&\ 4 \cdot \frac{
            -4\cos^4\frac{\theta}{3} \pr{4\cos^2\frac{\theta}{3} - 3}^2 +
            6 \pr{2\cos\frac{\theta}{3}\cos\frac{2\theta}{3} - \cos\frac{\theta}{3}}^2
        }{
            \pr{2\cos\frac{2\theta}{3} - 1}^2
        }
    \\=&\ \frac{
            -16\cos^4\frac{\theta}{3} \pr{2\cos\frac{2\theta}{3} - 1}^2 +
            24 \cos^2\frac{\theta}{3} \pr{2\cos\frac{2\theta}{3} - 1}^2
        }{
            \pr{2\cos\frac{2\theta}{3} - 1}^2
        }
    \\
    =& -16\cos^4\frac \theta 3 + 24\cos^2\frac{\theta}{3}
    \\
    =& -16 \pr{\cos^2\frac \theta 3 - \frac 3 4}^2 + 9 ;
\end{align}
onde
\begin{align}
    \gamma_\pm &= \gamma_1 \pm \gamma_2,
    \\
    \gamma_1 &= \frac 1 2 \pr{\cos\frac{2\theta}{3} + \cos\frac{4\theta}{3}}, \quad\text{e}
    \\
    \gamma_2 &= \frac{\sqrt{3}}{2} \pr{\cos\frac{4\theta}{3} - \cos\frac{2\theta}{3}}.
\end{align}

Sabendo que $0 \leq \theta < \pi/2 \implies 1 \geq \cos^2\frac{\theta}{3} > 3/4$,
e que $2\csc^2\pr{\pi/6} = 8$,
conclui-se que para $P = 3$,
\begin{align}
    f_\pi(\theta) \geq -16 \pr{\frac 1 4}^2 + 9 = 8 = 2\csc^2\frac \pi 6.
\end{align}

\subsection{Caso 2 -- P maior ou igual a 4}
Denote $\theta_\pm = \frac{\pi}{2P} \pm \frac{\theta}{P}$.
Deseja-se demonstrar, para $P \geq 4$, que
\begin{align}
    f_\pi(\theta) = \cos^2(\theta) \cdot \frac{
            \sin^2\pr{\theta_+} + \sin^2\pr{\theta_-}
        }{
            \sin^2\pr{\theta_+} \sin^2\pr{\theta_-}
        }
    \geq 2\csc^2\pr{\frac{\pi}{2P}}.
\end{align}
Usando algumas identidades trigonométricas,
é possível reescrever o numerador como
\begin{align}
    \sin^2\pr{\theta_+} + \sin^2\pr{\theta_-}
        =& \pr{ \sin\frac{\pi}{2P}\cos\frac{\theta}{P} +
            \cos\frac{\pi}{2P}\sin\frac{\theta}{P}}^2
        + \notag \\
        &\pr{ \sin\frac{\pi}{2P}\cos\frac{\theta}{P} -
            \cos\frac{\pi}{2P}\sin\frac{\theta}{P}}^2
    \\
    &= 2\sin^2\frac{\pi}{2P} \cos^2\frac{\theta}{P} +
            2\cos^2\frac{\pi}{2P} \sin^2\frac{\theta}{P}
    \\
    &= 2\sin^2\frac{\pi}{2P} \pr{1 -  \sin^2\frac{\theta}{P}} +
            2\cos^2\frac{\pi}{2P} \sin^2\frac{\theta}{P}
    \\
    &= 2\sin^2\frac{\pi}{2P} + 2\sin^2\frac{\theta}{P} \pr{
            \cos^2\frac{\pi}{2P} - \sin^2\frac{\theta}{P}
        }
    \\
    &= 2\sin^2\frac{\pi}{2P} + 2\sin^2\frac{\theta}{P} \pr{
            2\cos^2\frac{\pi}{2P} - 1
        }
    \\
    &= 2\sin^2\frac{\pi}{2P} + 2\sin^2\frac{\theta}{P}\cos\frac{\pi}{P}.
\end{align}
Usando as mesmas identidades trigonométricas,
é possível reescrever a raíz do denominador como
\begin{align}
    \sin\pr{\frac{\pi}{2P} + \frac{\theta}{P}} \sin\pr{\frac{\pi}{2P} - \frac{\theta}{P}}
        &= \frac{1}{2}\pr{\cos\frac{2\theta}{P} - \cos\frac{\pi}{P}}
    \\
    &= \frac{1}{2}\pr{\cos\frac{2\theta}{P} + 1 - 1 - \cos\frac{\pi}{P}}
    \\
    &= \cos^2\frac{\theta}{P} - \cos^2\frac{\pi}{2P}
    \\
    &= - \sin^2\frac{\theta}{P} + \sin^2\frac{\pi}{2P}.
\end{align}
Sendo assim, reescreve-se
\begin{align}
    f_\pi(\theta) &= \frac{
            2\sin^2\frac{\pi}{2P} + 2\sin^2\frac{\theta}{P}\cos\frac{\pi}{P}
        }{
            \pr{\sin^2\frac{\pi}{2P} - \sin^2\frac{\theta}{P}}^2
        } \cos^2\theta.
\end{align}

Para continuar a demonstração,
será necessária a utilização das identidades auxiliares
\begin{align}
    \sin\frac\theta P \geq \frac{2\theta}{\pi} \sin\frac{\pi}{2P} \label{idaux:sin},
\end{align}
e
\begin{align}
    \frac{2\sqrt{2}\pi^2\theta^2 + \pi^4}{\pr{\pi^2 - 4\theta^2}^2}
        \cos^2\theta
    \geq 1 \label{idaux:cos},
\end{align}
demonstradas na seção \ref{sec:anex-identidade-auxiliares}.
Usando as identidades \ref{idaux:sin} e \ref{idaux:cos}; e
notando que $\sin\frac{\pi}{2P} \geq \sin\frac{\theta}{P}$ no domínio e
$\cos\frac\pi P \geq \cos\frac\pi 4$; obtém-se
\begin{align}
    f_\pi(\theta) &\geq \frac{
            2\sin^2\frac{\pi}{2P} + 2 \pr{\frac{2\theta}{\pi} \sin\frac{\pi}{2P}}^2
            \cos\frac{\pi}{4}
        }{
            \pr{\sin^2\frac{\pi}{2P} - \pr{\frac{2\theta}{\pi}  \sin\frac{\pi}{2P}}^2}^2
        } \cos^2\theta
    \\
    &= \frac{2\sin^2\frac{\pi}{2P}}{\sin^4\frac{\pi}{2P}} \cdot \frac{
            1 + 2\sqrt 2\ \theta^2/\pi^2
        }{
            \pr{1 - 4\theta^2/\pi^2}^2
        }\cos^2\theta
    \\
    &= 2\csc^2\frac{\pi}{2P} \cdot \frac{
            \pi^4 + 2\sqrt 2\ \theta^2 \pi^2
        }{
            \pr{\pi^2 - 4\theta^2}^2
        }\cos^2\theta
    \\
    &\geq 2\csc^2\frac{\pi}{2P}.
\end{align}

\subsection{Identidades Auxiliares}
\label{sec:anex-identidade-auxiliares}
Essa Seção dedica-se à demonstração das identidades descritas pelas Eqs.
\ref{idaux:sin} e \ref{idaux:cos}.

\subsubsection{Identidade \ref{idaux:sin}}
A identidade
\begin{align}
    \sin\frac\theta P \geq \frac{2\theta}{\pi}  \sin\frac{\pi}{2P}
\end{align}
ilustra o limite inferior dado por uma reta que passa por baixo de $\sin\frac\theta P$
coincidindo nos pontos $\theta = 0$ e $\theta = \pi/2$:
\begin{align}
    \sin\frac{0}{P} - \frac{0}{\pi}\sin\frac{\pi}{2P} =
    \sin\frac{\pi}{2P} - 1 \cdot \sin\frac{\pi}{2P} = 0 .
\end{align}
O restante da demonstração segue da derivada segunda, já que
\begin{align}
    \frac{\partial^2}{\partial \theta^2} \pr{
        \sin\frac{\theta}{P} - \frac{2}{\pi}\theta \sin\frac{\pi}{2P}
    }
    = -\frac{1}{P^2} \sin\frac{\theta}{P}
    < 0
\end{align}
no domínio $0 \leq \theta < \pi/2$.
Logo, a identidade \ref{idaux:sin} possui concavidade para baixo no domínio e
\begin{align}
    \sin\frac\theta P - \frac{2\theta}{\pi}  \sin\frac{\pi}{2P} > 0
\end{align}
quando $0 < \theta < \pi/2$, concluindo a demonstração.

\subsubsection{Identidade \ref{idaux:cos}}
Deseja-se demonstrar que a identidade
\begin{align}
    \frac{2\sqrt{2}\pi^2\theta^2 + \pi^4}{\pr{\pi^2 - 4\theta^2}^2}
        \cos^2\theta
    \geq 1
\end{align}
é verdadeira no domínio $0 \leq \theta < \pi/2$.
Note que a expressão é igual a 1 quando $\theta = 0$.
Para o restante da demonstração, utiliza-se a identidade.
\begin{align}
    \cos\theta \geq 1 -\frac{\theta^2}{2} + \frac{\theta^4\pr{2\pi^2 - 16}}{\pi^4},
    \label{idaux:cos-aux}
\end{align}
demonstrada a seguir.

Considere o domínio $\theta \in [0, \pi/3]$.
Expandindo $\cos\theta$ em Série de Taylor em torno de $\theta = 0$,
\begin{align}
    \cos\theta = 1 - \frac{\theta^2}{2} + \frac{\theta^4}{24} - \frac{\theta^6}{720} +
        \frac{\theta^8}{40320} + O\pr{\theta^{10}}.
\end{align}
Como o termo de oitava ordem é positivo,
o somatório até o termo de sexta ordem resulta num limite inferior de $\cos\theta$:
\begin{align}
    \cos\theta \geq 1 - \frac{\theta^2}{2} + \frac{\theta^4}{24} - \frac{\theta^6}{720}.
\end{align}
Usando esse limite inferior e
denotando o lado direito de Eq. \ref{idaux:cos-aux} por $\gamma_\theta$,
\begin{align}
    \cos\theta - \gamma_\theta &\geq \frac{\theta^4}{24} - \frac{\theta^6}{720} -
        \frac{\theta^4\pr{2\pi^2 - 16}}{\pi^4}
    \\
    &= 2\theta^4 \pr{ -\frac{x^2}{1440} - \frac{1}{\pi^2} + \frac{1}{48} + \frac{8}{\pi^4} }.
\end{align}
Por conta do termo $2\theta^4$, conclui-se que essa expressão tem raíz em $\theta = 0$.
Já o termo entre parênteses possui raíz no domínio em $\theta \approx 1.537 > \pi/3$.
Avaliando a expressão entre parênteses para $\theta = 0$,
obtém-se um valor positivo $\approx 3.3 \cdot 10^{-3}$.
Logo, a expressão entre parênteses possui concavidade para baixo e
a Eq. \ref{idaux:cos-aux} é verdadeira para $\theta \in [0, \pi/3]$.

Considere o domínio $\theta \in (\pi/3, \pi/2)$.
Expandindo $\cos\theta$ em Série de Taylor em torno de $\theta = \pi/2$,
\begin{align}
    \cos\theta = -\pr{\theta - \frac \pi 2} + \frac{1}{6} \pr{\theta - \frac \pi 2}^3 -
        \frac{1}{120}\pr{\theta - \frac \pi 2}^5 + O\pr{\pr{\theta - \frac \pi 2}^7}.
\end{align}
Note que para $\theta < \pi/2$, o termo de terceira ordem é negativo,
enquanto o termo de quinta ordem é positivo.
Sendo assim, considere o limite inferior dado por
\begin{align}
    \cos\theta \geq -\theta + \frac \pi 2 + \frac{1}{6} \pr{\theta - \frac \pi 2}^3.
\end{align}
Usando esse limite inferior e
denotando o lado direito de Eq. \ref{idaux:cos-aux} por $\gamma_\theta$,
\begin{align}
    \cos\theta - \gamma_\theta &\geq
        -\frac{x^4(2\pi^2 - 16)}{\pi^4} + \frac{\pr{x - \frac\pi 2}^3}{6}) +
        \frac{x^2}{2} - x - 1 + \frac\pi 2.
\end{align}
Essa expressão tem raízes reais nos pontos $\theta \approx 0.88 < \pi/3$
e $\theta = \pi/2$.
Avaliando a expressão no ponto $\pi/3$ obtém-se um valor positivo
$\approx 1.8 \cdot 10^{-2} > 0$.
Logo, a Eq. \ref{idaux:cos-aux} é verdadeira no domínio $\theta \in [0, \pi/2)$.

Para demonstrar a Eq. \ref{idaux:cos}, basta mostrar,
utilizando a Eq. \ref{idaux:cos-aux} que
\begin{align}
    \frac{2\sqrt{2}\pi^2\theta^2 + \pi^4}{\pr{\pi^2 - 4\theta^2}^2}
        \pr{1 -\frac{\theta^2}{2} + \frac{\theta^4\pr{2\pi^2 - 16}}{\pi^4}}^2 - 1
    \geq 0
\end{align}
no domínio.
Usando a propriedade distributiva, obtém-se
\begin{align}
    \frac{x^2}{\pi^6} \pr{ c_4 x^4 + c_2 x^2 + c_0},
    \label{eq:lower-idaux-cos}
\end{align}
onde
\begin{align}
    c_4 &= 8\sqrt{2}\pr{4 - \pi^2} + \frac{\pi^4}{\sqrt 2} ,
    \\
    c_2 &= 2\pi^2\pr{8 - \pi^2\sqrt 2}\pr{1+\sqrt{2}} + \frac{\pi^6}{4},
        \quad \textnormal{e}
    \\
    c_0 &= \pi^4 \pr{8 + 2\sqrt 2 -\pi^2}.
\end{align}
A expressão \ref{eq:lower-idaux-cos} é igual a 0 quando $\theta = 0$.
Resta saber se ela é positiva no restante do domínio -- $0 < \theta < \pi/2$.
Considere apenas o termo entre parênteses.
Esse termo possui raízes nos pontos
$\theta \approx \pm 1.58$ e $\theta \approx \pm 3.89$.
Como $\pi/2 < 1.58$, resta saber se o termo entre parênteses é positivo
quando $-\pi/2 < \theta < \pi/2$.
Avaliando para $\theta = 0$, obtém-se um valor positivo $\approx 93.4$.
Portanto, a identidade da Eq. \ref{idaux:cos} é verdadeira.
\chapter{Projeção da Sobreposição Uniforme das Arestas nos Autovetores}
\label{apendice:proj-arestas-autovecs}

Resta calcular os valores de $\braket{\Sigma_\pm^* | D}$,
$\braket{\Delta_\pm | D}$ e $\braket{\Delta_\pm^* | D}$.
Relembre que $\ket D$ é dado pela Eq. \ref{eq:sup-uni-arestas-geral},
os autovetores são dados pelas Eqs. \ref{eq:evecs-Sigma} e \ref{eq:evecs-Delta}
e seus complexos conjugados.
Então,

\begin{itemize}
    \item Para $\braket{\Sigma_\pm^* | D}$, tem-se que
    \begin{align}
        \braket{\Sigma_\pm^*|D} &= \pr{ \braket{D|\Sigma_\pm^*} }^*
        = \braket{D|\Sigma_\pm}.
    \end{align}
    Logo,
    \begin{align}
        \card{\braket{\Sigma_\pm^*|D}}^2 &= \card{\braket{\Sigma_\pm|D}}^2 .
    \end{align}
    \item Para $\braket{\Delta_\pm | D}$, tem-se que
    \begin{align}
        \braket{\Delta_\pm | D} &= \pr{\braket{D | \Delta_\pm}}^*
        \\
        &= \frac{1}{4\sqrt{N_1 N_2}} \pr{
            \pm e^{\ii\Sigma} \xi + \xi^*
        }^*,
    \end{align}
    onde
    \begin{align}
        \xi &= \sqrt{k_1 k_2} + \ii\sqrt{k_1 (N_2 - k_2)} +
        \ii\sqrt{(N_1 - k_1)k_2} - \sqrt{(N_1 - k_1)(N_2 - k_2)} \\
    &= \paren{\sqrt{k_1} + \ii\sqrt{N_1 - k_1}} \paren{\sqrt{k_2} + \ii \sqrt{N_2 - k_2}}.
    \end{align}
    Logo,
    \begin{align}
        \card{\braket{\Delta_\pm | E}}^2 &= \frac{1}{16 N_1 N_2}
        \pr{\pm e^{\ii\Sigma} \xi + \xi^*} \pr{\pm e^{\ii\Sigma} \xi + \xi^*}^*
        \\
        &= \frac{1}{16 N_1 N_2} \pr{2\xi\xi^* \pm 2 \Re\pr{e^{\ii\Sigma}\xi^2}}
        \\
        &= \frac{1}{8}\pr{ 1 \pm \Re\pr{\frac{e^{\ii\Sigma}\xi^2}{N_1 N_2}} },
    \end{align}
    e usando
    \begin{align}
        \frac{e^{\ii\Sigma}\xi^2}{N_1 N_2} &= e^{\ii\Sigma}
        \paren{ \frac{2k_1 - N_1 + 2\ii\sqrt{N_1 - k_1}}{N_1} }
        \paren{ \frac{2k_2 - N_2 + 2\ii\sqrt{N_2 - k_2}}{N_2} }
        \\
        &= e^{\ii\Sigma} \paren{-\cos\theta_1 +\ii\sin\theta_1} \paren{-\cos\theta_2 +\ii\sin\theta_2}
        \\
        &= e^{\ii\Sigma} e^{-\ii\theta_1} e^{-\ii\theta_2}
        \\
        &= e^{-\ii\Sigma}
    \end{align}
    resulta em
    \begin{align}
        \card{\braket{\Delta_\pm | D}}^2 = \frac{1}{8} \pr{1 \pm \cos\Sigma}.
    \end{align}
    \item Para $\braket{\Delta_\pm^* | E}$, tem-se que
    \begin{align}
        \braket{\Delta_\pm^*|D} &= \pr{ \braket{D|\Delta_\pm^*} }^*
        = \braket{E|\Delta_\pm}.
    \end{align}
    Logo,
    \begin{align}
        \card{\braket{\Delta_\pm^*|D}}^2 &= \card{\braket{\Delta_\pm|D}}^2 .
    \end{align}
\end{itemize}

Juntando todos esses resultados e usando as identidades trigonométricas
$1 + \cos\theta = 2\cos^2\paren{\theta/2}$ e $1 - \cos\theta = 2\sin^2\paren{\theta/2}$,
obtém-se a Tabela \ref{tab:prob-autovals}.


\end{apendicesenv}


\begin{anexosenv}






\end{anexosenv}

\phantompart
\printindex

\end{document}